\documentclass{aa}  
\usepackage[usenames]{color}
\usepackage{graphicx}
\usepackage{txfonts}
\usepackage{color}
\usepackage{footnote}
\def\herschel{{\it Herschel}}
\def\spitzer{{\it Spitzer}}
\def\hubble{{\it HST}}
\def\kin{{\it Kinemetry}}
\def\cifull{[C\,{\sc i}] $^3P_2 \longrightarrow ^3P_1$}
\def\cishort{[C\,{\sc i}] 2--1}
\def\co7-6{CO J:7--6}
\def\h2ofull{H$_2$O 2$_{11} - 2_{02}$}
\def\water{H$_2$O}
\def\alphaco{$\alpha_{\mathrm CO}$}

\def\350mic{350\,$\mu$m}
\def\micron{~$\mu$m}

\def\msun{\hbox{$\rm ~M_{\odot}$}}

\def\arcdeg{$^{\circ}$}
\def\H0{{\rm ~km\,s^{-1}\,Mpc^{-1}}}

\def\kms{km\,s$^{-1}$}
\def\src{HATLAS\,J084933.4+021443}

\begin{document}

   \title{A resolved warm/dense gas Schmidt-Kennicutt relationship in a binary HyLIRG at $z=2.41$}

   \subtitle{ALMA observations of H-ATLAS\,J084933.4+021443}

   \author{Jonathan~S.~G\'omez
          \inst{1,2}
          \and 
          Hugo~Messias
          \inst{3}
          \and
          Neil~M.~Nagar
          \inst{1}
          Gustavo~Orellana
          \inst{1,4}
          \and
          R.\,J.~Ivison 
          \inst{5,6}
          \and
          Paul~van~der~Werf
          \inst{7}        
          }

		\institute{Astronomy Department, Universidad de Concepci\'on, Concepci\'on, Chile\\
                  \email{jonathangomez@udec.cl}
        		  \and
				  Departamento Astronom\'ia y Astrof\'isica, Pontificia Universidad Cat\'olica de Chile, Av. Vicu\~na Mackenna 4860, Stgo., Chile
        		  \and
        		  Joint ALMA Observatory, Alonso de C\'ordova 3107, Vitacura 763-0355, Santiago, Chile
				  \and
				  Instituto de F\'isica y Astronom\'ia, Universidad de Valpara\'iso, Avda. Gran Breta\~na 1111, Valpara\'iso, Chile
        		  \and
                  European Southern Observatory, Karl-Schwarzschild-Stra{\ss}e-2, 85748 Garching bei M{\"u}nchen, Germany
                  \and
                  Institute for Astronomy, University of Edinburgh, Royal Observatory, Blackford Hill, Edinburgh EH9 3HJ, UK
				  \and
				  Leiden Observatory, Leiden University, P.O. Box 9513, NL-2300 RA Leiden, The Netherlands}       

   \date{Received June 5, 2018; accepted XXX 2018\\
   Submitted to Astronomy \& Astrophysics}

 
  \abstract
  {}  
   {Hyperluminous infrared galaxies (HyLIRGs; star-formation rates of up to $\approx 1000$ \msun\,yr$^{-1}$) -- while 
   rare -- provide
   crucial `long lever arm' constraints on galaxy evolution. \src, a $z=2.41$ binary HyLIRG with at least two additional
   luminous companion galaxies, is thus an optimal test-ground for studies of star formation and galaxy evolution during `cosmic noon'.}
   {We have used ALMA to obtain \textit{resolved} imaging and kinematics of atomic and molecular emission lines, and
    rest-frame 340- to 1160~GHz continuum emission, for the known luminous component galaxies in H-ATLAS\,J084933.4+021443: W, T, M, C.}
   {All four component galaxies are spatially ($\sim$0\farcs3 or 2.5\,kpc) resolved in CO J:7--6, 
    \cishort, \water\  and the millimetre (mm) to sub-mm continuum.  Rotation-dominated gas kinematics is comfirmed in W and T. 
    The significant extension to component T, in gas and continuum, along its kinematic minor axis, is attributable to its lensing magnification. Spatially 
    resolved sub-mm spectral energy distributions (SEDs) reveal that component W is well fit with greybody emission from dust 
    at a single temperature over the  full extent of the galaxy, despite it containing a powerful AGN, while component T requires an additional component of 
    hotter nuclear dust and additional sources of emission in the mm.
    We confirm that \cishort\ can be used as a rough tracer of warm/dense molecular gas in extreme systems, though
    the \cishort/CO luminosity ratio increases sub-linearly. 
    We obtain an exquisite and unprecedented  resolved (2.5-kpc-scale) 'warm/dense molecular gas' Schmidt-Kennicutt (SK) relationship for 
    components 
    W and T.  Gas exhaustion times for all apertures in W (T) are 1--4\,Gyr (0.5--2\,Gyr). 
    Both W and T follow a  resolved `warm/dense gas' SK relationship with power law $n\sim 1.7$, significantly 
    steeper than the $n\sim 1$ found previously via `cold' molecular gas in nearby `normal' star-forming galaxies.}
   {}


   \maketitle
\section{Introduction}
\label{sectintro}

Sub-millimetre surveys have greatly advanced our understanding of galaxy evolution by
uncovering a population of heavily dust-obscured galaxies at high redshift, the so-called
sub-millimetre galaxies (SMGs).  Although the first of these to be
discovered \citep{ivison98} was a
hyperluminous infrared (IR) galaxy (HyLIRG; $L_{\rm IR}\, \geq \, 10^{13}\, $L$_\odot$,
where the IR luminosity is measured across $\lambda_{\mathrm{rest}} = 8$--$1000$\micron),
the vast majority of the numerous SMGs uncovered thereafter were
ultraluminous infrared galaxies (ULIRGs; $L_{\rm IR}\, \geq \, 10^{12}$\,L$_\odot$, forming stars at $\geq 100$\msun\,yr$^{-1}$ -- see, e.g., \citealt{bla02, casey14,
fudaet17}). The importance of SMGs in studies of galaxy formation 
and evolution has been underlined by {\it Spitzer} and {\it Herschel Space Observatory}
results which show that SMGs contribute significantly to the total amount of star
formation in the early Universe (e.g.\ \citet{maget09}; \citet{gle10}). 

The IR luminosity of a HyLIRG implies an extreme star-formation rate \citep{ken98}, 
SFR$\, \geq 10^3$ M$_{\odot}$ yr$^{-1}$, in the absence of a significant contribution
to $L_{\rm IR}$ from an active galactic nucleus (AGN) and for a
`normal' \citep{chab03} stellar initial
mass function (IMF), although \citet{zhang18}
showed recently that the IMF in dusty starbursts must be top heavy.
Naively, this suggests starburst lifetimes of only 
$\sim 100$ Myr, unless star formation migrates around an extended gas reservoir.
While rare and extreme, HyLIRGs are excellent laboratories with which to confront 
the most recent hydrodynamic simulations \citep{bahe17, baret17, benoet17, suket17, charet17, lagos18} of isolated and merging galaxies (e.g.\ \citet{hayet11}),
test star-formation `laws', and the effect of feedback, driven by AGN (e.g.\ \citet{bouet10,davet11,davet12,cicet14})
and/or intense star formation (e.g., \citet{sheos12,lillet13}) on the evolution of a galaxy.

Of specific relevance to the results presented here are the resolved Schmidt-Kennicutt (SK) relationship \citep{schmidt59,ken89},
i.e.\ the power-law relationship between the surface density of gas and star formation 
($\mathrm{\Sigma_{SFR} \propto \Sigma_{H_2}^N}$) on $\leq$kpc scales within
a galaxy, and the use of the sub-mm C\,{\sc i} lines instead of CO \citep{flolau85, dosol98, yang10, bolet13, carwal13, rodri14}
or HCN \citep{gao97, gaosol04, shiet17, oteet17}
to estimate the total molecular gas of a galaxy \citep{walet11, israet15, qianet17}.
While the galaxy-integrated SK relation, i.e.\
the surface densities of the galaxy-integrated star formation 
and cold molecular gas mass (hereafter the galaxy-integrated SK relationship),
has been extensively studied -- e.g.\ \citet{youet86}, \citet{solsage88}, \citet{buatet89}, \citet{gaosol04},
\citet{bouet07}, \citet{krutho07}, \citet{dadet10}, \citet{genet10},
\citet{ken_ev12}, \citet{chelet16} -- the resolved SK relationship
(at $\lesssim$1-kpc resolution within galaxies; hereafter kpc-scale SK) 
has been constrained in relatively few nearby \citep{wobli02,ken07,biget08,
leret08,krumet09,biget10,biget11,boqet11,momet13,royet15}  
and  high-redshift \citep{freet13,thomet15} galaxies. 
\citet{biget08} and \citet{biget11} find that 1~kpc-scale resolved
SK relationship in their
`normal' star-forming galaxies is consistent with an exponent of $N\sim 1$.
In contrast, \citet{momet13} find a super-linear slope ($N=1.3$,
and even up to 1.8) for the 
resolved SK relationship in their sample of nearby spirals.
A super-linear slope ($N=1.5$) was also
found by \citet{royet15} in H\,{\sc i}-dominated regions of nearby spiral and dwarf galaxies.

The \cifull\ and [C\,{\sc i}] $^3P_1 \longrightarrow ^3P_0$ lines -- both are required
to determine the excitation temperature of C\,{\sc i} \citep{stuet97}
-- can be used to robustly determine the 
mass of the neutral carbon \citep{stuet97,weiet03,weiet05}.
In ultraviolet (UV) or cosmic ray-dominated regions, 
the (typically optically thin) C\,{\sc i} emission lines are expected to be primarily produced in the dissociated surfaces of molecular 
clouds, though observations show that they are present throughout the cloud \citep[e.g.][]{gloet15}
and have been argued to be a  better tracer of total molecular gas mass than 
the CO line \citep{weiet03,weiet05,biset15,glocla16}. Our detections of both CO and \cishort\ (even if only the
higher of the two C\,{\sc i} sub-mm lines) in
four independent galaxies (spatially resolved in two) allow us to compare both species as molecular gas tracers.
Molecular lines from e.g.\ H$_2$O, HCN and CS have a much higher critical density 
and therefore probe the dense molecular star-forming phase. 
\citet{omoet13} find a relation between the far-infrared (FIR) and
H$_2$O luminosities for a sample of high-redshift starburst galaxies. 
The H$_2$O detections for this sample are all associated with underlying
FIR emission, implying that the H$_2$O emission traces star-forming regions. 
However, the H$_2$O molecules can also be excited in the dissipation of supersonic turbulence in molecular gas 
or by slow shocks (e.g.\ \citealt{flopin10}). 
In the case of purely shock-excited H$_2$O, it is unlikely that underlying FIR
emission would be detected in regions of strong H$_2$O emission (e.g.\ \citealt{goiet15,andet13}).



The HyLIRG, HATLAS\,J084933.4+021443 (hereafter HATLAS\,J084933) at $z=2.41$, though identified only in 2012,
is now one of the few well-studied HyLIRGs 
(\citet{iviet13}; hereafter I13).  It has a relatively brief literature history: 
identified in the \herschel\ ATLAS imaging survey \citep{ealet10} as a \350mic\
peaker, then
CO J:1--0 spectroscopy with the Greenbank Telescope (GBT) constrained its redshift to 2.410
\citep{harret12}.
I13 presented a detailed study of the molecular gas and rest-frame sub-mm emission
of this source using the Jansky Very Large Array (CO J:1--0), CARMA (CO J:3--2), IRAM PdBI (CO J:4--3) together with continuum imaging from the SMA, \herschel, \spitzer, VISTA, and \hubble,
and optical spectroscopy from Keck. They determined that HATLAS\,J084933 comprises
at least four starburst galaxies scattered across a $\sim 100$-kpc region at $z=2.41$. The two brightest 
galaxies, dubbed W and T, both HyLIRGs in their own right, are separated by $\sim 85$~kpc on the sky. Of these, T is
amplified modestly by a foreground galaxy with a lensing magnification of $\sim 2\times$. The molecular gas
reservoirs of W and T, each $\sim 3$~kpc in size, are rotation-dominated, and counter-rotate. 
These two components have CO line strengths and widths typical of the brightest SMGs and
lie among SMGs in the `global' (i.e.\ galaxy-wide) equivalent of the Schmidt-Kennicutt relationship.
Two other components, dubbed M and C, are ULIRGs, though they lie relatively close to the HyLIRG cutoff. 
Their estimated molecular gas masses (but not their estimated dynamical masses) are almost a factor
$10\times$ lower than those of W and T.

With the goal of a more comprehensive study of the physics in HATLAS\,J084933 -- particularly with the goal of probing the resolved Schmidt-Kennicutt relationship at the highest SFRs and gas densities -- but also of exploring the
resolved kinematics (rotation vs.\ dispersion), and searching for evidence of outflows via
kinematics and P~Cygni profiles (e.g.\ molecular outflows as seen by \citealt{cicet14}, \citealt{feruet10}), we have obtained new observations using the Atacama Large millimetre/sub-millimetre Array
(ALMA) of this
source. These new observations include resolved imaging  
of the CO J:3--2 and CO J:7--6 emission lines (tracers of the dense molecular gas), the 
\cifull\ fine-structure line ($\nu_{\mathrm{rest}} = 809.34$~GHz; 
hereafter \cishort), and the \h2ofull\ emission line 
($\nu_{\mathrm{rest}} = 752.03$~GHz, hereafter H$_2$O). 
These lines are detected and resolved towards all four principal components of 
HATLAS\,J084933:
W, T, C, and M, thus allowing a significant improvement in our understanding of the source
as compared to I13. 
We additionally obtained resolved continuum
imaging at rest-frame 341~GHz,  750~GHz,  808~GHz, and 1160~GHz, 
which can be used as a tracer of the resolved star-formation rate and dust mass, amongst other quantities.

In this work we present the first set of results based on the new ALMA data. A forthcoming paper will
address the resolved dust properties and the resolved (atomic and molecular) gas to dust ratios.
In Section 2 we detail our ALMA sub-mm  observations. The results of these observations are presented in 
Section 3.  In Section 4 we analyse and discuss the results and present our  conclusions.  
We adopt a cosmology with $H_0$ = 71 km s$^{-1}$ Mpc$^{-1}$ , $\Omega_m$ = 0.27, and 
$\Omega_\Lambda$ = 0.73, so 1\arcsec\ is equivalent to 8.25 kpc at $z = 2.41$.

\section{Observations and Data Processing}

HATLAS\,J084933 was observed by ALMA as part of Project 2012.1.00463.S (P.I.: G.~Orellana) 
during ALMA Cycle 2. Four of the five
approved "science goals"  of the project were observed over the period 2015 August to September,
as detailed below. The fifth science goal, designed to observe the CO J:9--8 line (of both $^{12}$CO and 
$^{13}$CO) plus nearby water vapour and OH+ lines, was not observed.

We used the Common Astronomy Software Applications (CASA 4.4.0) software for all data calibration and
imaging steps, thus obtaining the final data cubes, continuum maps, emission-line only cubes, 
and moment 0, 1, and 2 maps of the individual emission lines. Further processing and analysis
was performed with our Interactive Data Language (IDL) and Python codes.

\subsection{CO J:3--2 and rest-frame 341~GHz (880\micron) continuum imaging}

Approximately 35\,min of on-source integration was obtained in Band 3 with thirty four 12-meter 
during 2015 August 30.
The Band 6 receivers were to the CO J:3--2 line ($\nu_{rest}=345.795991$ GHz \citep{mornor94}, 
redshifted to $\nu_{obs}= 101.465$ GHz) in one of the four spectral windows (SPWs). The second SPW in the 
upper sideband (USB) was set to  partially cover the CS J:7--6 line and the two SPWs in 
the lower sideband (LSB) did not cover any known strong lines and were thus intended to detect
continuum emission.
To maximise sensitivity, the ALMA correlator was used
in `continuum' or Time Division Multiplexing (TDM) mode, which gave a spacing of
$\sim 46$ km s$^{-1}$ per channel and a total velocity coverage of $\sim 5000$ \kms\ per baseband.

Using a Briggs weighting with a robust parameter of +0.5, and the intrinsic
spectral resolution, the CO line was imaged
with a synthesised (FWHM) beam of $1\farcs13 \times 0\farcs51$ with a position angle 
(P.A.) of 115\arcdeg.
The r.m.s.\ noise was $\sigma = 0.27$ mJy beam$^{-1}$ in each 46 km s$^{-1}$ channel.
Line-free channels were used to create a continuum map at observed-frame $\sim 100$~GHz (
rest-frame $\sim 341$~GHz); here the
synthesised beam was $1\farcs 23 \times 0\farcs 56$ beam FWHM (P.A., 116\arcdeg) and the
r.m.s.\ noise was $\sigma = 0.08$ mJy beam$^{-1}$.

\subsection{CO J:7--6, \cifull, and rest-frame 808~GHz (370\micron) continuum imaging}

Approximately 21 min of on-source integration time was obtained in Band 6 with thirty four 12-meter 
antennas in 2015 September. The last scan of the observation was not on the phase-calibrator, but rather on the target. Nevertheless, the target is bright enough to enable self-calibration, hence making that scan usable. 
Two overlapping spectral windows in the USB were tuned  to contiguously cover both
the CO J:7--6 line ($\nu_{rest}=806.651801$ GHz; \citet{mornor94},
redshifted to an observed frequency of $\sim 236$~GHz),
and the \cifull\ (\cishort)  line ($\nu_{rest}=809.34197$ GHz; \citet{mullet01}, redshifted to
an observed frequency of $\sim 237$~GHz). The other two spectral windows were set to cover
a water vapour line and neighbouring continuum (see next subsection). Once more, for maximum sensitivity, the
correlator was set to TDM mode resulting in a spectral spacing of 19.7 km s$^{-1}$ per channel
and a total velocity coverage of $\sim 2500$ \kms\ per SPW for all SPWs.

Image cubes of the CO J:7--6 and \cishort\ line were made at the intrinsic spectral resolution
(19.7 km s$^{-1}$ channels) and 
using natural weighting, resulting in a synthesised (FWHM) beam of $0\farcs 27 \times 0\farcs 24$ 
at P.A., 3.9\arcdeg. 
The r.m.s.\ noise in each channel was 0.31 mJy beam$^{-1}$ for the CO J:7--6 and \cishort\ line cubes. 
Line-free channels were used to make a map of the continuum emission near observed frequency
237~GHz (corresponding to rest frequency 808~GHz).
Here, maps made with Briggs robust parameter of +0.5 resulted 
in a synthesised beam of $0\farcs 26 \times 0\farcs 24$ at P.A., 2.9\arcdeg and an r.m.s.\ noise
of $\sigma = 0.24$ mJy beam$^{-1}$.

\subsection{\h2ofull\ Imaging, and 750~GHz (400\micron) continuum imaging}

In the science goal described in the previous Section (in which the USB was used to cover the
CO J:7--6 and \cishort\ line), the two LSB SPWs were set to 
220.538 GHz, the redshifted frequency of the \h2ofull\ (H$_2$O) line 
($\nu_{rest}=752.033143$ GHz; \citet{dioet80}), and the neighbouring continuum. The spectral spacing
was 21.2 km s$^{-1}$ per channel. 
The image cube of the \h2ofull\ line, made with natural weighting,  had an r.m.s.\ noise of 
0.34 mJy beam$^{-1}$ per native channel 
and a synthesised (FWHM) beam size of $0\farcs 29 \times 0\farcs 26$ at P.A., 13.7\arcdeg. 
Line-free channels in these two LSB SPWs were used to make a map of the continuum emission near 
observed frequency, 220~GHz (corresponding to rest-frame frequency, 750~GHz).
Here, maps made with Briggs robust parameter of +0.5 resulted in 
a synthesised beam of $0\farcs28 \times 0\farcs26$ at P.A., 179\arcdeg and an r.m.s.\ noise
of $\sigma = 0.17$ $\mu$Jy beam$^{-1}$.

\subsection{Observed frame 340~GHz (rest-frame 1160 GHz or 260\micron) Continuum Imaging}

Approximately 53 min of on-source integration was obtained with 38 working 12-meter antennas 
during 2015 June 5. The spectral set-up was designed to detect four different water vapour lines,
one in each SPW, none of which were detected. 
This dataset was thus used only to obtain a continuum image near
observed-frame 340~GHz (corresponding to rest frame 1160~GHz).
The receivers were tuned such that the USB was centred at 1160 GHz (260\micron). 
A weighting scheme with Briggs robust parameter of +0.5 resulted in a synthesised beam of
$0\farcs 34 \times 0\farcs 3$ (P.A., 83.6\arcdeg) and a noise level of 
$\sigma = 0.26$ mJy beam$^{-1}$ in the continuum image.

\section{Results}
\label{secresults}

Continuum emission at all observed frequencies is strongly 
detected and resolved towards galaxy components  W and T and relatively weakly detected in components C and M. 
The emission lines CO J:3--2, CO J:7--6, and \cishort\ are also strongly detected and resolved in W an T 
and weakly (but reliably) detected and resolved in components C and M. 
Further, the  H$_2$O line is detected, and spatially resolved, in components W an T.  No other emission
lines or components (i.e.\ galaxies) were reliably detected in our continuum or line imaging.

In the following subsections, we first present galaxy-integrated measurements 
(both continuum and emission line)
for each of the four components, then the resolved continuum and emission line maps,  
the resolved galaxy kinematics, 
and finally other galaxy resolved quantities including resolved emission line ratios and a 
resolved 'warm/dense gas' Schmidt-Kennicutt relationship.

%

\subsection{Galaxy-integrated properties: line profiles and ratios, continuum emission, and other measured quantities} 
\label{secglobal}

	\begin{figure*}[ht]
	\centering
    \hspace{-13.1cm} \includegraphics[scale=0.44]{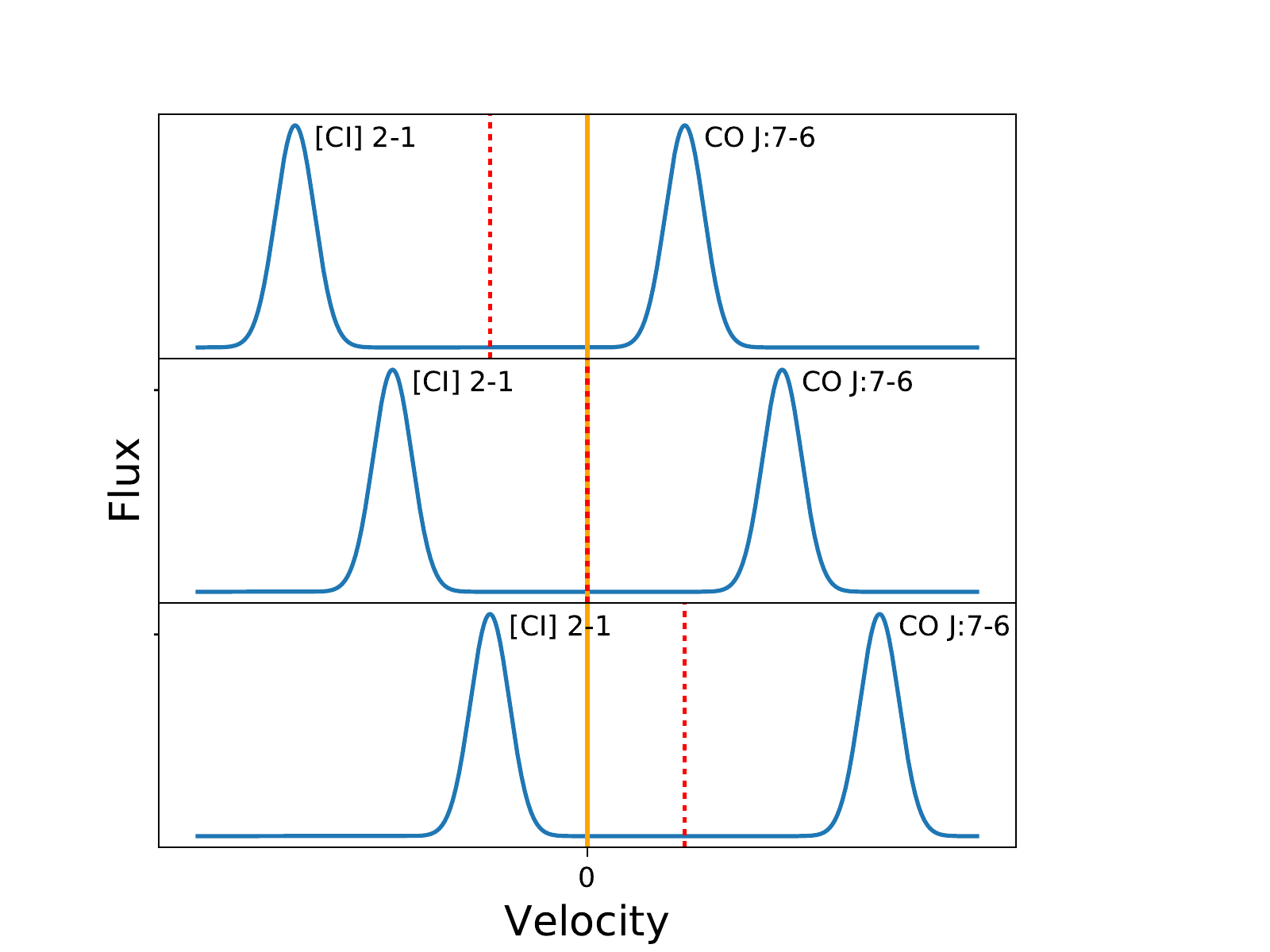}\\ \vspace{-5.4cm}
	\hspace{-1.63cm} \includegraphics[scale=0.44]{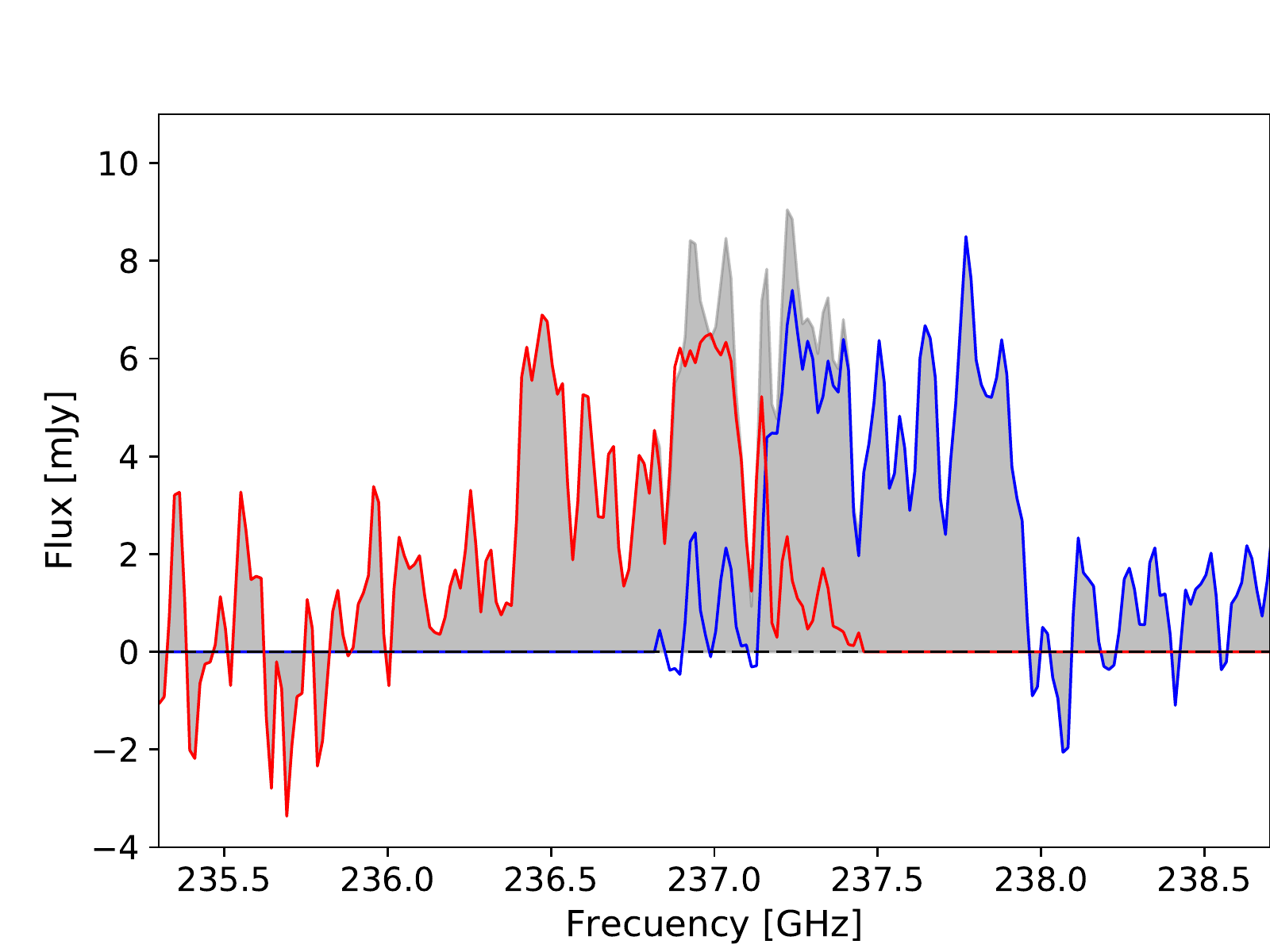}\\ \vspace{-5.4cm}
	\hspace*{11.9cm}  \includegraphics[scale=0.44]{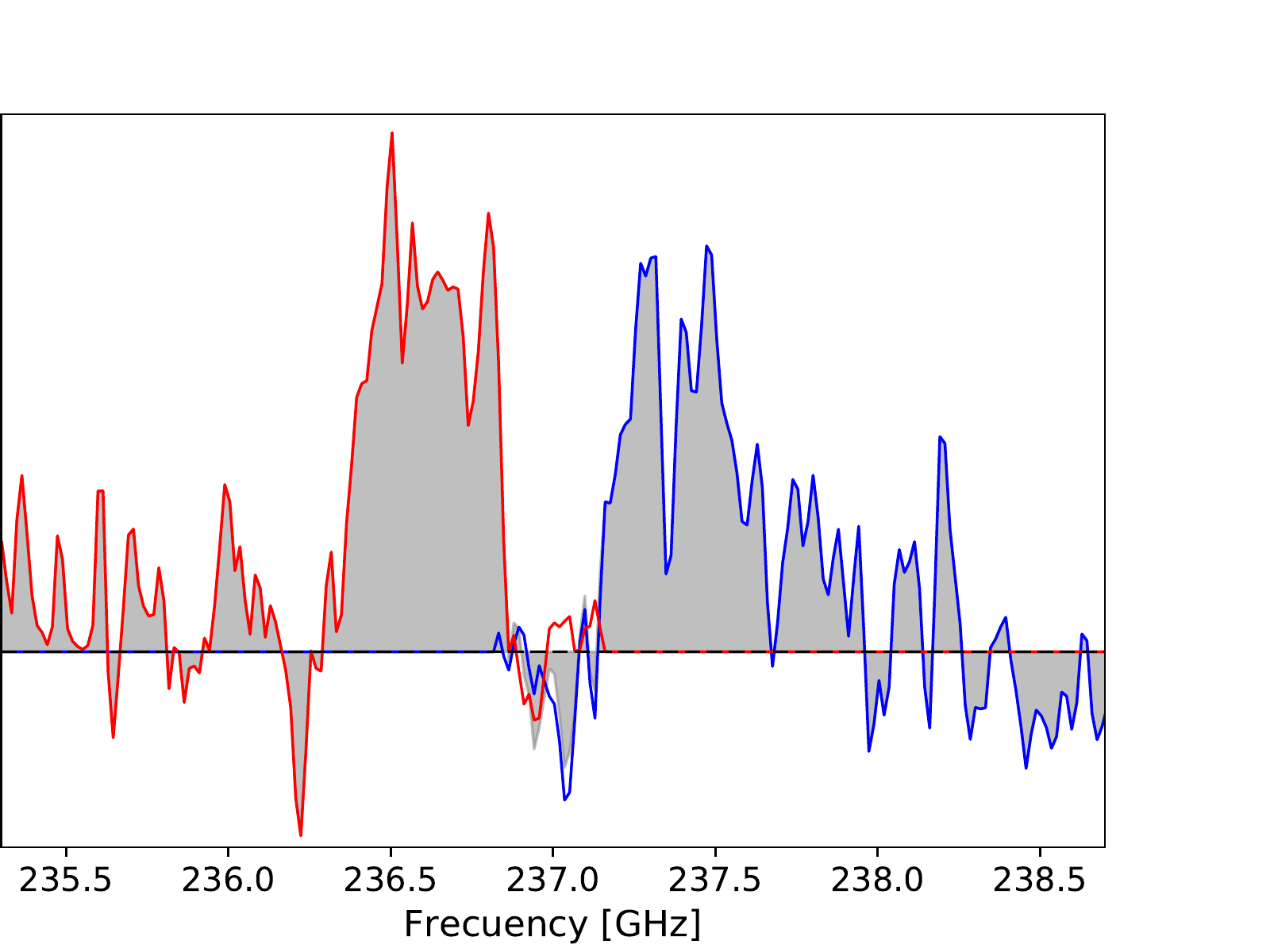}\\ 
	\caption{Left panel: an illustration of the method used to separate the \co7-6\ and \cishort\ lines
        in individual channels of the datacube (see text). The dashed red line shows our choice for the dividing frequency
        (based on the fitted gas rotation model) and the solid yellow line shows the systemic velocity of the galaxy 
        (see Section~\ref{secglobal}). 
        Middle and right panels: 
        the observed (grey filled regions) and 
        rotation-model-separated profiles (see text) of the 
        CO J:7--6 (red) and \cishort\ (blue) emission lines in component W (middle panel) 
        and component T (right panel). 
	}
	\label{figcociprofile}
	\end{figure*}

The emission lines of CO J:3--2, CO J:7--6, \cifull\ and \h2ofull\ are detected in the four previously
identified galaxies of the HATLAS\,J084933 system (W, T, M and C); of these, the latter has the weakest lines. 
The observed galaxy-integrated spectra of CO J:7--6 and \cishort\ in components W and T are shown 
in Fig.~\ref{figcociprofile}.
Unfortunately, the relatively close rest frequencies of CO J:7--6 and \cishort\ 
(rest-frame $\Delta \nu$ = 2.7~GHz or $\Delta$ v = 1050 \kms) and the relatively broad 
velocity profile of the lines in component W results in a super-position of these two lines
in the galaxy-integrated spectra of W; a similar (albeit less blended) situation is also seen in 
component T (see Fig.~\ref{figcociprofile}). 

Given that CO J:7--6 and \cishort\ show relatively similar ordered velocity fields, 
and our argument that these velocity fields are best interpreted as rotation (see Section~\ref{sectrot}), 
we are able to separate the emission from the two lines.  
That is, the two lines can be cleanly separated in individual (19.7 \kms) velocity channels of our data cube: the
width of each line is $\sim$100--250~\kms\ in a given velocity channel while the separation between the lines
is $\sim$1000 \kms. This clear separation between the two lines in individual velocity channels was visually
verified in the data cube.
We thus use our rotation-only velocity models (see Section~\ref{sectrot}) to predict, for each
spatial pixel, 
the velocity (or channel) which would most cleanly separate the CO J:7--6 and \cishort\ lines.
The left panel of Fig.~\ref{figcociprofile} illustrates our choice of the dividing frequency
(dashed red line) for a given systemic ruled by the rotation of the galaxy, blue- or red-shifted spatial pixel of the galaxy. 
The rotation-model-separated profiles are shown
with blue and red lines in the middle and right panels of Fig.~\ref{figcociprofile} for components W and
T, respectively. Note that this separation is valid only if all
the gas participates in rotation; a weak outflow component with velocities $\gtrsim$ 500 \kms\ may not be
validly separated by our method.  

The galaxy-integrated profiles of all detected emission lines in W and T are shown in the right-most
panels of Figs~\ref{figmomw} and \ref{figmomt}, respectively, and those corresponding to 
components C and M (where de-blending was not necessary)
are shown in the right-most panels of Fig.~\ref{figmomcm}.

   \begin{figure}[h]
   \vspace{0.7cm}
   \hspace{-0.5cm}
   \includegraphics[scale=0.63]{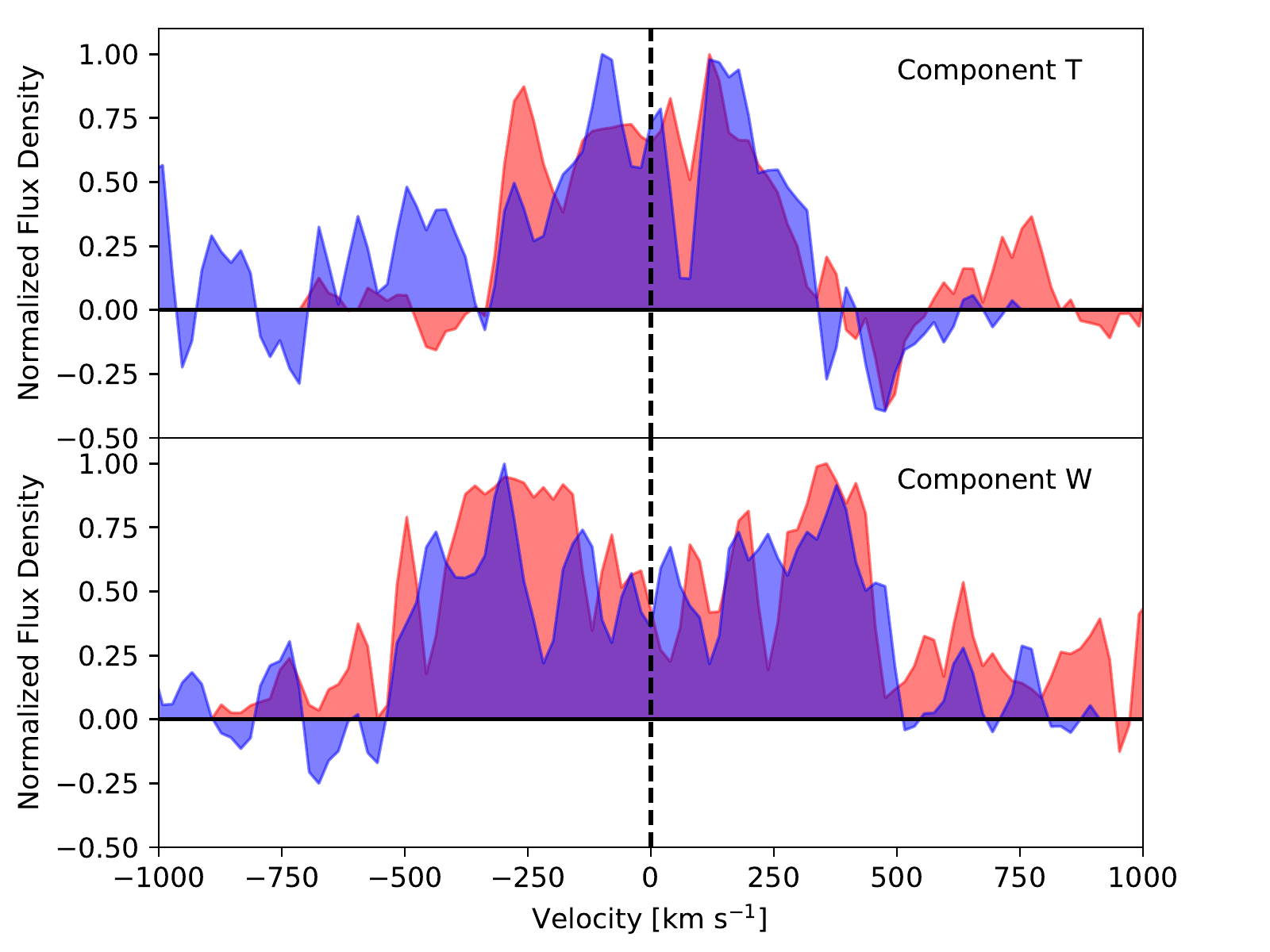}
   \caption{A comparison of the galaxy-integrated normalized spectral profiles 
   of the  CO J:7--6 
   (red) and [C\,{\sc i}] 2--1 (light blue) lines for component T (top panel) and W (bottom panel). 
   Overlap regions of the two spectra appear darker blue.
   The dashed black line shows our adopted zero velocity. }
   \label{figcompa_profile-tw}
   \end{figure}

A direct comparison of the galaxy integrated line profiles of CO J:7--6 and \cishort\ in components
W and T are shown in
Fig.~\ref{figcompa_profile-tw}. In W, both profiles show a double-horned
structure, either a result of the global galaxy kinematics or  attributable to higher optical depths at velocities 
closer to systemic. 
Additionally, while the redshifted emission
from the two lines are relatively well matched (in a normalised sense), the CO J:7--6 emission 
shows a significant excess over \cishort\ at blue-shifted velocities. Both of the above points, at a less
significant level, could also be claimed  to be true in  component T. In component T we also call attention 
to the potential absorption feature at $\sim$ 400 \kms\ seen in both species; this is discussed in more detail in
Section~\ref{secoutflows}.


	\begin{table*}[h]
	\caption{HATLAS J084933.4+021443: line fluxes and luminosities, observed continuum fluxes, and derived properties}
	\centering
	\begin{tabular}{l c c c c c}
	\hline\hline
	Property                   & unit of measurement          &        W        &        T       &        C       &       M         \\
	\hline\hline
	$S_{\mathrm{CO\, J:3--2}}$  &[Jy km s$^{-1}$]              & 4.29$\pm$0.53   & 4.22$\pm$0.46  & 0.95$\pm$0.28  & 0.63$\pm$0.16    \\
	$S_{\mathrm{CO\, J:7--6}}$  &[Jy km s$^{-1}$]              & 2.704$\pm$0.73  & 2.72$\pm$0.51  & 0.51$\pm$0.36  & 0.46$\pm$0.27    \\
	$S_{\mathrm{[C\,{\sc i}]\, 2-1}}$  &[Jy km s$^{-1}$]              & 2.07$\pm$0.82   & 1.33$\pm$0.86  & 0.21$\pm$0.19  & -                \\
	$S_{\mathrm{H_2O}}$        &[Jy km s$^{-1}$]              & 0.69$\pm$0.23   & 0.87$\pm$0.09  & -              & -                \\
	\hline
	$L'_\mathrm{CO\, J:3--2}$   &[$10^9$ K km s$^{-1}$ pc$^2$] & 134$\pm$10      & 133$\pm$10     & 29.9$\pm$8.7   & 19.8$\pm$4.9     \\
	$L'_\mathrm{CO\, J:7--6}$   &[$10^9$ K km s$^{-1}$ pc$^2$] & 15.5$\pm$2.2    & 15.6$\pm$1.4   & 2.93$\pm$2.0   & 2.64$\pm$1.5     \\
	$L'_\mathrm{[C\,{\sc i}] 2-1}$     &[$10^9$ K km s$^{-1}$ pc$^2$] & 11.8$\pm$2.6    & 7.6$\pm$1.9    & 1.2$\pm$1.0    & -                \\
	$L'_\mathrm{H_2O}$         &[$10^9$ K km s$^{-1}$ pc$^2$] & 4.58$\pm$2.2    & 5.77$\pm$1.4   & -              & -                \\
	\hline
	$\mathrm{S_{\nu obs 100GHz}}$  &[mJy]                         & 0.18 $\pm$0.08  & 0.11$\pm$0.08  & -              & -                \\
	$\mathrm{S_{\nu obs 220GHz}}$  &[mJy]                         & 5.14 $\pm$0.17  & 3.63$\pm$0.17  & 0.41$\pm$0.17  & 0.36$\pm$0.17    \\
	$\mathrm{S_{\nu obs 237GHz}}$  &[mJy]                         & 5.79 $\pm$0.24  & 4.39$\pm$0.24  & 0.38$\pm$0.24  & 0.51$\pm$0.24    \\
	$\mathrm{S_{\nu obs 340GHz}}$  &[mJy]                         & 22.9 $\pm$0.26  & 21.4$\pm$0.55  & 2.59$\pm$0.55  & 1.8$\pm$0.5      \\
	\hline
	log$M_\mathrm{dyn}$        &[$M_\odot$]                   & 11.86$\pm$0.7   & 11.51$\pm$0.9  & 10.55$\pm$0.13 & 10.22$\pm$0.16   \\
	
    (T$_{\rm ex}$=30~K): log$M_\mathrm{CI}$ &[$M_\odot$]      & 8.41            & 8.22           &  7.42          & -                \\
    (T$_{\rm ex}$=40~K): log$M_\mathrm{CI}$ &[$M_\odot$]      & 8.28            & 8.1            &   7.29         & -                \\
 	log$M_\mathrm{\rm H_2}$  (from CI; T$_{\rm ex}$=40~K) &[$M_\odot$]           & 12.0            & 11.8           & 11.0           & -                \\
    $\mathrm{T_{dust}^a}$      &[K]                           & 39.8$\pm$1.0    & 36.2$\pm$1.1   & -              & -                \\
	$\mathrm{log\, L_{{IR}}^a}$&[L$_\sun$]                    & 13.52$\pm$0.04  & 13.16$\pm$0.05 & 12.9$\pm$0.2   & 12.8$\pm$0.2     \\
    log$M_\mathrm{H_2 + He}^a$  (from CO) &[$M_\odot$]        & 11.04$\pm$0.12  & 11.10$\pm$0.13 & 10.25$\pm$0.15 & 10.11$\pm$0.14   \\   
    \hline
     $^a$ : from Ivison et al. (2013).
	\end{tabular}
    \label{tabglobal}
	\end{table*}   

Galaxy-integrated (for each component W, T, C and M) measurements of the continuum flux densities, emission-line
fluxes and luminosities, dynamical mass estimates and molecular gas masses,  obtained from our new
dataset, are listed in Table~\ref{tabglobal}. For easy reference, the estimated dust temperature and 
the total IR luminosity obtained from detailed global SED modelling by I13 are also listed in the table. 
The quantities listed in Table~\ref{tabglobal} were derived as follows: 
the dynamical mass is calculated under the assumption that the galaxy velocity field is 
rotation-dominated (see Section~\ref{sectrot}): $M_\mathrm{dyn} = [R \times (v_\mathrm{max}/ \sin(i))^2]/G 
= [R \times V_{\mathrm{asym}}^2]/G$, 
where G is the gravitational constant, $v_\mathrm{max}$ is the maximum projected rotational velocity, $i$ is the galaxy disk inclination, 
$V_\mathrm{asym}$ is the asymptotic inclination-corrected rotational velocity 
listed in Tables~\ref{tabw} and \ref{tabt}, 
and R was taken to be the deconvolved half-light radius in CO J:3--2.

Emission-line luminosities ($L'_{\mathrm{CO}}$; in units of in K km s$^{-1}$ pc$^2$)
are calculated using:
\begin{equation}
L'_{\mathrm{CO}}=3.25 \times 10^7 \times S_{\mathrm{CO}} \Delta v \dfrac{D_L^2}{(1+z)^3 \nu^2_{obs}}
\end{equation}
where $S_{\mathrm{CO}} \Delta v$ is the measured flux of the emission line in Jy km s$^{-1}$, 
$D_L$ is the luminosity distance in Mpc, $z$ is the redshift, 
and $\nu_{obs}$ is the observed frequency of the line in GHz \citep{soldowrad92}. 

The molecular gas mass is typically estimated from the CO J:1--0 (hereafter CO) 
luminosity \citep{solvan05} using
\begin{equation}
M_{\mathrm{mol}} = \alpha_{\mathrm{CO}} \times L'_{\mathrm{CO}}
\end{equation}
where $M_{\mathrm{mol}}$ has units of $M_\odot$ and 
$\alpha_{\mathrm{CO}}$ has units of M$_\odot$/(K km s$^{-1}$ pc$^2$ ).
There remains significant debate on the value of $\alpha_{\mathrm{CO}}$. While for nearby
`normal' star-forming galaxies a value of $\alpha_{\mathrm{CO}} \sim 4.3$ M$_\odot$ (K km s$^{-1}$ pc$^2$ )$^{-1}$
is recommended \citep{bolet13}, the value of $\alpha_{\mathrm{CO}}$ is often lower in starburst galaxies.
In a seminal analysis of CO radiative transfer 
and gas dynamics in the starburst nuclei of low-redshift ULIRGs on scales $<$1 kpc, 
\citet{dosol98} found a characteristic value of the conversion factor for H$_2$ + He of
$\alpha_{\mathrm{CO}} \sim 0.8$ M$_\odot$ (K km s$^{-1}$ pc$^2$ )$^{-1}$ in these systems. 
This value, which implies more CO emission per unit molecular gas mass, is commonly adopted for IR-luminous 
starbursts,  where the gas is not in virialised individual clouds \citep{bolet13}. 
Given the extreme starburst nature of the four components of \src, and for consistency with I13, we use $\alpha_{\mathrm{CO}} \sim 0.8$
in this work, while noting that a higher value may be more appropriate. In Table~\ref{tabglobal} we list the galaxy-integrated 
molecular gas mass obtained
using the CO J:1--0 luminosity of I13. In future sections we use our resolved CO J:7--6 maps, together
with the global CO J:7--6 to CO J:1--0 ratios for each component, to  obtain resolved molecular 
gas masses within each component.

The neutral carbon gas mass has been estimated following \citet{weiet03}, assuming that the \cishort\ line
is optically thin. Since we have only observed the
\cishort\ line (and not the [C\,{\sc i}] $^3P_1 \longrightarrow ^3P_0$ line) we must assume an
excitation temperature (T$_{\rm ex}$) for C\,{\sc i}. When both C\,{\sc i} lines have been observed, 
the derived $T_{\rm ex}$ is $\lesssim T_{\rm dust}$ \citep{stuet97,weiet05,popet17}. Table~\ref{tabglobal}
thus lists the estimated \cishort\ masses
for two assumed values of $T_{\rm ex}$ which bracket the estimated dust temperatures (I13) of components
W and T. 
As mentioned in Section~\ref{sectintro} the \cishort\ mass has been used to
trace the total molecular gas mass, and this estimate is likely to be more reliable than
that via CO in low-metallicity galaxies and in environments with strong UV or cosmic ray 
dissociation \citep{biset15}. 
The conversion of \cishort\ gas mass to total molecular gas mass is still under debate and 
here we follow \citet{weiet03} in using
$X_{\rm [C\,{\sc i}]} = M_{\rm [C\,{\sc i}]}/(6 M_{\rm H2}) = 3 \times 10^{-5}$; note that this assumes an adopted
carbon abundance 50\% higher than the Galactic abundance. Note also that the molecular gas masses estimated
from \cishort\ in Table~\ref{tabglobal} are derived from the \cishort\ gas masses calculated assuming 
$T_{\rm ex} = 40$~K.

\begin{figure}
   \hspace{-0.5cm}
   \includegraphics[scale=0.63]{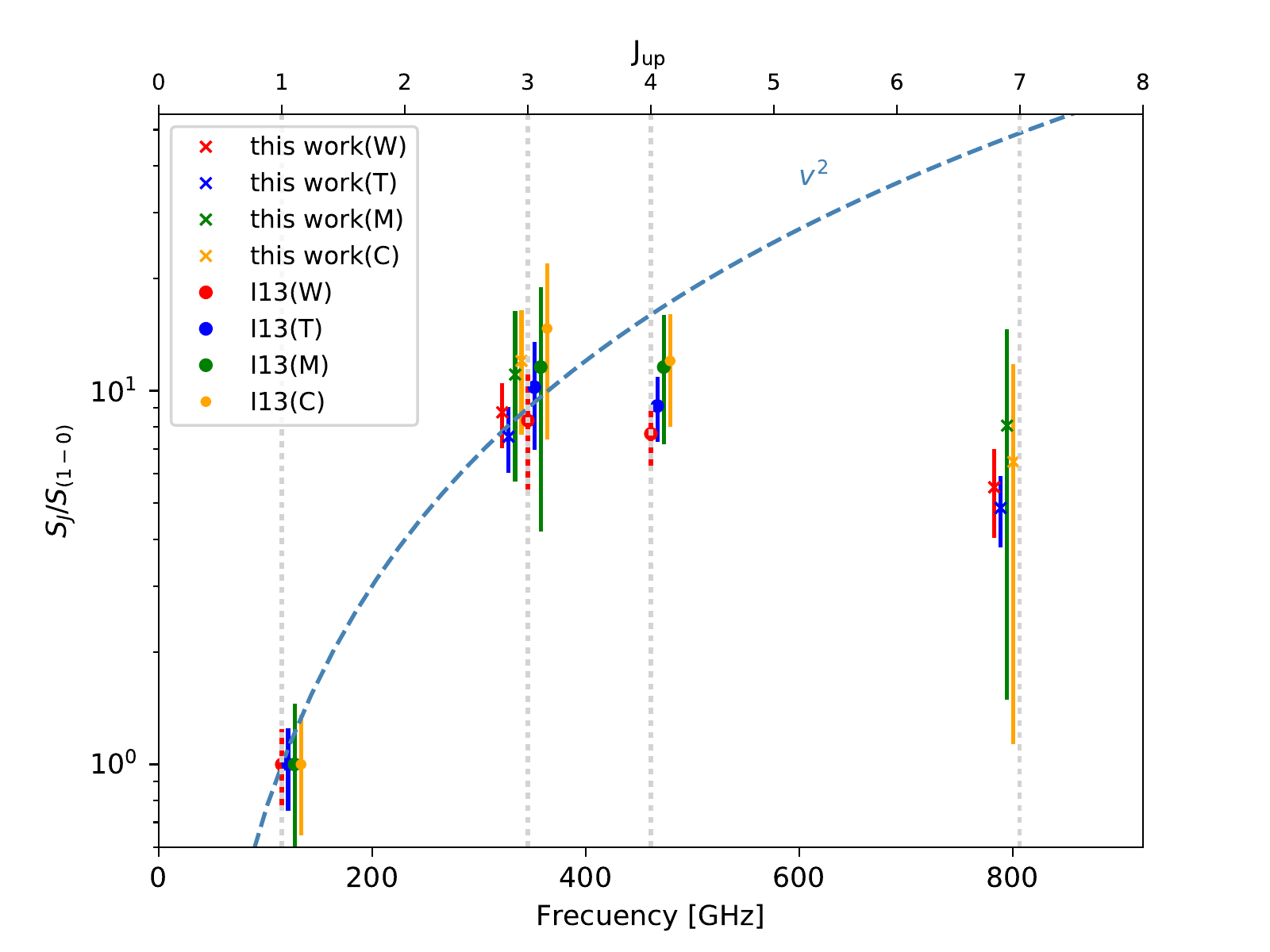}\\ 
   \caption{The CO ladder for all (four) components of \src:
   W (red), T (blue), M (purple), and C (light green).
   The $x$-axis is labelled with the rotational quantum number of the upper state (top; 1, 3, 4,
   and 7; small  displacements in the $x$-axis are used for the different points to avoid confusion) 
   and the rest frequency of the line (bottom) and 
   the $y$-axis is the line flux normalised to that of CO J:1--0.
   Line fluxes from this work are shown with solid circles and those from
   I13 are shown with diamonds.
   The dashed line shows the CO ladder expectation for the case of a constant brightness temperature 
   on the Rayleigh-Jeans (RJ) scale, i.e.\ $S \sim \nu^2$ (note that the RJ approximation is not valid for 
   high $J$).}
   \label{figcoladderT}
   \end{figure}

The variation of the galaxy-integrated CO line luminosity with CO rotational transition (the `CO ladder') for
each of the four components of \src\ is shown in Fig.~\ref{figcoladderT}. Component W appears to
peak at CO J:3--2, while T, C, and M are roughly flat between CO J:3--2 and CO J:4--3. Note that the
CO J:4--3 data of I13 come from observations with IRAM PdBI (6 antennas) in its most extended
configuration: the relatively limited $uv$ spacing as compared to our ALMA CO J:3--2 observation
could mean that the former images missed some line emission. Thus, conservatively, we can
only claim that the CO ladder peaks somewhere between CO J:3--2 and CO J:6--5 for all four components. 
A peak at J$\sim$5 in these galaxies
would be consistent with the CO ladders seen in typical SMGs \citep{carwal13}.
Note that the uncertainty in the location of the peak of the CO ladder does not cause additional uncertainties 
in our conversion between CO J:7--6 and CO J:1--0 luminosities in later sections.

\subsection{Resolved Continuum: Maps and Spectral Shape} 
\label{seccont}

Continuum emission at all observed frequencies, corresponding to rest-frame frequencies of
341~GHz (880\micron), 750~GHz (400\micron), 808~GHz (370\micron) and 1160~GHz (260\micron) at $z=2.41$, 
is strongly detected and resolved towards galaxy components  W and T and relatively weakly detected 
in components C and M. 
Of the four frequencies, the observed 340~GHz (corresponding to rest-frame 1160~GHz or 260\micron) 
continuum maps yielded both the highest resolution and the highest signal-to-noise detections; these
are shown in Fig.~\ref{fig1160ghz}. 
Interestingly, the continuum emission of component T is primarily extended along its kinematic minor axis 
(thin white line in the panel), though at the faintest emission levels it is also extended along the major axis
especially to the SE. The brighter continuum emission in component W appears extended along PA $\sim$
100\arcdeg, an angle between the major and minor kinematic axes; at these brighter continuum emission levels
W does not show any clear extension along its kinematic major axis. Only when considering the fainter continuum
emission is W more extended along its major axis. Component C shows a potential extension in PA $\sim$105\arcdeg: to
the W as a connected extension and to the E as a lower significance separated knot. 
The distribution of the continuum emission in W and T along the major and minor kinematic axes
can also be appreciated in  Fig.~\ref{figflsize}, where these are compared to the emission line flux
distributions. 

In the following subsections, we use the observed-frame 340~GHz maps (rest frame 260\micron\ emission) to 
constrain  the global and resolved star formation rates of each component.

   \begin{figure*}[h]
   \centering
   \hspace*{-12.1cm}\includegraphics[scale=0.4]{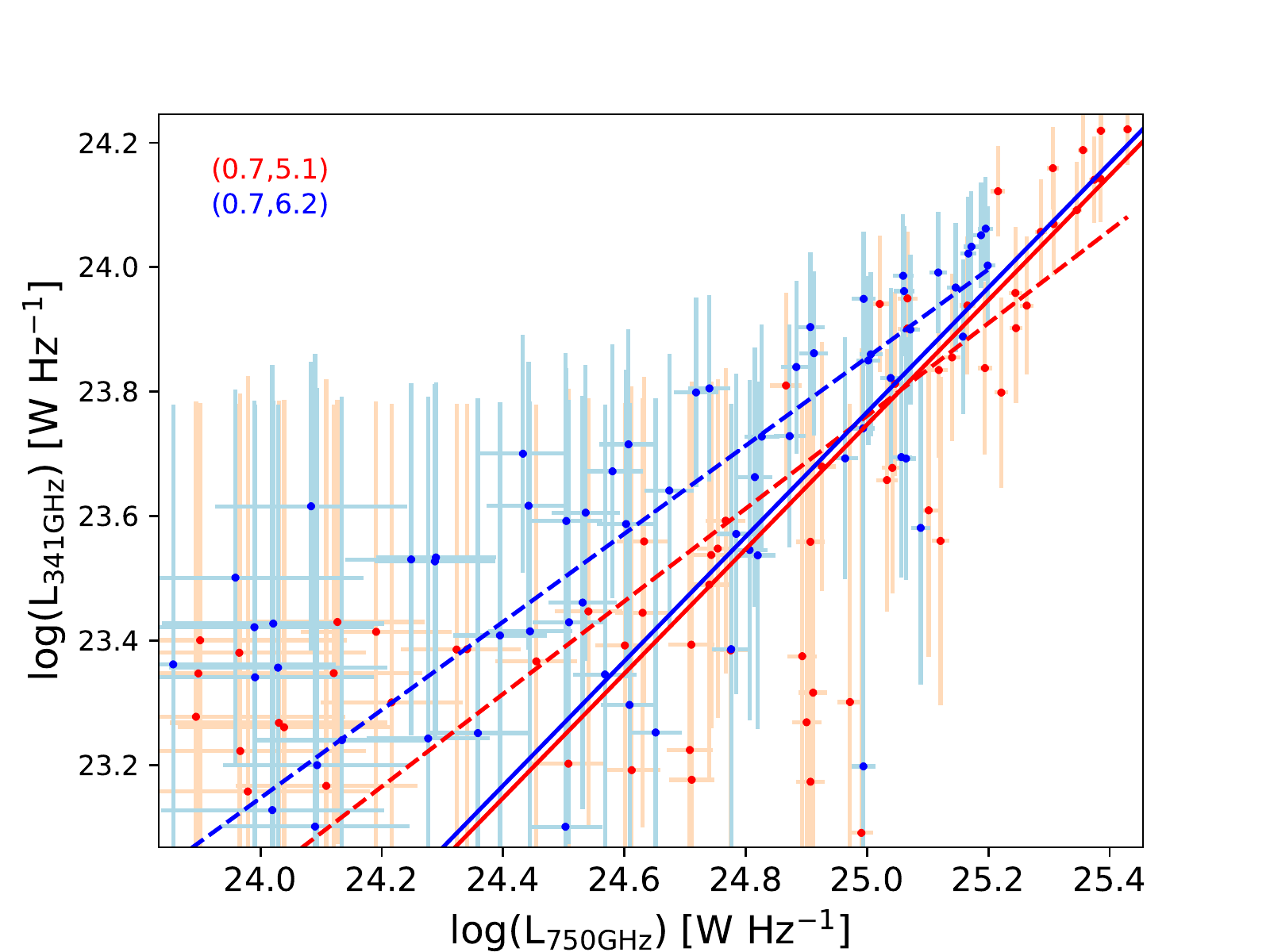}\\ \vspace{-4.92cm}   \hspace*{0.1cm} 
   \includegraphics[scale=0.4]{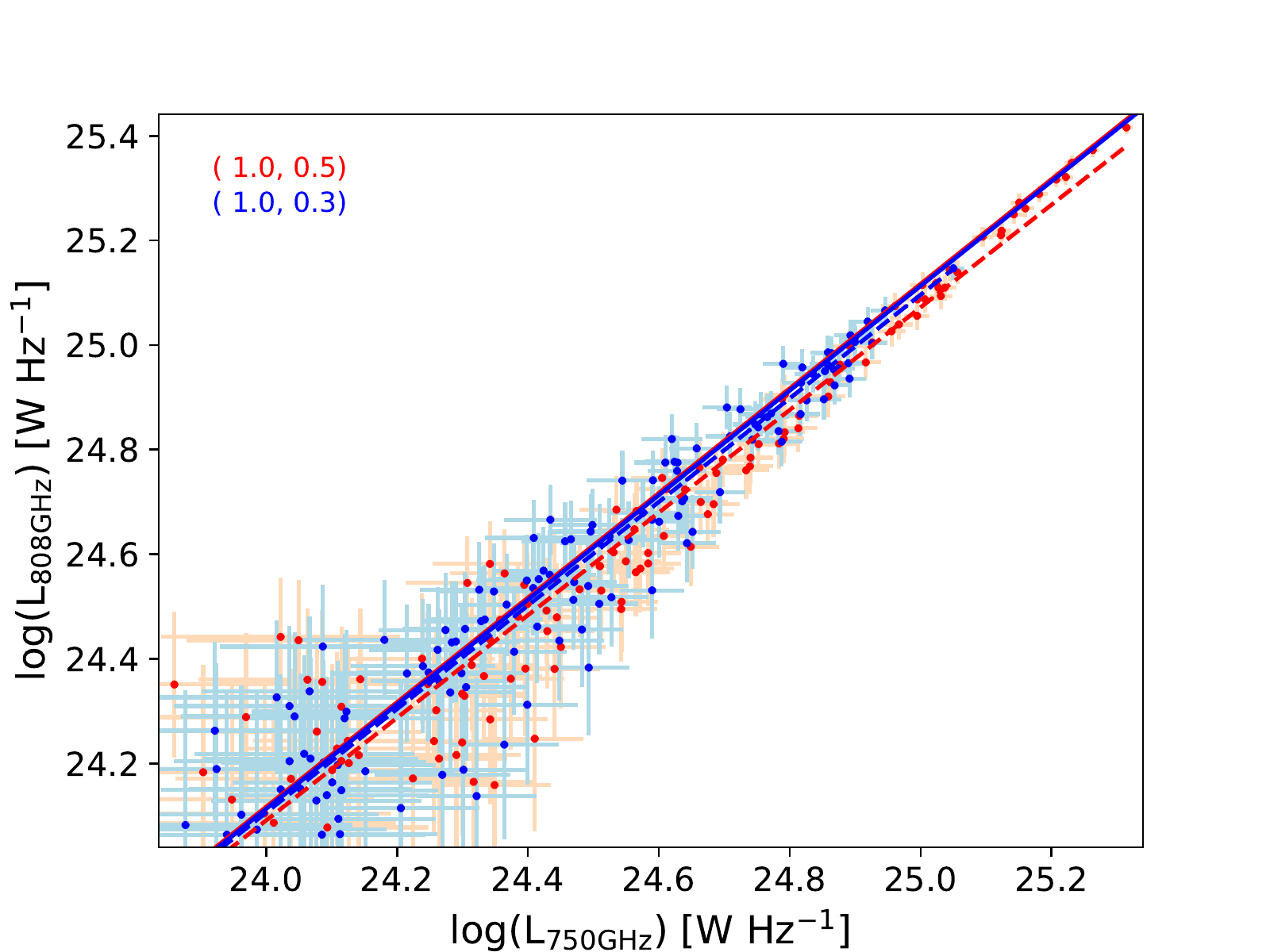}\\ \vspace{-4.92cm} \hspace*{12.1cm}
   \includegraphics[scale=0.4]{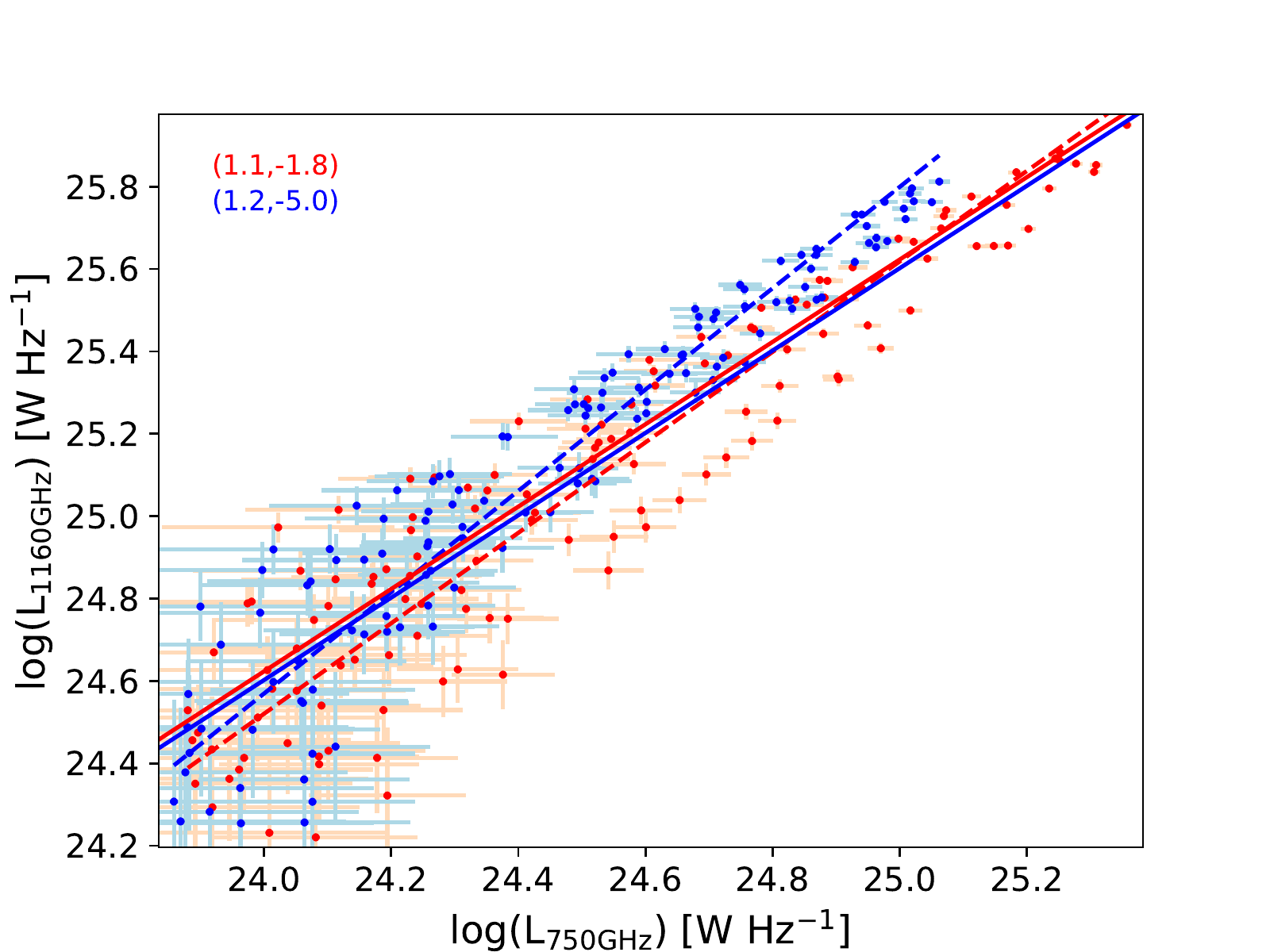}\\ 
   \caption{The relationships between the (resolved) 
   continuum luminosities at rest-frame 341~GHz (880\micron), 750~GHz (400\micron), 
   808~GHz (370\micron), and 1160~GHz (260\micron) 
   for components W (red points) and T (blue points). 
   The two maps used for each panel were convolved to a common resolution.
   Each data point was calculated over an aperture equivalent to the FWHM of the synthesised beam,
   with apertures spaced by  half a synthesised beam; 
   i.e.\ roughly a quarter of the data points are independent measurements.
   Error bars for each point are shown in light red and light blue
   for component W and T, respectively. The 
   dashed lines in the corresponding colour delineate the linear fits 
   to the data; the coefficients of these fits (slope, intercept)
   are listed in the panel in the corresponding colour.  The solid lines show the expected
   relationships for greybody emission (with $\beta=2.0$) for dust at temperature, $T_{\rm dust}=40$~K (red)
   and 36~K (blue), the estimated dust temperatures derived by I13 for W and T, respectively.}
   \label{figratecont}
   \end{figure*}

We explore the resolved continuum spectral slopes of components W and T by comparing
the four observed continuum luminosities in multiple resolved apertures (Fig.~\ref{figratecont}).
To obtain the data in each panel, the higher resolution image was convolved with a Gaussian to degrade the resolution to that of the lower-resolution image. 
The data in the figure were then
extracted over apertures with sizes equivalent to the FWHM of the synthesised beam of the lower resolution image, with half a beam spacing 
between apertures.
Thus roughly a quarter of the data points are independent.  
I13 fitted the observed galaxy-wide sub-mm SEDs of W and T with a model of greybody (assuming emissivity, $\beta=2.0$)
emission from dust at a single temperature, obtaining dust temperatures of $39.8 \pm 0.1$~K and $36.1 \pm 1.1$~K for
W and T, respectively. 
The predictions of greybody emission with these temperatures and $\beta=2.0$ are plotted in all panels
of Fig.~\ref{figratecont}.
Between 370 and 400\micron\ (middle panel of Fig.~\ref{figratecont}) the resolved dust temperatures
are roughly in agreement with the globally-derived values, though the nuclear apertures of both W and T 
(the highest luminosities) have flux density ratios which imply dust temperatures higher by several tens of degrees. 
Both at longer wavelengths (880\micron; left panel of Fig.~\ref{figratecont}) and at the shortest observed
wavelengths, which are closest to the IR peak (260\micron; right panel of 
Fig.~\ref{figratecont}), component W continues to follow the global temperature predictions. 
Component T, however, shows significant excess emission at both the longest and shortest  wavelengths:
all apertures of T show a consistent flux excess at 880\micron\ (left panel) attributable to cooler dust temperatures
or an additional source of 880\micron\ emission; the nuclear apertures of T show a consistent excess of
260\micron\ emission implying significantly higher temperatures for a single dust component or the presence of
a second hotter component of dust. Both excesses could be explained by the presence of an AGN: the nuclear dust
would be heated by the AGN, leading to the short-wavelength excess, and synchrotron and thermal 
emission from a powerful AGN jet could explain the excess emission seen at ~mm wavelengths.

The relationship between the resolved continuum spectral slopes (and thus dust mass and temperature) 
and the atomic and molecular gas emission (and thus mass and excitation), is deferred to a future work.


\subsection{Component Sizes and Emission-Line Maps}
   
\begin{figure}[h]
   \hspace*{-0.48cm} 
   \includegraphics[scale=0.64]{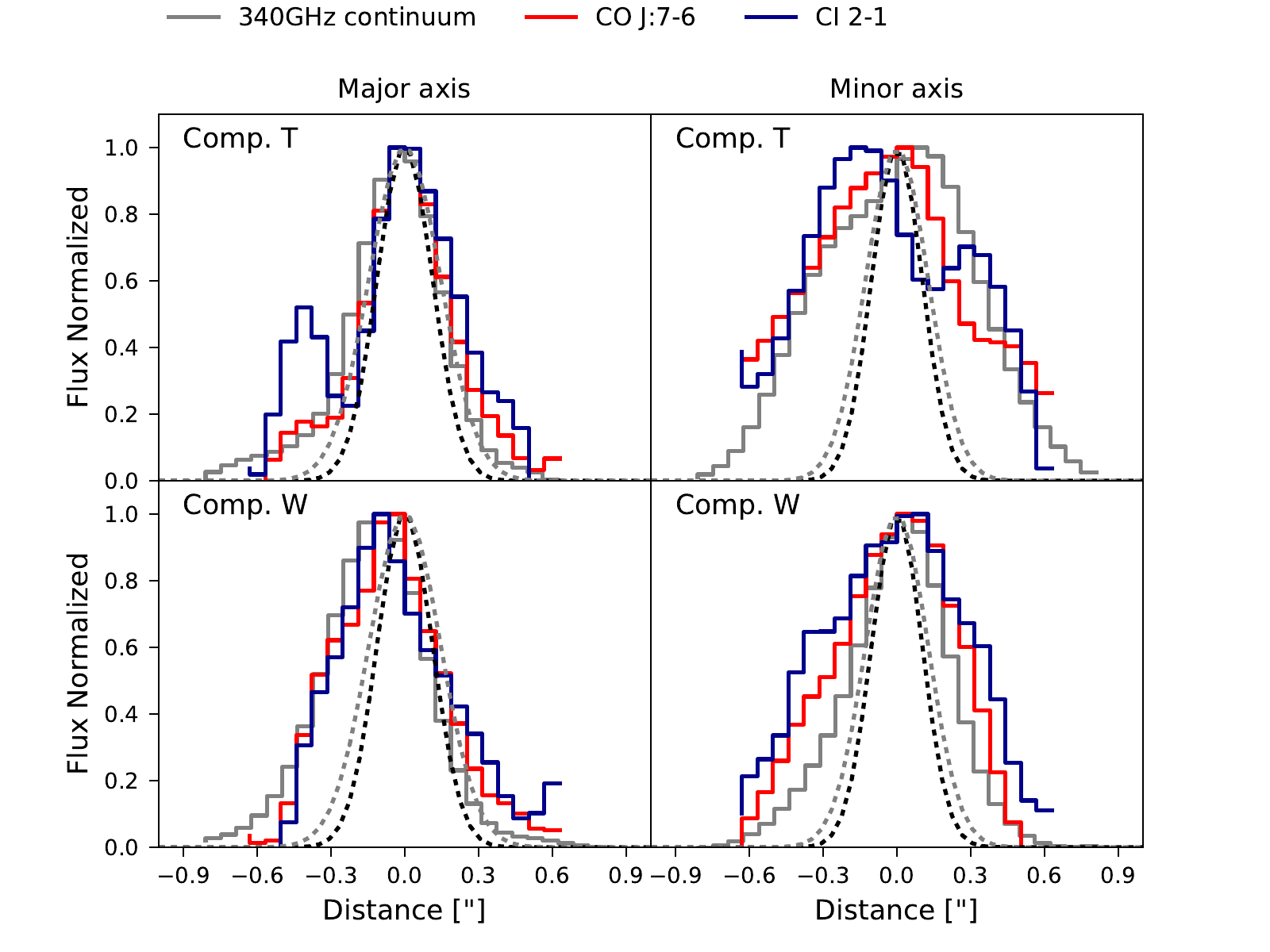} \\ \vspace{-0.4cm}
   \caption{The spatial extent of the emission line gas (\co7-6\ and \cishort)
   and rest-frame 1160-GHz continuum emission  
   in component T (top row) and component W (bottom row) along the kinematic major 
   (PA=135\arcdeg\ for T and PA=55\arcdeg\ for W) and kinematic minor axes (PA=225\arcdeg\ for T 
   and PA=145\arcdeg\ for W). These major and minor axes are illustrated in 
   Figs~\ref{fig1160ghz}, ~\ref{figmomw} and ~ \ref{figmomt}. 
   Each panel shows the normalised flux, extracted along a one-pixel-wide slit oriented along the corresponding
   PA, of the continuum emission (grey solid line; obtained from the observed 340-GHz continuum map),
   the \co7-6\ emission line (dark red solid line) 
   and the \cishort\ emission line (dark blue solid line). 
   The major axis of the synthesised beam of the \co7-6\ and \cishort\ moment 0 maps (340-GHz continuum map) 
   is shown by the grey (black) normalised Gaussian. Both synthesised beams are close to circular.
   Both lines and continuum are clearly resolved in all panels, with similar extensions in both emission
   lines and the continuum. The deconvolved sizes along these PAs are listed in Tables~\ref{tabsizecont}
   and \ref{tabsizeco76}. }
   \label{figflsize}
   \end{figure}

\begin{figure*}[h]
   \centering
   \vspace{-1.cm} 
   \hspace*{-14cm}\includegraphics[scale=0.25]{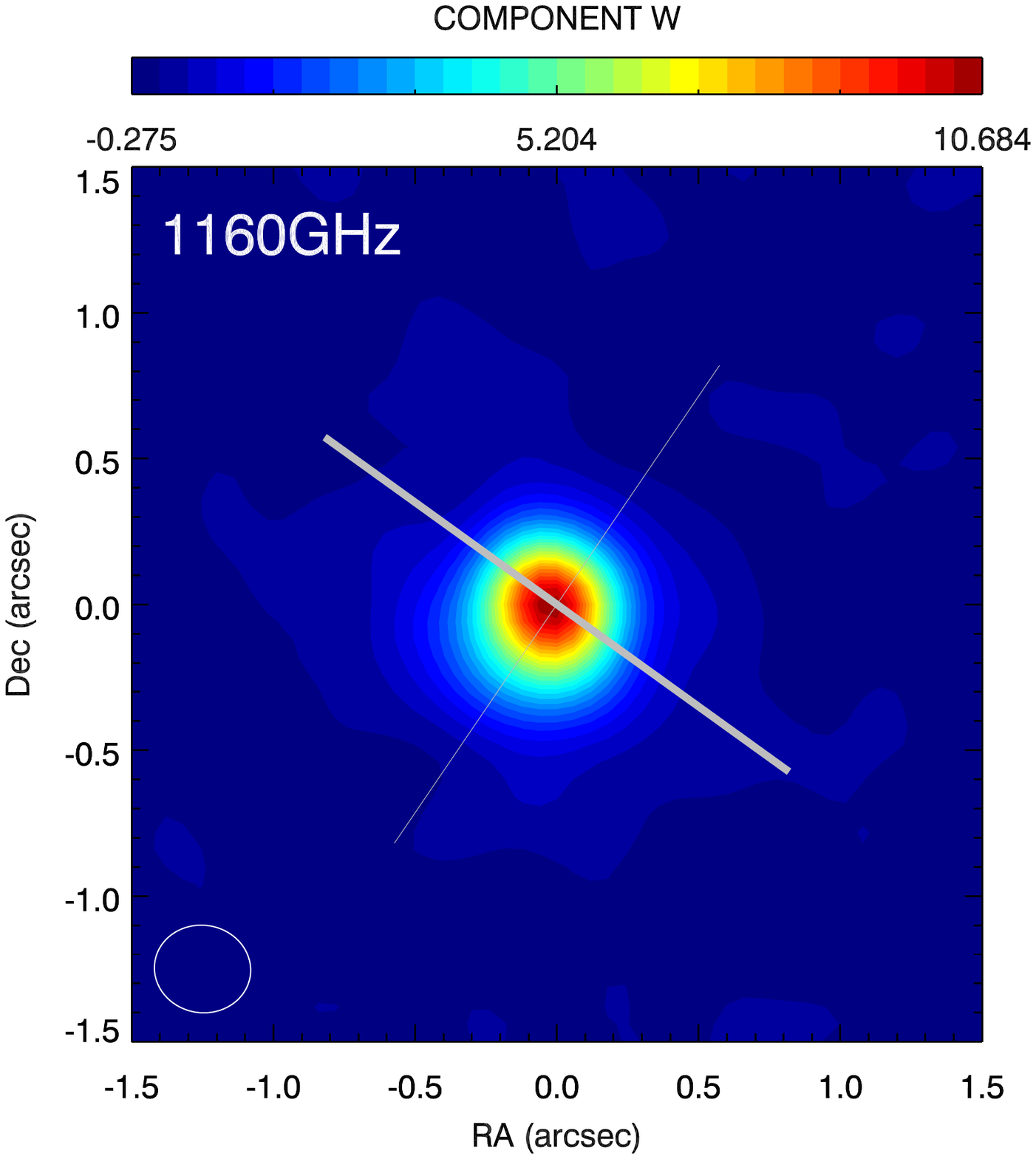}\\ \vspace{-7.005cm}    
   \hspace{-4.8cm} \includegraphics[scale=0.25]{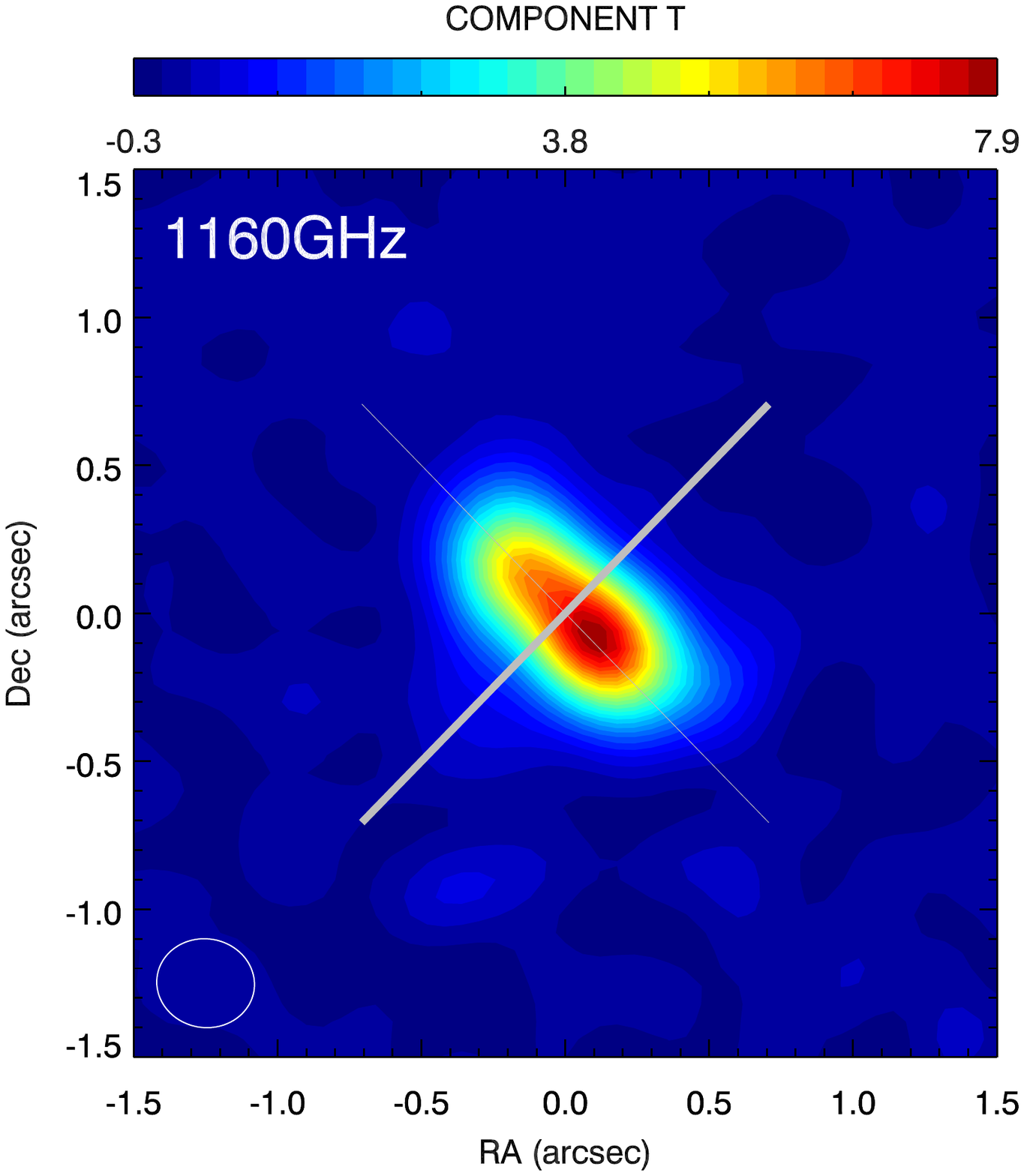}\\   \vspace{-7.005cm}
   \hspace{4.8cm} \includegraphics[scale=0.25]{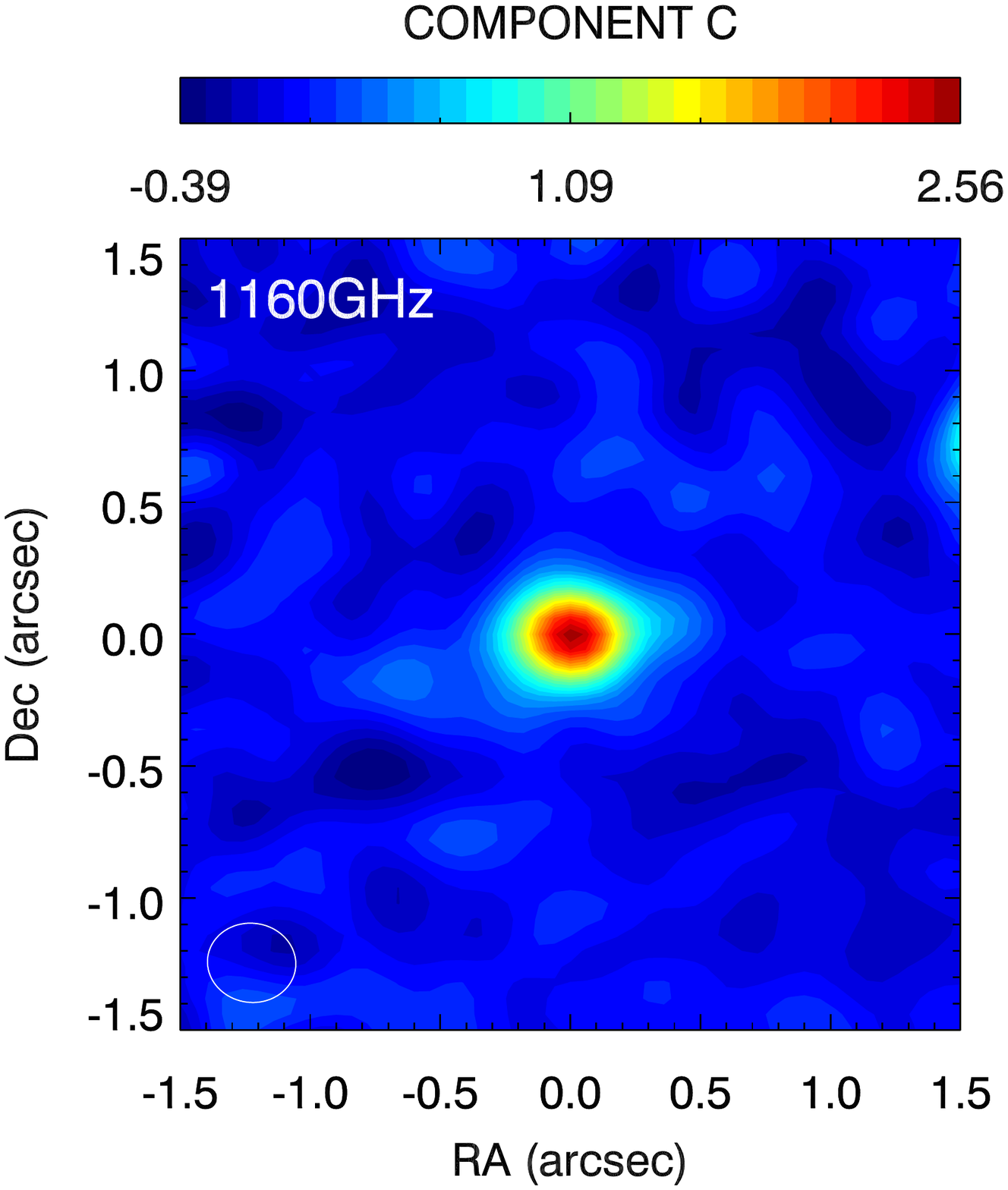}\\   \vspace{-7.005cm} 
   \hspace*{13.5cm} \includegraphics[scale=0.25]{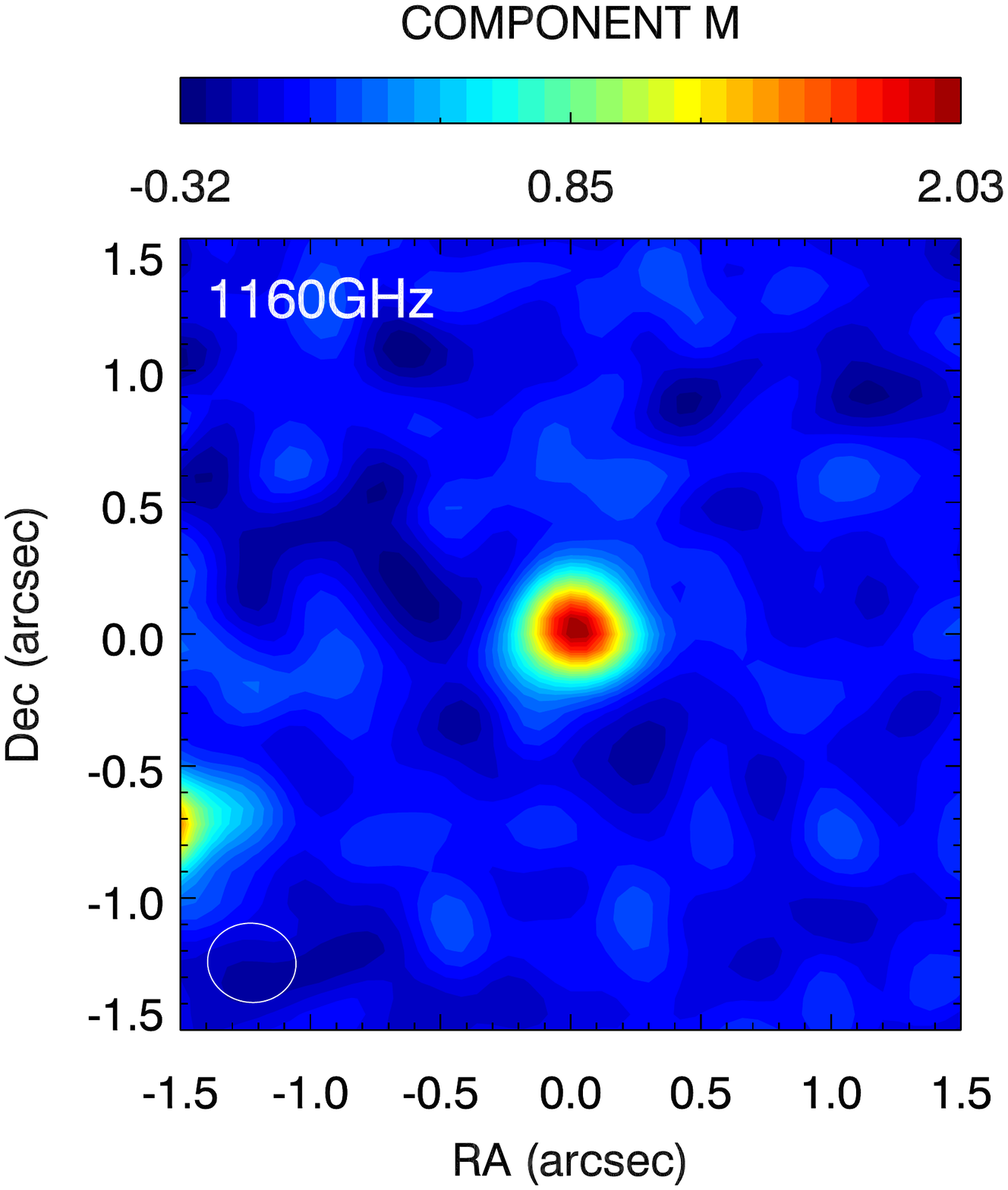}\\
   \vspace{-0.8cm}
   \caption{Maps of the observed-frame 340-GHz (corresponding to rest-frame 260\micron) continuum emission in 
   (left to right)
   components W, T, C and M. Continuum fluxes are in mJy $\mathrm{beam^{-1}}$ following
   the colour bar above each panel. Each panel is 3\arcsec\ $\times$ 3\arcsec\ in size and the axes,
   in arcsec, are centred on the kinematic centre of each component, as obtained from \textit{Kinemetry}. These
   kinematic nuclear positions, hereafter used as the galaxy positions, are:
   W: 08:49:33.6XX, +02:14:44.68;  
   T: 08:49:33.0XX, +02:14:39.69; 
   C: 08:49:33.9XX, +02:14:44.86;
   M: 08:49:33.8XX, +02:14:45.58.
   For components W and T, the kinematic major and minor axes are shown with  thick and thin grey lines, respectively.
   }
   \label{fig1160ghz}
   \end{figure*}

All detected emission lines -- CO J:3--2, CO J:7--6, \cishort\ and H$_2$O --
are resolved across several synthesised beams in W and T, and the CO J:7--6 and \cishort\ lines
are partially resolved in C and M. 
Components W and T show the largest spatial extents in both continuum and line emission.
Comparisons of the spatial distribution of the \co7-6, \cishort, and rest-frame
1160~GHz continuum emission along two position angles (PAs) on the sky  in components
W and T are shown in Fig.~\ref{figflsize}. These PAs 
represent the major and minor kinematic axes under the assumption that the kinematics
in W and T are rotation-dominated (Section~\ref{sectrot}).

Within the noise, the distributions of the gas and 
continuum emission are roughly similar in component W, though there is some indication that the
gas (especially \cishort) emission is more extended than the continuum emission along the kinematic minor axis.
Additionally, and unexpectedly, the emission from both atomic and molecular gas is more extended along the minor axis, 
rather than the major axis.
Even more significant differences are seen in component T. Here too the gas and dust is more
extended along the minor instead of the major axis. 
Along the minor axis there are significant offsets between the locations of the peak CO J:7--6, \cishort and 
continuum emission.  The continuum peak (the `SW knot') has a significantly
higher CO to \cishort\ ratio as compared to the NE extension; similar to \cishort, the H$_2$O
emission is also stronger in the NE extension as compared to the SW knot (see the next Section). 
The continuum-to-gas ratios are highest in the NE extension and lowest in the SW knot. 
Along the kinematic major axis, the gas and continuum are relatively similarly distributed, except
for a \cishort-rich region 0\farcs4 from the nucleus.
The deconvolved sizes of components W and T along the PAs shown
in Fig.~\ref{figflsize} are listed in Tables~\ref{tabsizecont} and \ref{tabsizeco76}.

	\begin{table}[h]
	\caption{Observed-frame 340-GHz continuum sizes}
	\centering
	 \begin{tabular}{c c c}
	 \hline
     Component     & kinematic  & kinematic             \\
	               & minor-axis  & major-axis             \\
	 \hline
	 W             & 0.46\arcsec(3.80 kpc)    & 0.45\arcsec(3.71 kpc)  \\
	 T             & 0.75\arcsec(6.19 kpc)    & 0.39\arcsec(3.22 kpc)  \\
	 \hline
	 \end{tabular}
    \label{tabsizecont}
	\end{table}
    
	\begin{table}[h]
	\caption{CO J:7--6 size}
	\centering
	 \begin{tabular}{c c c}
	 \hline
     Component     & kinematic  & kinematic             \\
	               & minor-axis  & major-axis             \\
	 \hline
	 W             & 0.58\arcsec(4.79  kpc)    & 0.53\arcsec(4.37 kpc)  \\
	 T             & 0.75\arcsec(6.19 kpc)     & 0.36\arcsec(2.97 kpc)  \\
	 \hline
	 \end{tabular}
    \label{tabsizeco76}
	\end{table}

   \begin{figure*}
   \centering
   \vspace{-5.5cm} 
   \hspace*{-1.28cm} \includegraphics[scale=0.753]{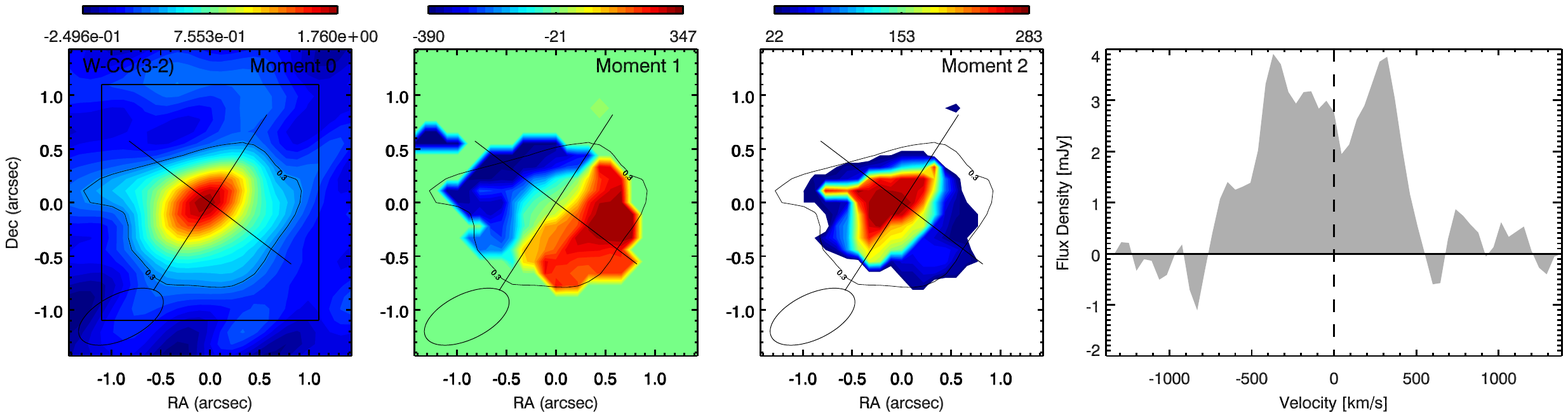}\\ \vspace{-11.1cm} 
   \hspace*{-1.28cm} \includegraphics[scale=0.753]{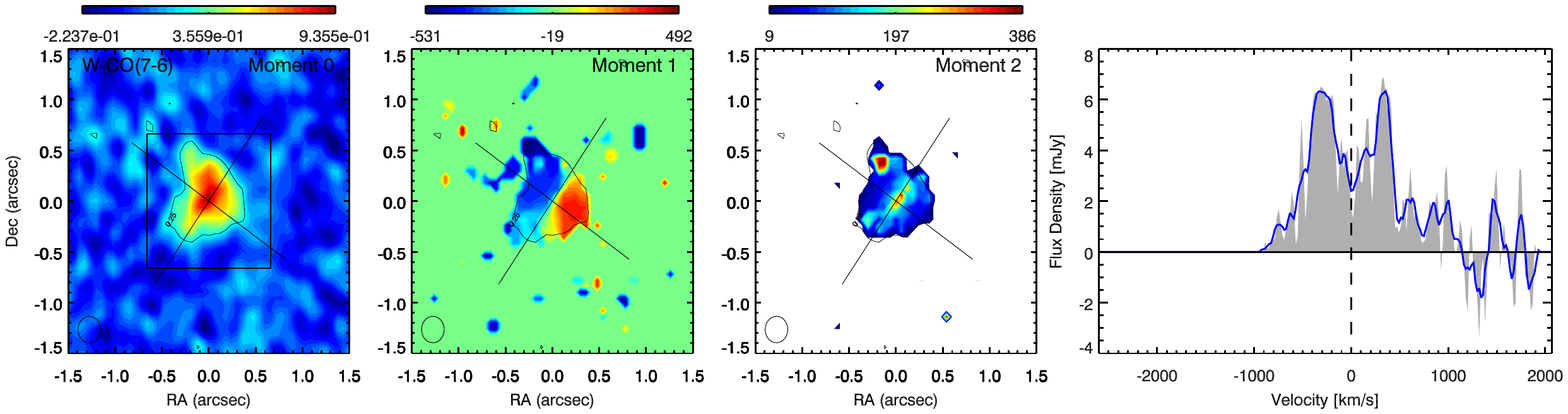}\\ \vspace{-11.1cm} 
   \hspace*{-1.28cm} \includegraphics[scale=0.753]{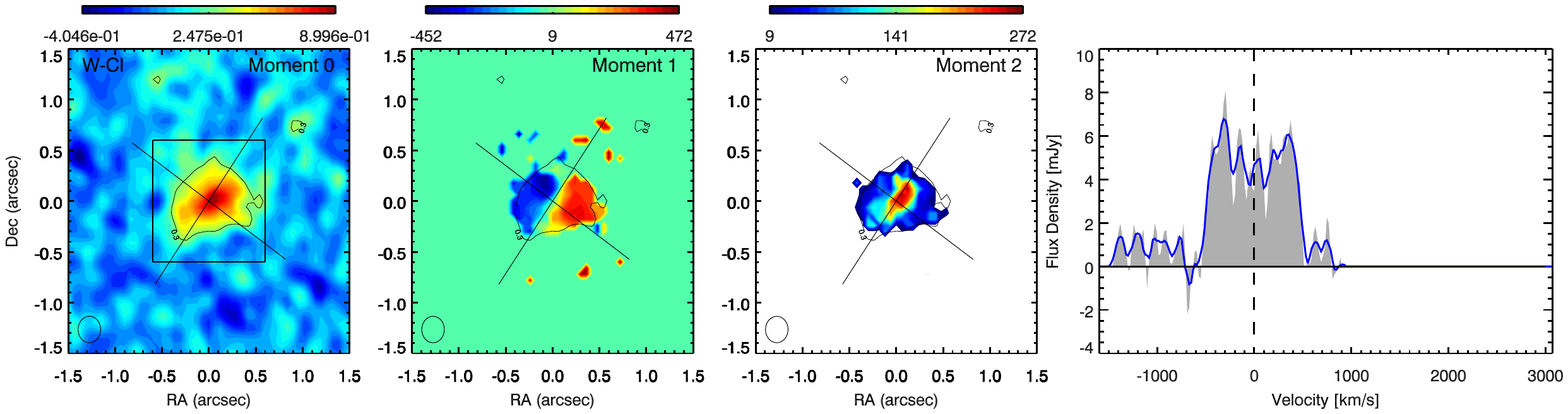}\\      \vspace{-11.1cm} 
   \hspace*{-1.28cm} \includegraphics[scale=0.753]{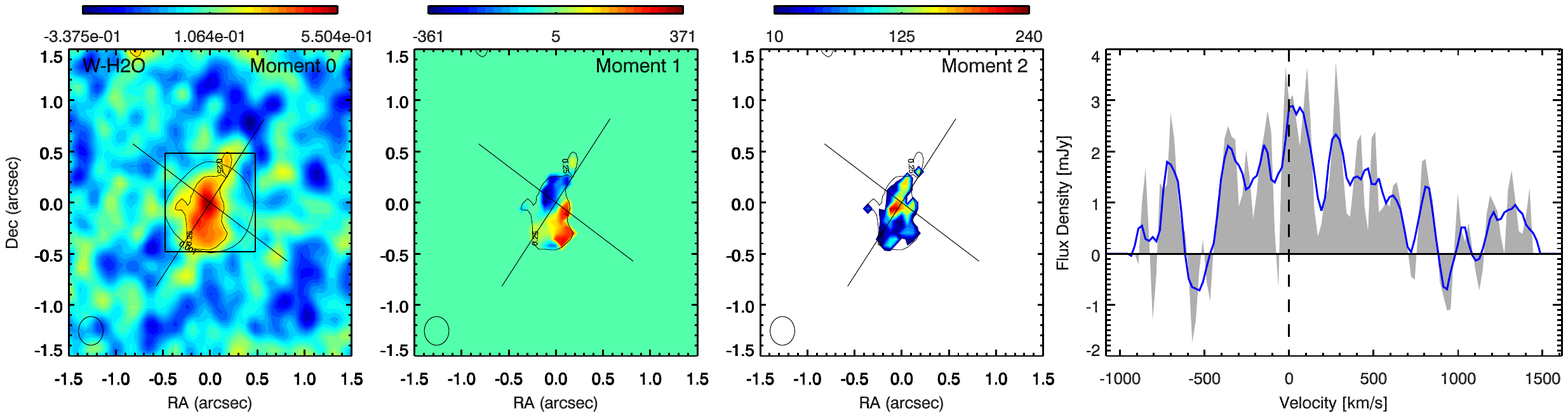}\\     \vspace{-5.5cm}   
   \caption{For component W, we show from left to right: maps of the integrated flux (moment 0; units of Jy \kms\ beam$^{-1}$), 
   intensity-weighted average velocity (moment 1; \kms\ relative to systemic), 
   and velocity dispersion (moment 2; \kms), and the 
   galaxy-integrated line profile, of the emission lines detected. 
   From top to bottom the lines are:  CO J:3--2, CO J:7--6, \cifull\ and \h2ofull.  
   In the three left-most columns the colours follow the respective colour bars,
   the synthesised beam is shown at the lower left, and 
   axes units are arcseconds with the same central position used in all panels. 
   For ease, the kinematic major and minor axes are shown in black solid lines and an illustrative
   single flux contour from the moment 0 image is shown in all panels of the same row.
   All moment maps were made from data cubes created with natural weighting. 
   The right-most column shows the corresponding galaxy-integrated line profile extracted within the 
   square apertures shown in the corresponding left-most panel (i.e.\ the moment 0 image). The
   line profiles are shown both at the observed spectral resolution (grey histograms; spectral
   resolutions of -- top to bottom --    46 km s$^{-1}$, 
   19.7 km s$^{-1}$, 19.7 km s$^{-1}$, and 21.2 km s$^{-1}$, per channel), and at a smoothed
   resolution (blue solid lines in the lower three panels) of $\sim$ 100 \kms. 
   The line profiles of CO J:7--6 and \cifull\ have been de-blended as explained in Section~\ref{secglobal}. 
   }
   \label{figmomw}
    \end{figure*}

   \begin{figure*}
   \centering
   \vspace{-5.5cm} 
   \hspace*{-1.28cm} \includegraphics[scale=0.753]{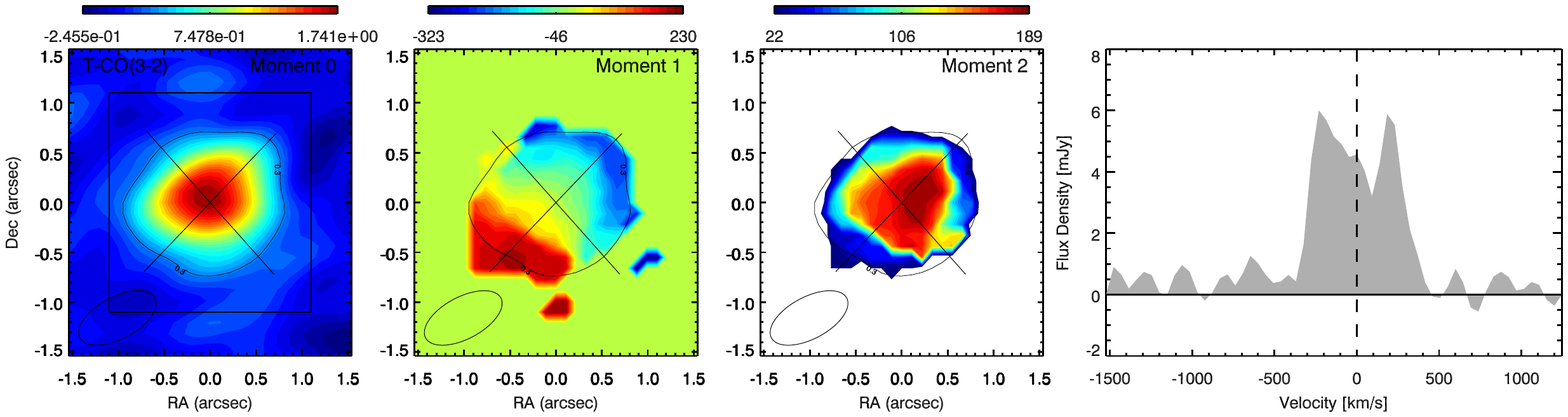}\\ \vspace{-11.1cm} 
   \hspace*{-1.28cm} \includegraphics[scale=0.753]{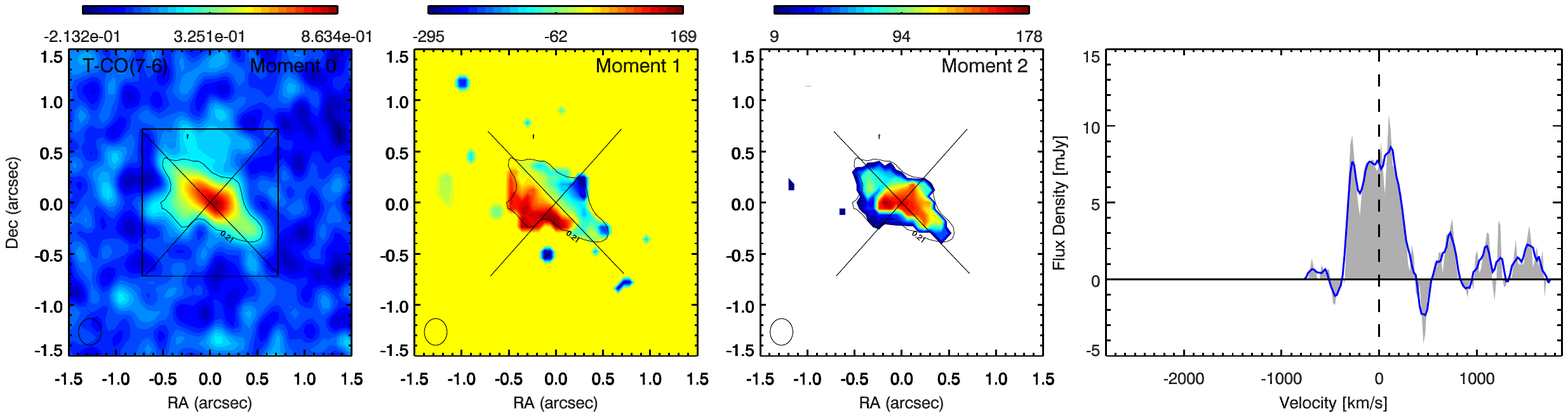}\\ \vspace{-11.1cm} 
   \hspace*{-1.28cm} \includegraphics[scale=0.753]{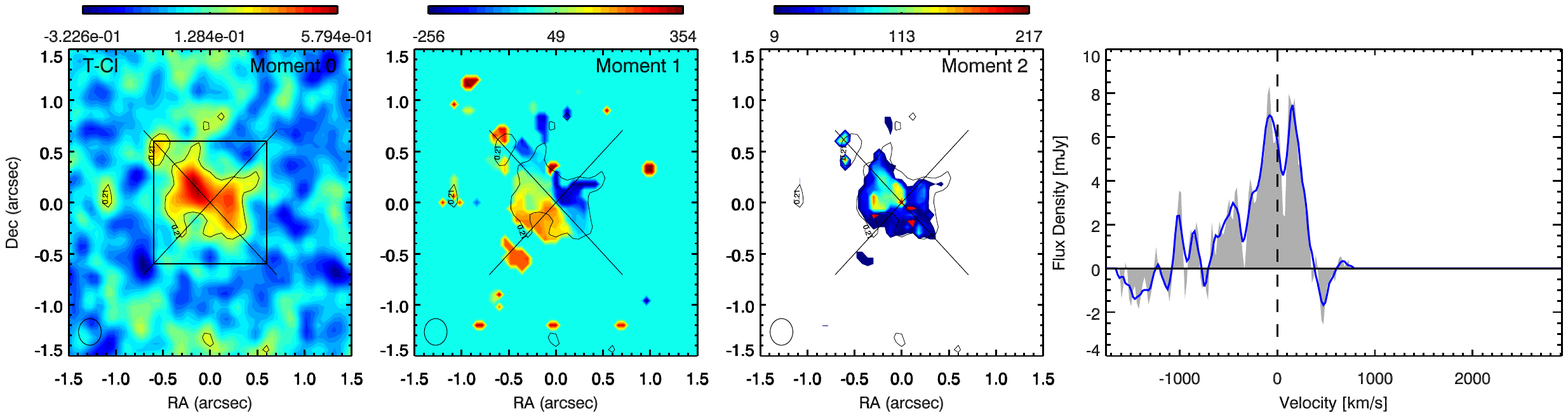}\\      \vspace{-11.1cm} 
   \hspace*{-1.28cm} \includegraphics[scale=0.753]{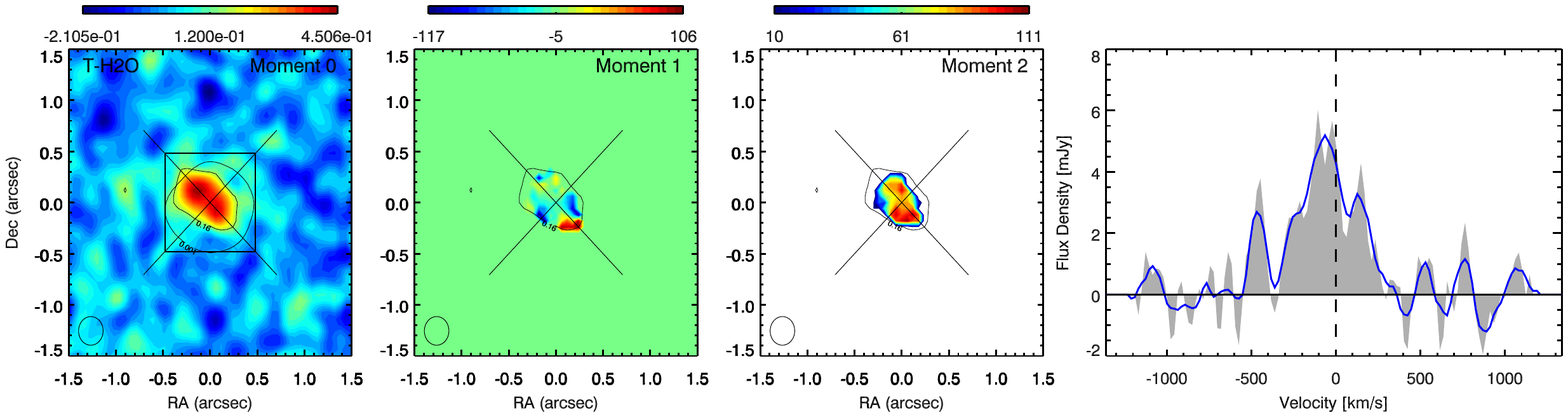}\\     \vspace{-5.5cm}  
   \caption{As in Fig.~\ref{figmomw} but for component T. 
    Here the relatively narrow line profiles, as compared to W, allow a cleaner separation of the 
    galaxy-wide CO J:7--6 and \cifull\ line profiles. 
   }
   \label{figmomt}
   \end{figure*}
%

   \begin{figure*}
   \centering
   \vspace{-5.5cm} 
   \hspace*{-1.28cm} \includegraphics[scale=0.753]{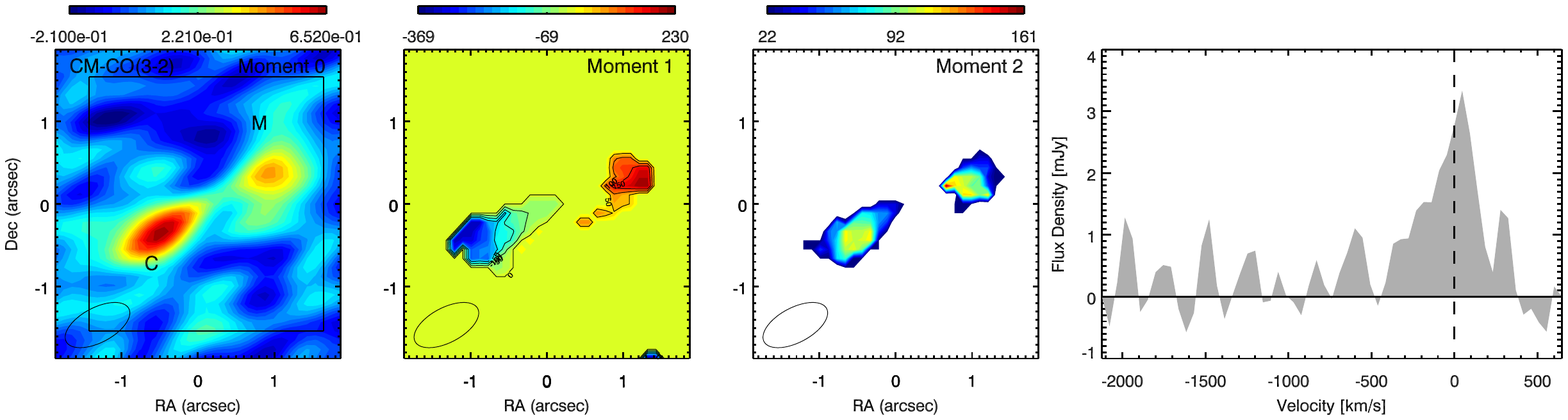}\\ \vspace{-11.1cm} 
   \hspace*{-1.28cm} \includegraphics[scale=0.753]{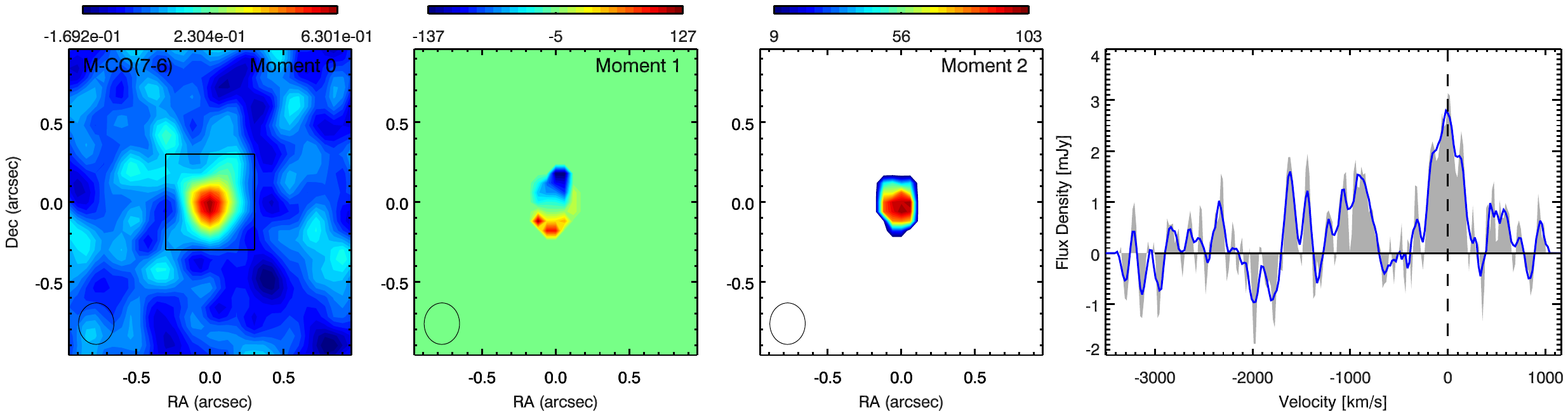}\\ \vspace{-11.1cm} 
   \hspace*{-1.28cm} \includegraphics[scale=0.753]{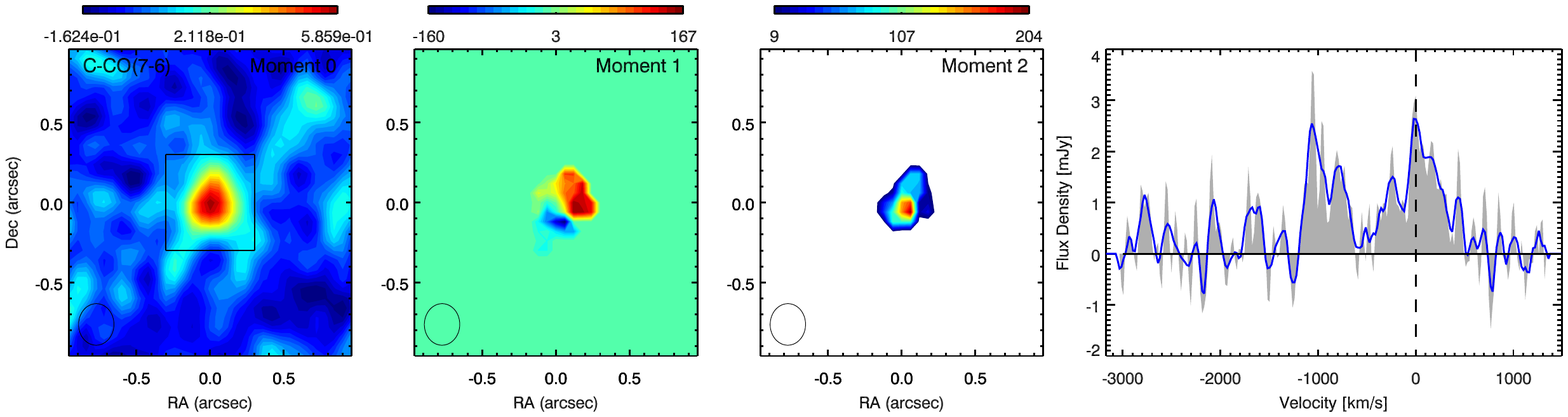}\\     \vspace{-5.5cm} 
   \caption{Similar to Fig.~\ref{figmomw} but for  components M and C.
   The top row shows the results for the CO J:3--2 line in both M and C together.   
   The middle and bottom rows show the results for the CO J:7--6 line 
   separately for components M and C, respectively. 
   In the right-most panels of the middle and bottom rows 
   (the galaxy-integrated profiles of the CO J:7--6 lines in M and C, respectively)
   the CO J:7--6 line covers a velocity range of approximately $\pm 300$ \kms\ and 500 \kms, 
   respectively. For both components the \cifull\ line is also
   clearly visible at lower ($\sim$1000 \kms\ to the blue) velocities. We do not show the equivalent
   moment maps for the \cifull\ line in M and C as they are significantly noisier. 
   }
   \label{figmomcm}
   \end{figure*}

The moment maps (0 = integrated flux, 1 = intensity weighted velocity map, 2 = velocity dispersion
map) of the CO J:3--2, CO J:7--6, \cishort\ and H$_2$O emission lines 
are presented in Figs~\ref{figmomw}, \ref{figmomt}, and \ref{figmomcm}, for
components, W, T, and C and M, respectively. 
The galaxy-integrated line profiles shown in the right-most column of these figures were extracted from the square apertures shown in the 
left-most column 
(moment 0) maps: a 3"x3" aperture centred on 
RA +2$^{\circ}$14' 44\farcs618 DEC 8$^h$ 49$^m$ 33$^s$.592   for all lines in W,
a 3"x3" aperture centred on 
RA +2$^{\circ}$14' 39\farcs697 DEC 8$^h$ 49$^m$ 32$^s$.960  for all lines in T.
The CO J:3--2 line profile of C and M together are extracted over a 
3\farcs8x3\farcs8 aperture centred on RA +2$^{\circ}$14' 45\farcs219 DEC 8$^h$ 49$^m$ 33$^s$.868.
The individual CO J:7--6 line profiles of C and M have been extracted over a 
0\farcs5x0\farcs5 aperture centred on each component. 

All four components show ordered velocity fields. In W and T, both   CO and \cishort show consistent blue- and red-shifted sides
and similar peak-to-peak velocities.
The molecular gas velocity fields are best appreciated in the CO J:7--6 (rather than CO J:3--2) 
velocity maps which have both the highest signal  to noise and the highest spatial resolution.
Components W and T show (projected) 
maximum velocities of roughly $\pm 500$ \kms\ and  $\pm270$ \kms, respectively.
Components M  and C have smaller velocity gradients,  
with $v_{max}$ = 132 km s$^{-1}$ (264 km s$^{-1}$ peak to peak) and 
$v_{max}$ = 164 km s$^{-1}$ (327 km s$^{-1}$ peak to peak), respectively.
Note that the internal kinematics of components C and M are resolved for the first time in our
CO J:7--6 and \cishort\ maps. Further,
the relatively low spatial resolution CO J:3--2 maps show an almost continuous  (in flux and velocity) bridge
between C and M. 
The CO and \cishort\mbox dispersion maps show higher dispersions near the nucleus, and along the
kinematic minor axis; the former can be partly due to the rotational velocity gradient along
the major axis. 
The observed velocity fields 
could be interpreted as either ordered rotation (Section~\ref{sectrot} or outflows in a galaxy
disk (Section~\ref{secoutflows}); in the following section we argue for the former scenario. 

The \water\ emission line, which admittedly has a lower signal to noise, appears to follow different
kinematics as compared to CO and \cishort. In W, the blue and redshifted components of \water\ can
be argued to be separated by the kinematic minor axis (i.e. the same as CO) or the kinematic major axis. In T the kinematics
is significantly more disordered, with blue shifted velocities seen along the kinematic major axis on both sides
of the kinematic minor axis, and an isolated red velocity component at the edge of the detected extent. 

While the CO and \cishort\ lines follow roughly similar kinematics in W, there are notable
differences in both their morphology and kinematics: e.g.\ CO J:7--6 shows a notable blueshifted highly-dispersive NE-clump which is
not present in \cishort. These differences are the primary reason for the different global emission line 
profiles presented in the previous section.

In component T, the CO and \cishort\ emission lines show relatively similar kinematics. 
While CO J:7--6 emission is roughly centred on the `nucleus' the \cishort\ and water vapour
line emission straddle the  nucleus, similar to what is seen in the (rest-frame) 1160~GHz  continuum emission. 
Note that component T (see the CO J:7--6 maps)  
is highly extended along its kinematic minor axis (see middle panel of Fig.~\ref{figmomt}); 
this was not obvious in the earlier I13 maps due to their lower resolution. 
The extension along the minor axis is best appreciated in Fig.~\ref{figflsize} and Tables \ref{tabsizecont}, 
and \ref{tabsizeco76}. If the velocity field in T is interpreted as due to rotation, then the ratio of the
minor/major axes in this galaxy implies that the lensing-produced spatial magnification of this component
is $\gtrsim 2\times$ along the direction of the minor axis.

Both C and M are also resolved kinematically in the CO J:7--6 line (Fig.~\ref{figmomcm}) and also in the \cishort\
line (not shown): if their velocity fields are due to rotating gas, then the major axis PAs of C and M are
$\sim -$30\arcdeg, and $\sim$160\arcdeg, respectively. The lower resolution CO J:3--2 maps
appear to show a continuous bridge, in integrated flux and in velocity, between the two components. Given their
small spatial and velocity separation, it is likely that these two components are interacting, which is known to be common amongst SMGs (e.g.\ \citealt{ivison07, engel10}).


The galaxy-integrated CO and \cishort\ emission line profiles in W and T can be interpreted as double-peaked or 
as having a central plateau. This spectral shape could be due to the intrinsic velocity distribution of gas, due to the emission lines being optically thick, or even due to the presence of superposed absorption line components which originate in outflowing atomic and molecular gas. 

\subsection{Modelling the Observed Velocities: Rotation}
\label{sectrot}

	\begin{table}[h]
	\caption{Results of \kin\ modelling of component W}
	\centering
	\begin{tabular}{c c c c}
	\hline\hline
	Property                          & CO J:3--2          & CO J:7--6         & \cifull          \\
	\hline
	PA.                               & $45^{\circ}$       & $55^{\circ}$     & $55^{\circ}$     \\
	Inclination                       & $48^{\circ}$      & $48^{\circ}$     & $48^{\circ}$     \\
	$V_{\mathrm{asym}}$ (km s$^{-1}$) & 464               & 525              & 525              \\
	\hline
	\end{tabular}
    Rows are (a) PA (N to E); (b) inclination (0$^{\circ}$ is face-on); and 
         (c) V$_{\rm asym}$, the peak to peak de-projected rotation velocity. 
        \label{tabw}
	\end{table}

	\begin{table}[h]
	\caption{Results of \kin\ modelling of Component T}
	\centering
	\begin{tabular}{c c c c}
	\hline\hline
	Property                           & CO J:3--2         & CO J:7--6         & \cifull          \\
	\hline
	PA.                                & $144^{\circ}$    & $136^{\circ}$    & $135^{\circ}$    \\
	Inclination                        & $49^{\circ}$     & $54^{\circ}$     & $49^{\circ}$     \\
	$V_{\mathrm{asym}}$ (km s$^{-1}$)  & 298              & 241              & 298              \\
	\hline
	\end{tabular}
        \noindent Rows are as in Table~\ref{tabw}. 
        \label{tabt}
	\end{table}


   \begin{figure*}[h]
   \centering
   \vspace{-4.9cm}
   \hspace*{-0.51cm} \includegraphics[scale=0.71]{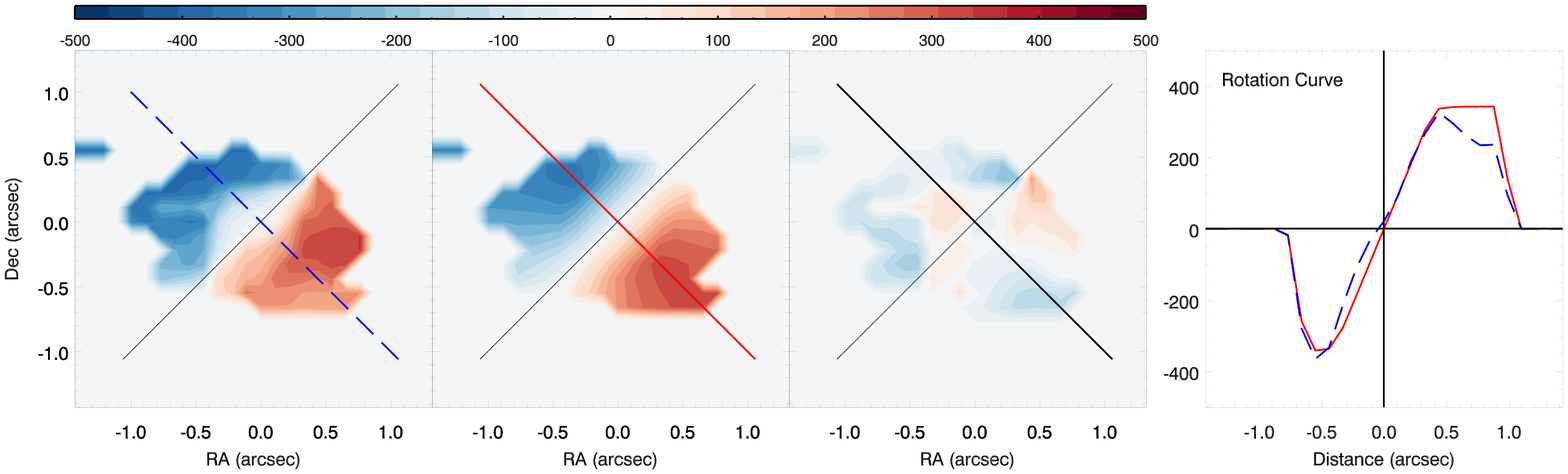} \\ \vspace{-10.55cm}
   \hspace*{-0.51cm} \includegraphics[scale=0.71]{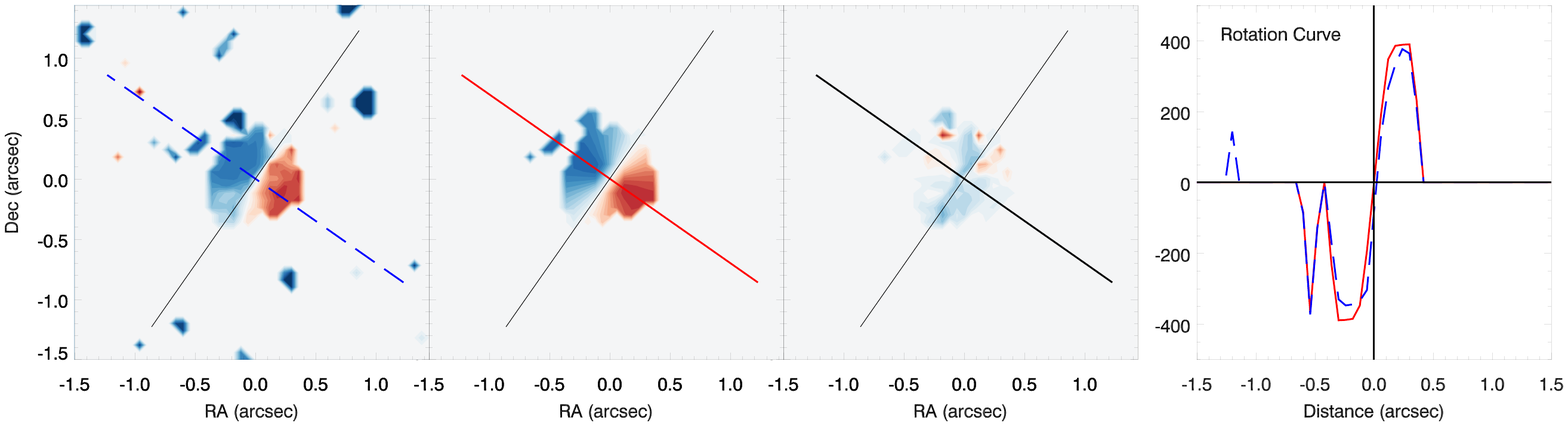} \\ \vspace{-11.08cm} 
   \hspace*{-0.51cm} \includegraphics[scale=0.71]{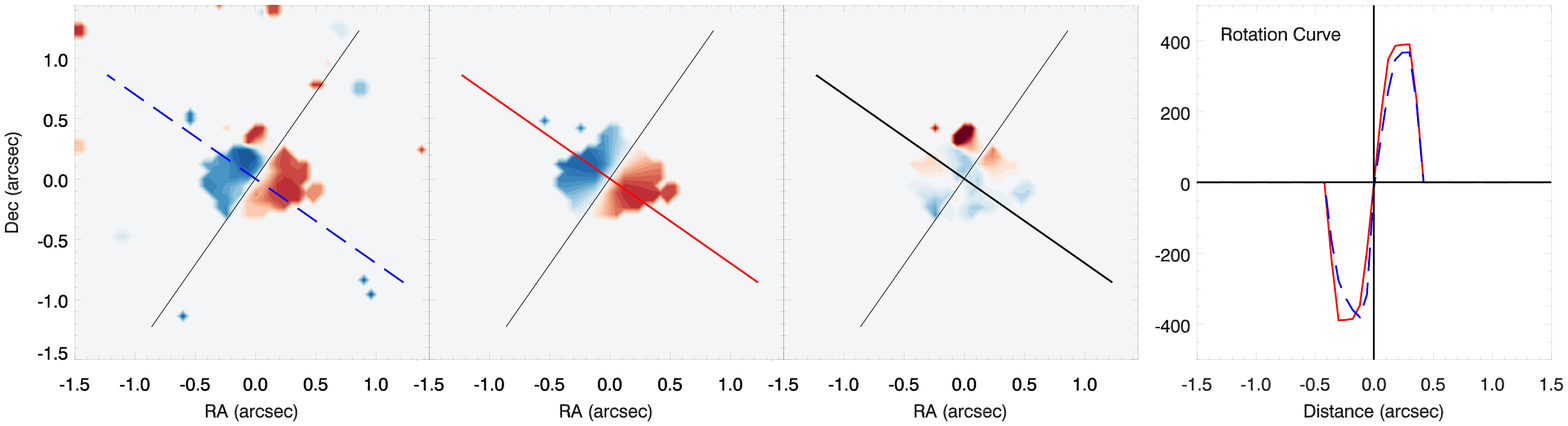}\vspace{-4.9cm}\\ 
   \caption{Results from the rotation-only modelling of the kinematics of component W.
    From left to right: the observed velocity (moment 1) maps, 
    our rotation-only model (based on parameters derived from \kin; see Section 3.2), 
    the velocity field residuals 
    (observed velocity field minus the rotation-only model), and 
    our derived rotation curves for, top to bottom: CO J:3--2,  CO J:7--6 and \cishort.
    Each observed velocity field was input to the \kin\ package to
    determine the kinematic PA and inclination (see text): these independently derived
    values are similar for all lines except CO J:3--2. 
    The pure rotation models were then constructed (by eye) using these
    values of PA and inclination together with a simple model of solid body rotation which
    changes to a flat rotation beyond a certain radius.
    In the left three columns, the colours follow the common colour bar (in km/s) at the top, and  
    areas with low signal to noise in the moment 0 map are masked. The 
    lines denote the major and minor axes, respectively. 
    The rotation curve panels show the observed radial (i.e.\ projected) velocity along the major axis
    in \kms\ on the $y$-axis (blue dashed line) and the model rotation curve (red solid line).
    }
   \label{figvelmodelw}
   \end{figure*}

   \begin{figure*}[h]
   \centering 
   \vspace{-4.9cm}
   \hspace*{-0.51cm} \includegraphics[scale=0.71]{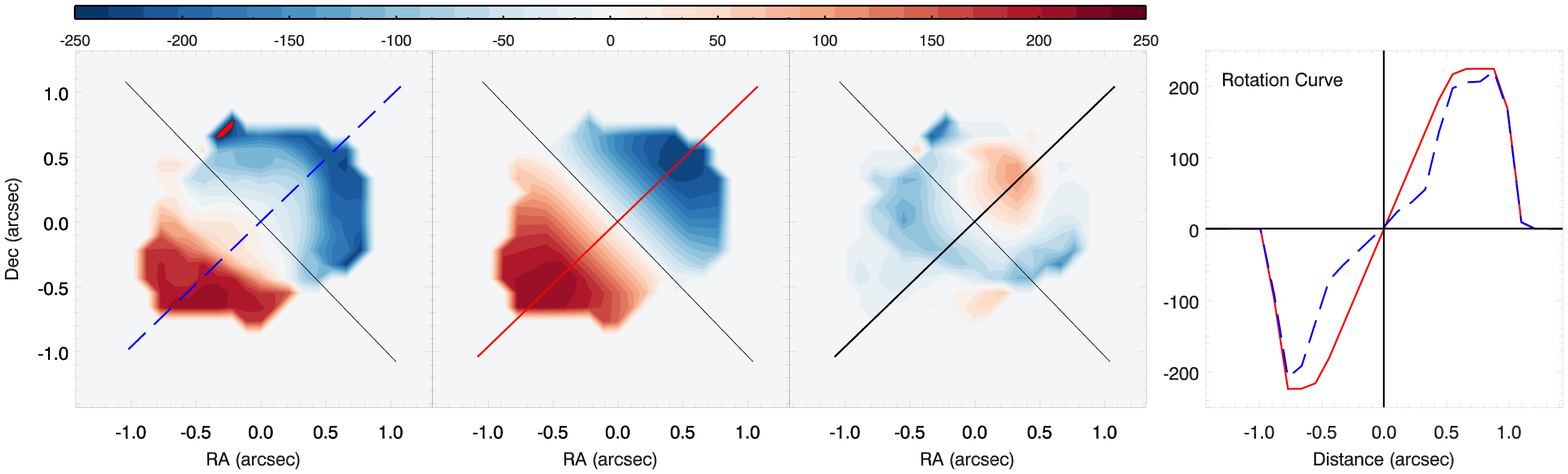}\\ \vspace{-10.55cm}
   \hspace*{-0.51cm} \includegraphics[scale=0.71]{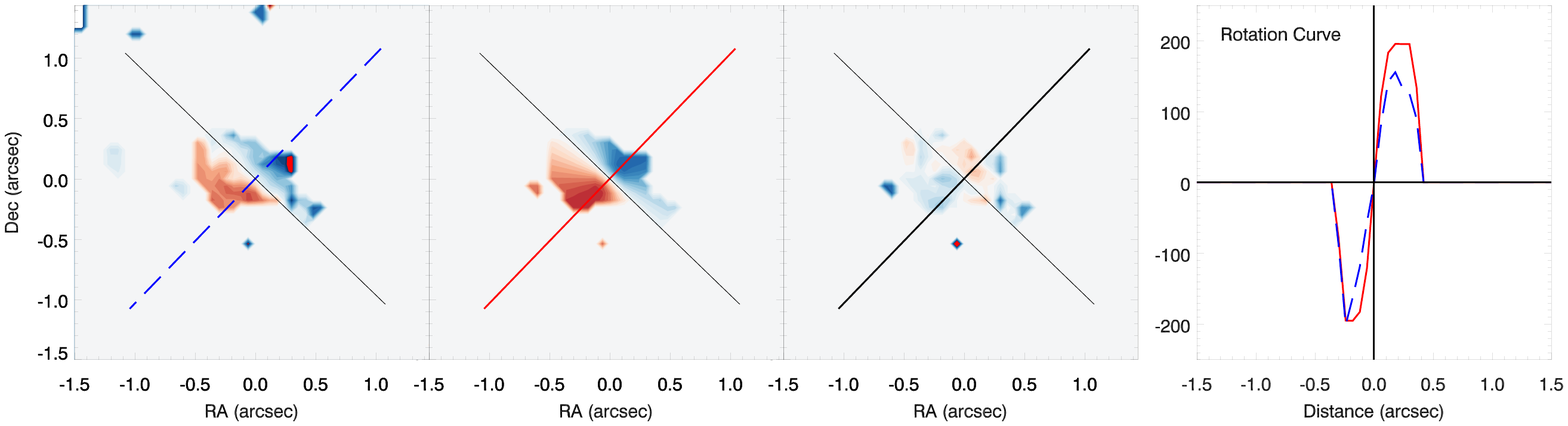}\\ \vspace{-11.08cm}
   \hspace*{-0.51cm} \includegraphics[scale=0.71]{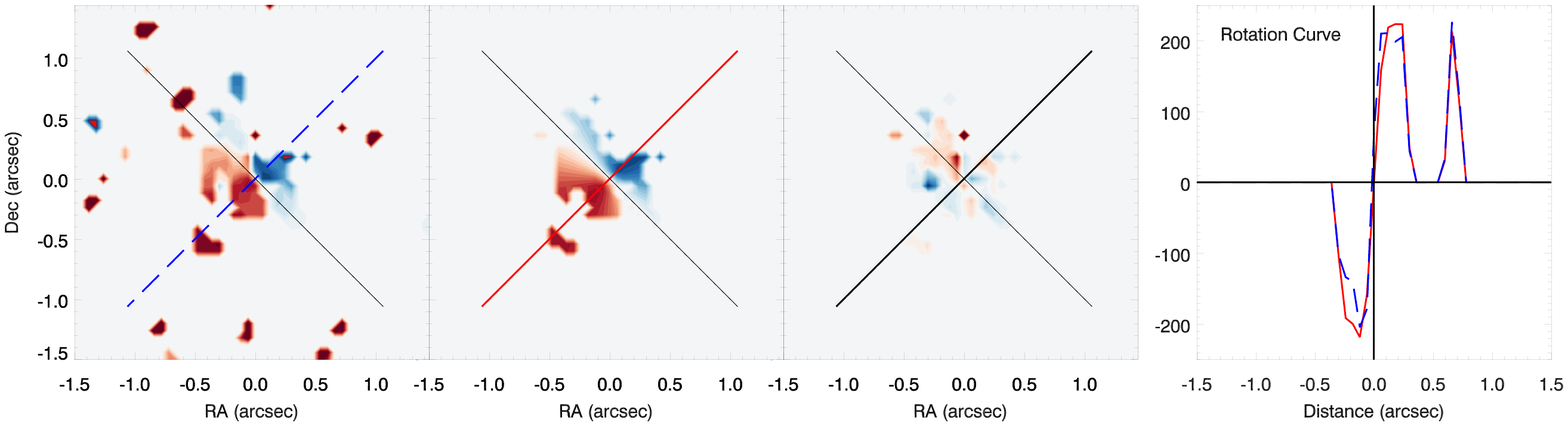}\\ \vspace{-4.9cm}
   \caption{As in Fig.~\ref{figvelmodelw} but for component T of  HATLAS J084933.4+021443
   }
   \label{figvelmodelt}
   \end{figure*}
   
As mentioned in the previous section, the ordered velocity fields in W and T could be interpreted
as either rotation or outflows within the disk. The rotation scenario -- the more likely one -- is supported
by the pattern seen in the position-velocity diagrams presented in Section~\ref{secoutflows}. 
The velocity fields of all strongly detected emission lines in components W and T were input to
the \textit{Kinemetry} package \citep{kraet06} in order to constrain the kinematic
centre, position angle
of the line of nodes, and the inclination, and the rotation curve of the galaxy, all under the
assumption that the velocity field is rotation-dominated. The initial guess for the nuclear 
position was that of the rest-frame 1160-GHz continuum peak.
The resulting parameters obtained by \kin\ are listed in Tables~\ref{tabw} and ~\ref{tabt}, and in
the caption of Fig.~\ref{fig1160ghz}.
Note that in the tables, $V_{\mathrm{asym}}$ is the asymptotic inclination-corrected rotational velocity.
Clearly within each component the \kin-determined
PA and inclination are similar for CO J:3--2, CO J:7--6, and \cishort, though the results from the
CO J:3--2 line are significantly different, most likely due to their significantly lower spatial resolution. 
We created a simple toy rotation model for each of W and T. This rotation model follows solid body rotation at small radii and a flat 
rotation curve at larger radii: \\
\begin{equation}
V_{\mathrm{radial}} = \left\{
\begin{array}{ll}
	V_{\mathrm{slope}} \cdot r \cdot \cos(\phi) \cdot \sin(i)      & \mathrm{if\ } r \le r_{\mathrm{flat}} \\
	V_{\mathrm{slope}} \cdot r_{\mathrm{flat}} \cdot \cos(\phi) \cdot \sin(i) & \mathrm{if\ } r >  r_{\mathrm{flat}}\\
\end{array}
\right.
\end{equation}
where $V_{\mathrm{radial}}$ is the observed radial velocity in \kms, 
$V_{\mathrm{slope}}$ is the slope of the solid body rotation in the inner region (\kms\ kpc$^{-1}$), 
$r$ (kpc) and $\phi$ (degrees) are the polar coordinates in the disk of the galaxy, 
and $r_{\mathrm{flat}}$ is the radius at which the solid body rotation changes to a flat rotation
curve. 
The model is initially constrained using the \kin-derived parameters 
(Tables \ref{tabw} and \ref{tabt}).
The kinematic centre derived by \kin\ was used for each galaxy, and we varied 
the values of $V_{\mathrm{slope}}$ and $r_{\mathrm{flat}}$ so as to get the best agreement (via visual
inspection) between the toy
model, the velocities seen along the major axis of each component, and the rotation curve produced
by \kin. We use this toy model for all further analysis (e.g.\ residual velocity maps) as it relatively well follows the
\kin-derived model but smooths out the small-scale variations of the latter (which are most likely caused
by signal-to-noise and resolution limitations, rather than being true physical features). 

The results of the rotation-only modelling of  components W and T are shown in 
Figs~\ref{figvelmodelw} and \ref{figvelmodelt}, respectively. For both components W and T,
the \kin\ analysis and thus the rotation-only models are very similar for both CO J:7--6
and \cishort. The models for CO J:3--2 also give similar PAs and inclinations, but the best-fit rotation
curve is relatively smooth, as expected given the lower spatial resolution of the data.
In Figs~\ref{figvelmodelw} and \ref{figvelmodelt}  we also present (in the third column) the residuals between 
the observed velocity field and the toy model velocity field, 
i.e.\ $V_{galaxy} - V_{model} $, in order to highlight deviations from pure rotation. 
The residuals are typically $\lesssim$50 \kms, and there are no immediately obvious patterns of ordered
non-circular motions. For example, outflows in the disk could show opposite colours along each half minor axis, a signature not seen in the residual velocity fields.

\subsection{Non-circular Disk Kinematics: testing or outflows and P~Cygni profiles}
\label{secoutflows}

\begin{figure*}
   \centering
   \hspace*{-15.7cm}  \includegraphics[scale=0.42]{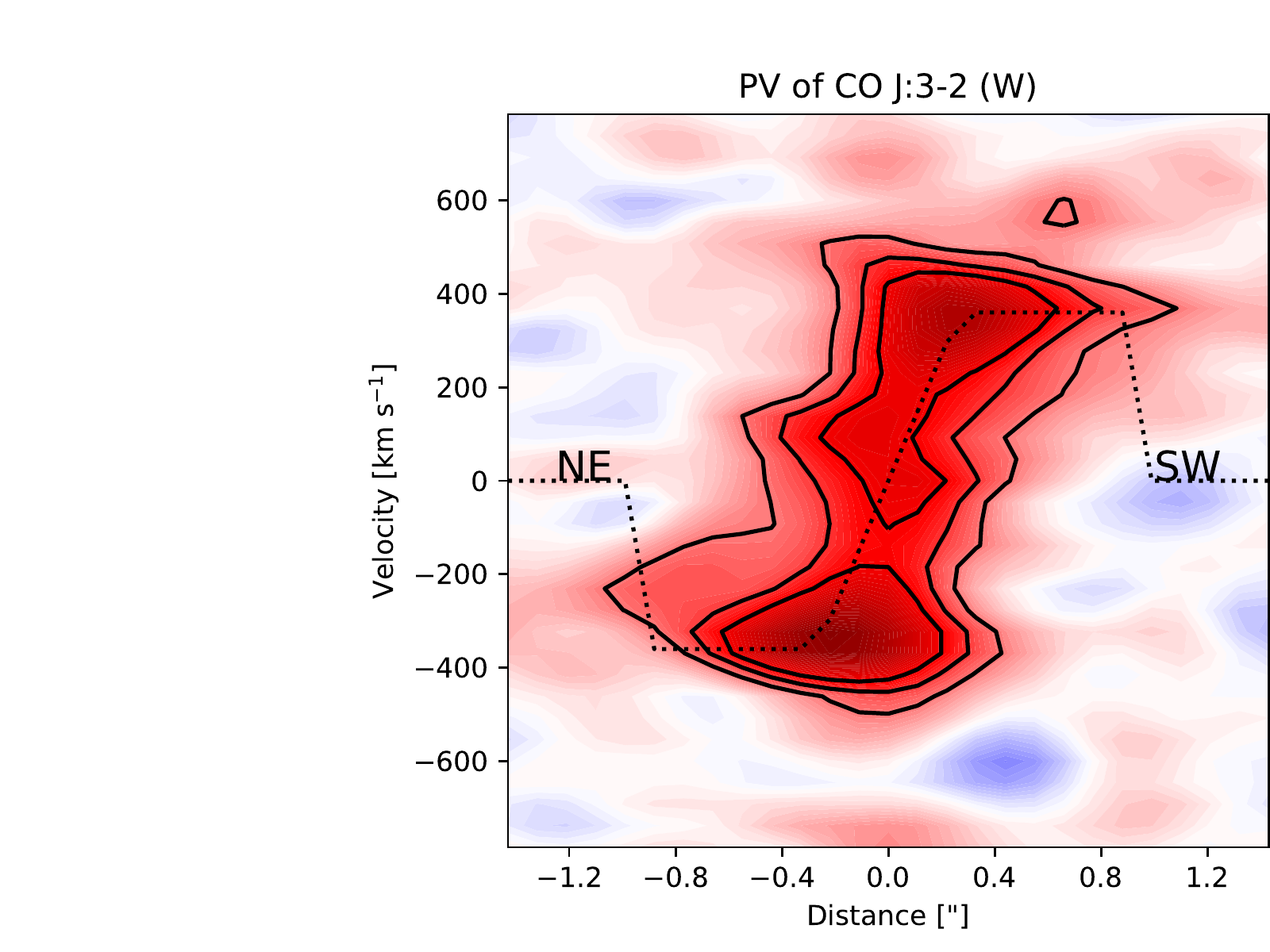}\\  \vspace{-5.158cm}    
   \hspace*{-2.08cm}  \includegraphics[scale=0.42]{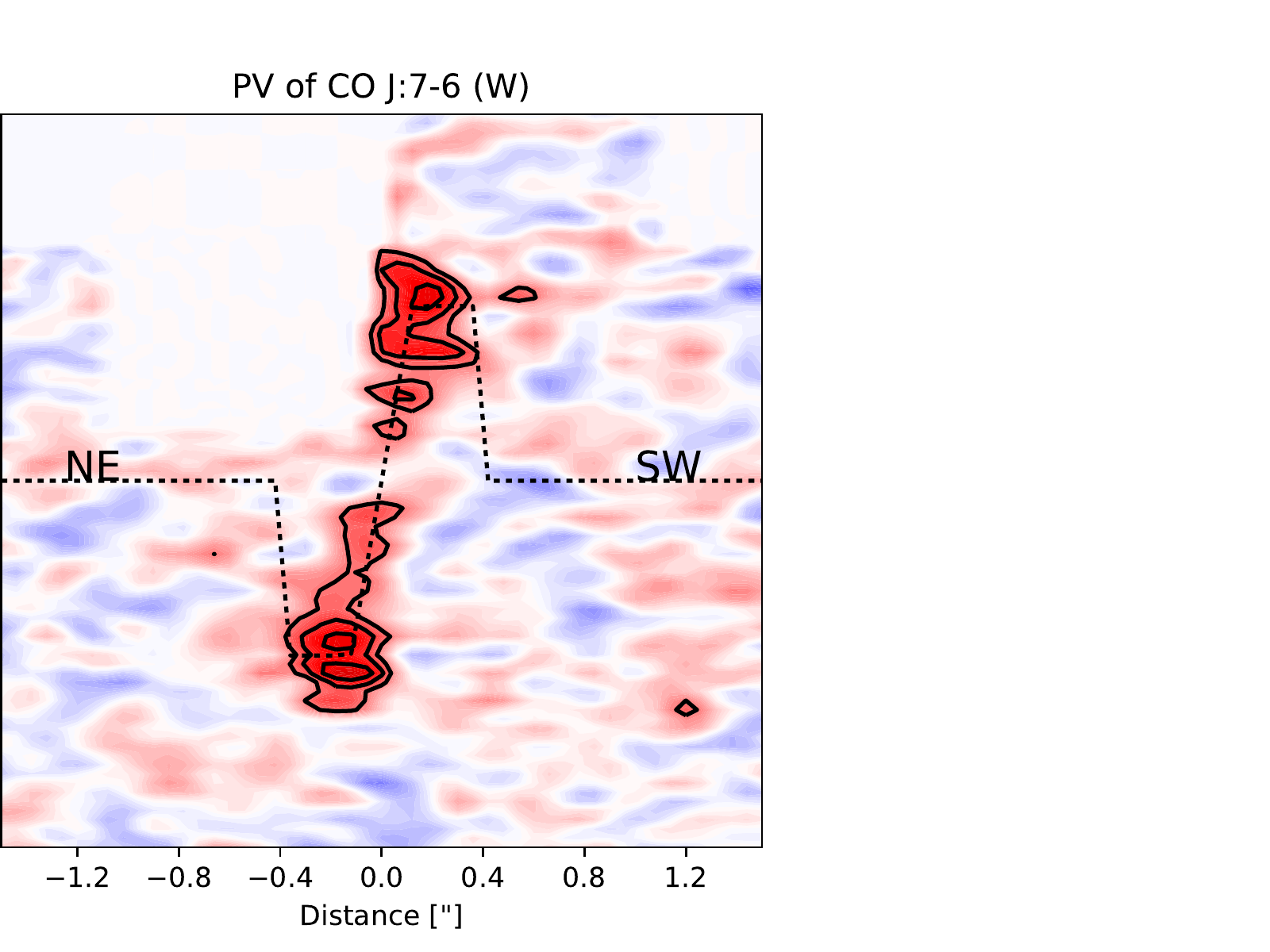}\\  \vspace{-5.158cm} 
   \hspace*{6.1cm}    \includegraphics[scale=0.42]{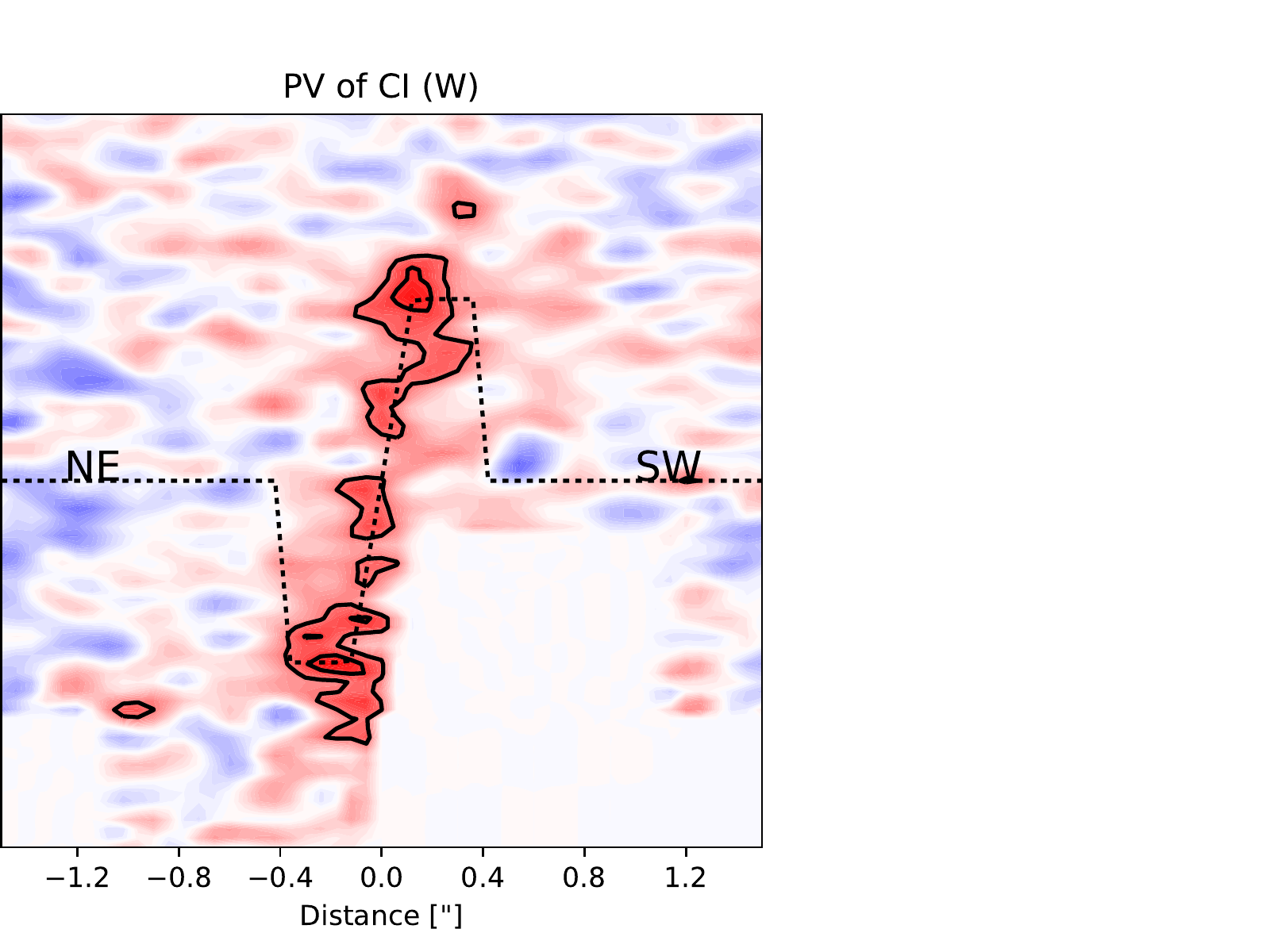}\\    \vspace{-5.158cm}
   \hspace*{12.9cm}   \includegraphics[scale=0.42]{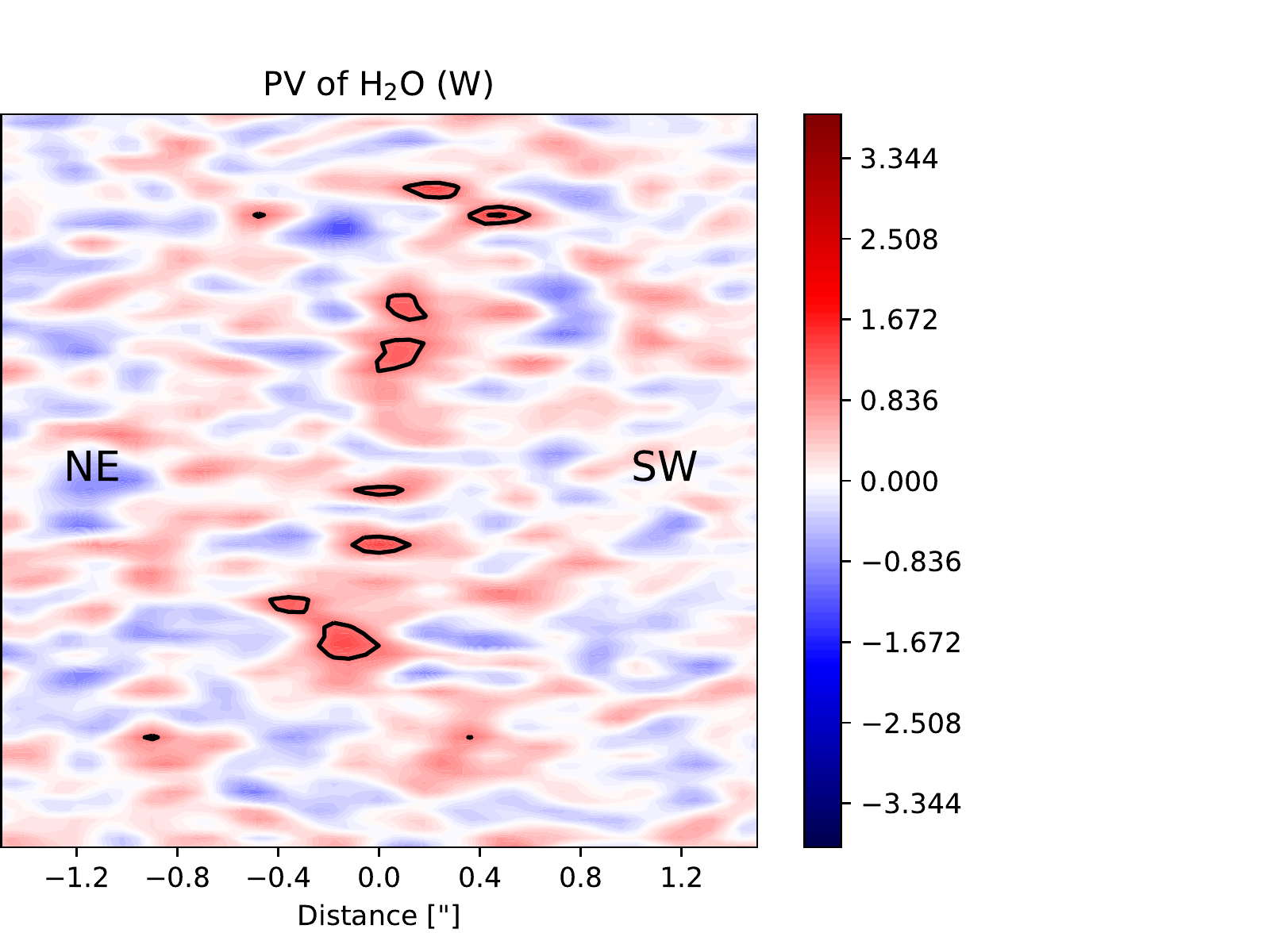}\\
   
   \hspace*{-15.7cm}  \includegraphics[scale=0.42]{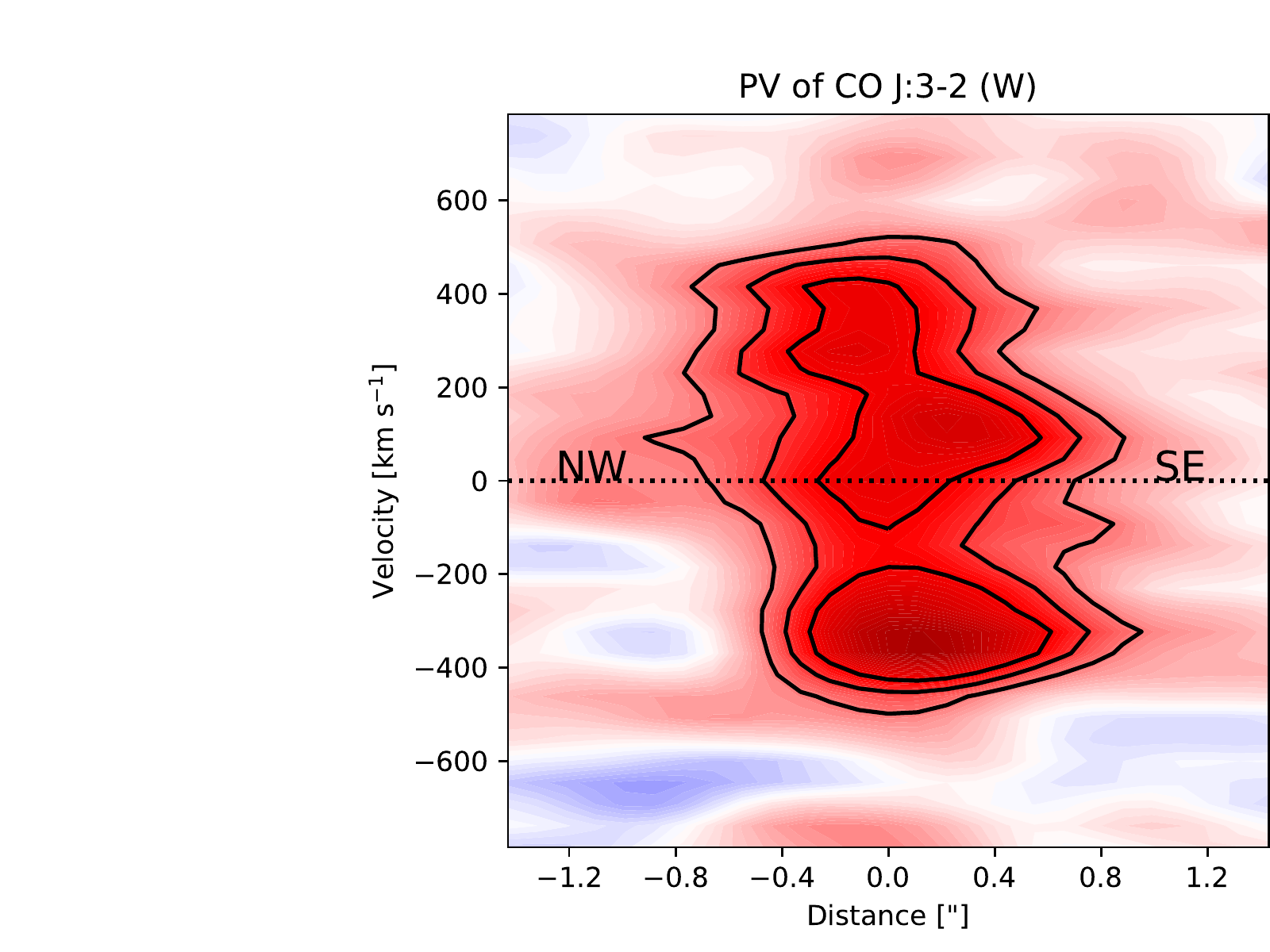}\\  \vspace{-5.158cm}
   \hspace*{-2.08cm}  \includegraphics[scale=0.42]{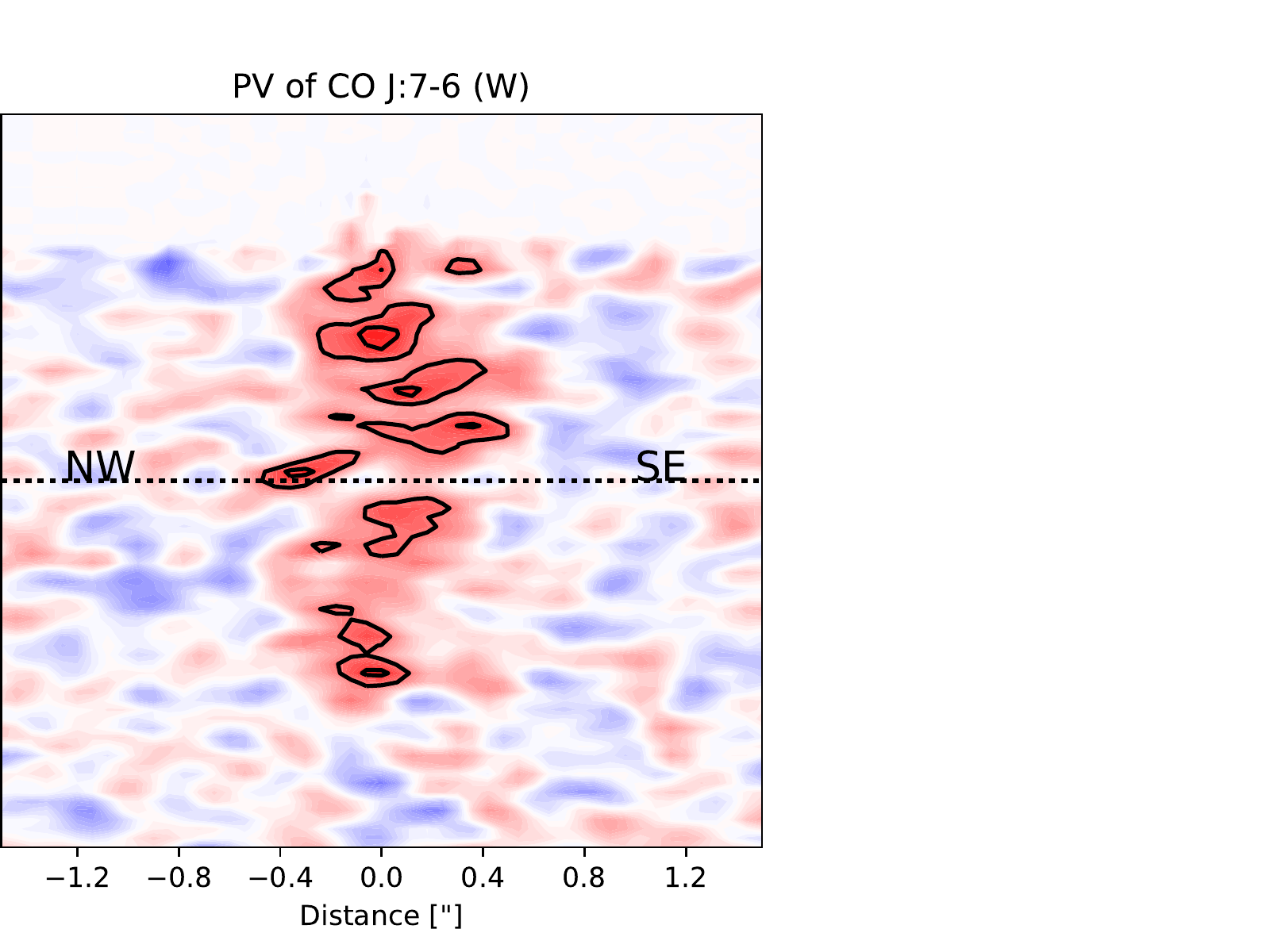}\\  \vspace{-5.158cm} 
   \hspace*{6.1cm}    \includegraphics[scale=0.42]{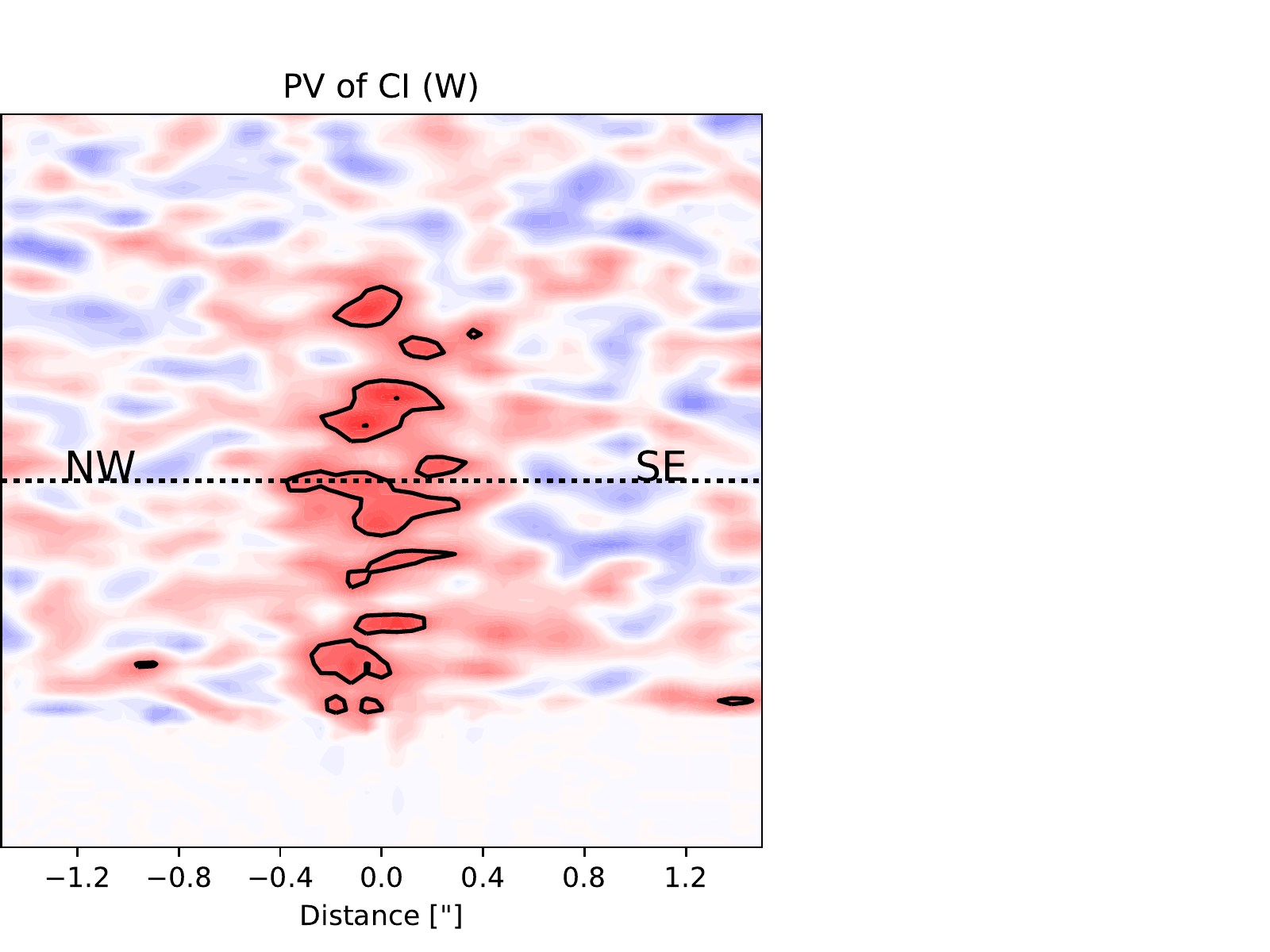}\\    \vspace{-5.158cm}
   \hspace*{12.9cm}   \includegraphics[scale=0.42]{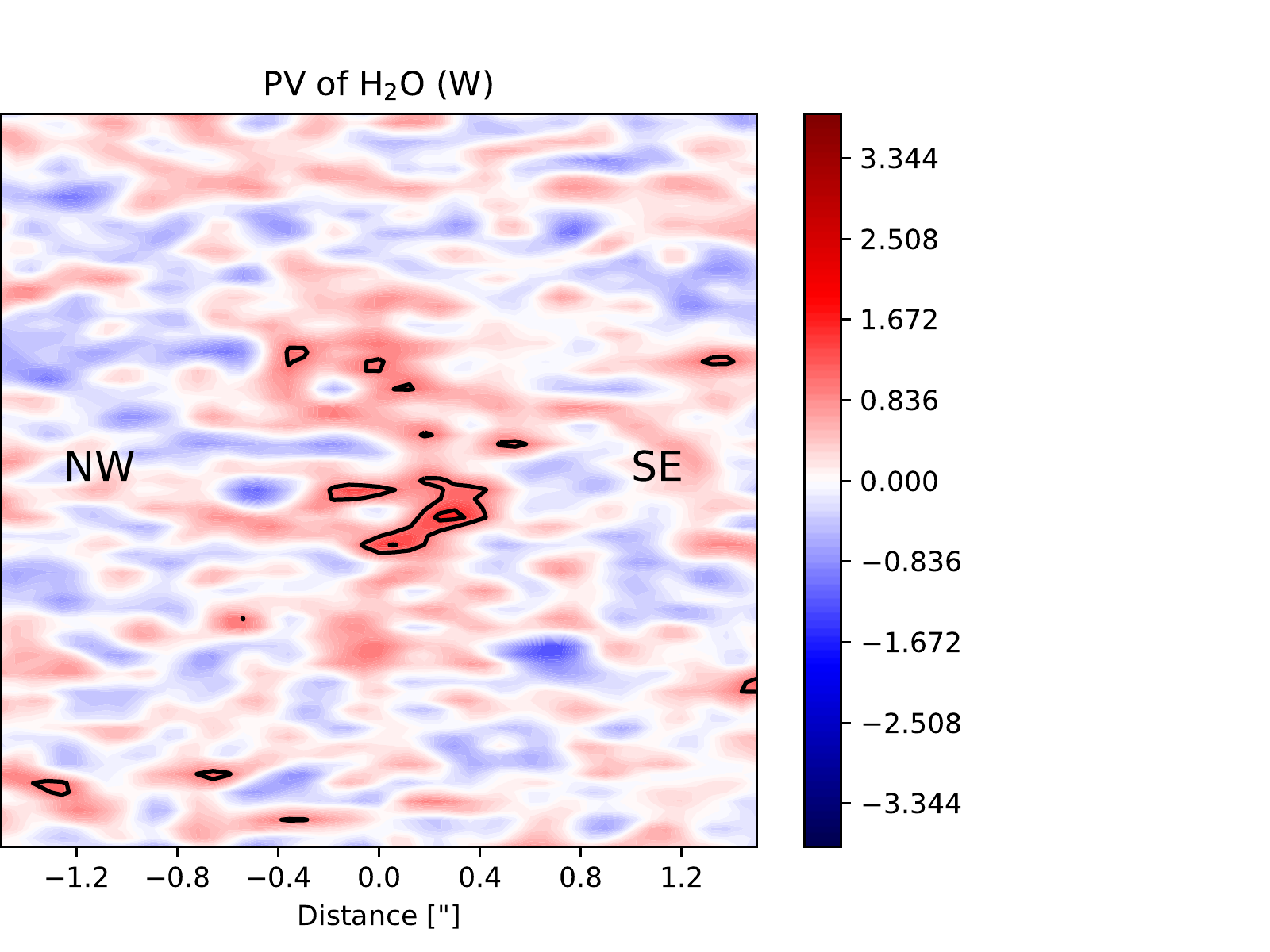}\\
   
   \caption{Position-velocity (PV) diagrams for (left to right) the CO J:3--2, CO J:7--6 and \cishort\  emission
   lines in component W are shown in colour following the common colour bar to the far right of each row (flux in units of mJy \kms\
   $\mathrm{beam^{-1}}$) and in black contours (1, 1.5 and 2 mJy \kms
   $\mathrm{beam^{-1}}$). 
   At our adopted distance for HATLAS\,J084933, $1''$ corresponds to 
   8.25 kpc. 
   The top (bottom) row shows the PV diagram along the major (minor) axis of component W.
   The dashed black lines show the predictions of our adopted rotation model (see Sect 3.3). }
   \label{figpvw}
   \end{figure*}
   
   \begin{figure*}
   \centering
   \hspace*{-15.7cm} \includegraphics[scale=0.42]{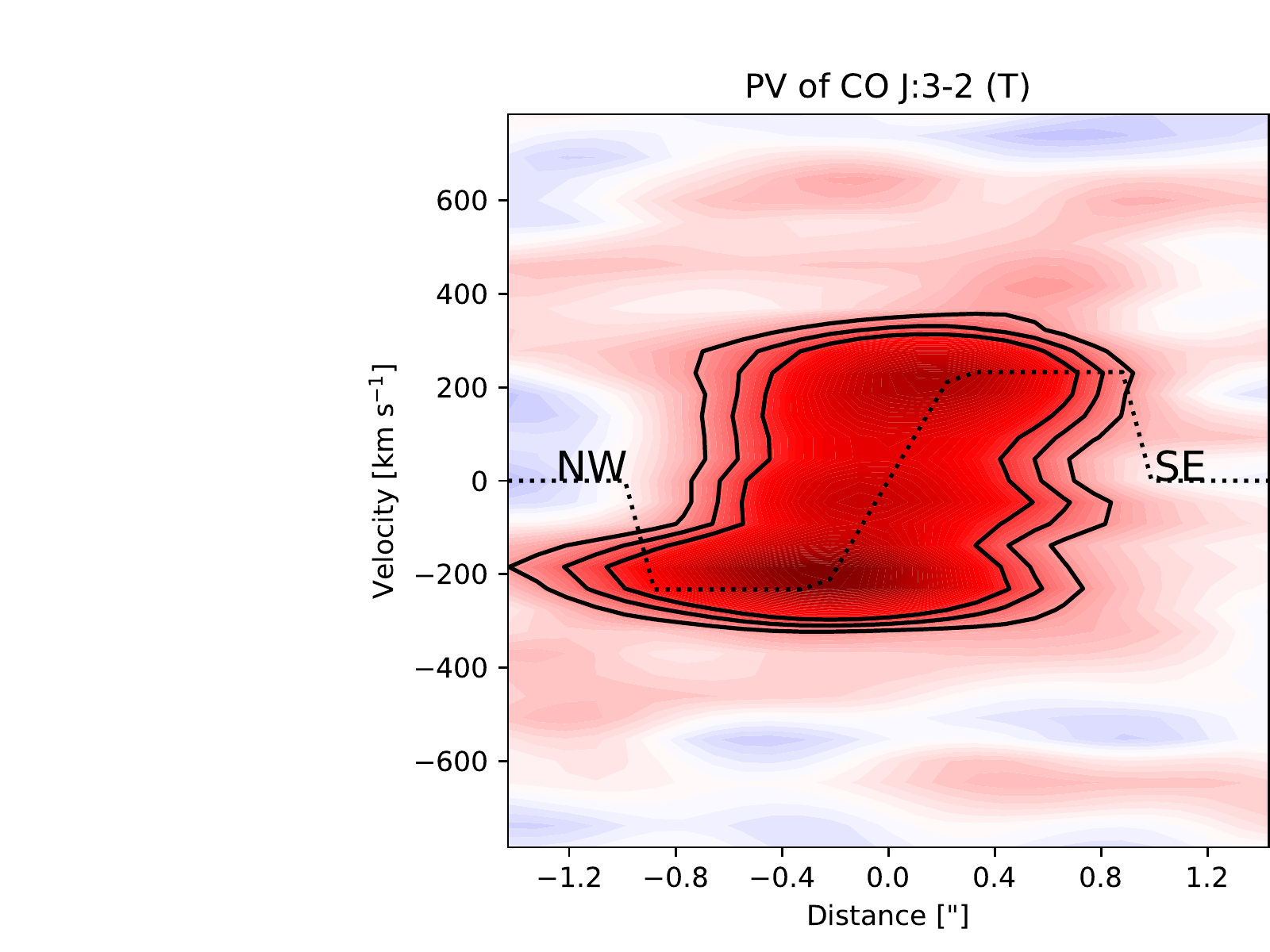}\\ \vspace{-5.158cm}    
   \hspace*{-2.08cm} \includegraphics[scale=0.42]{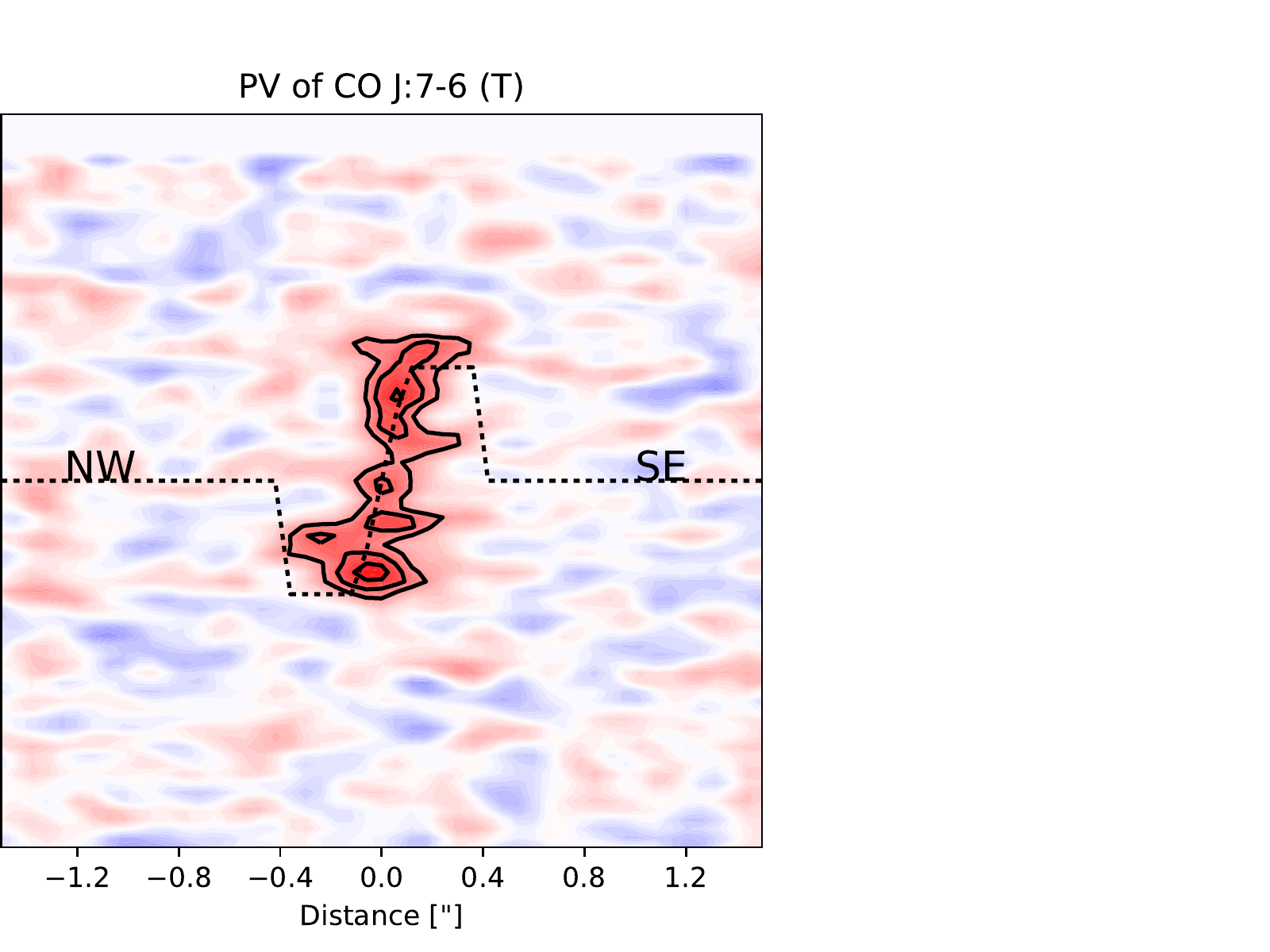}\\ \vspace{-5.158cm} 
   \hspace*{6.1cm}   \includegraphics[scale=0.42]{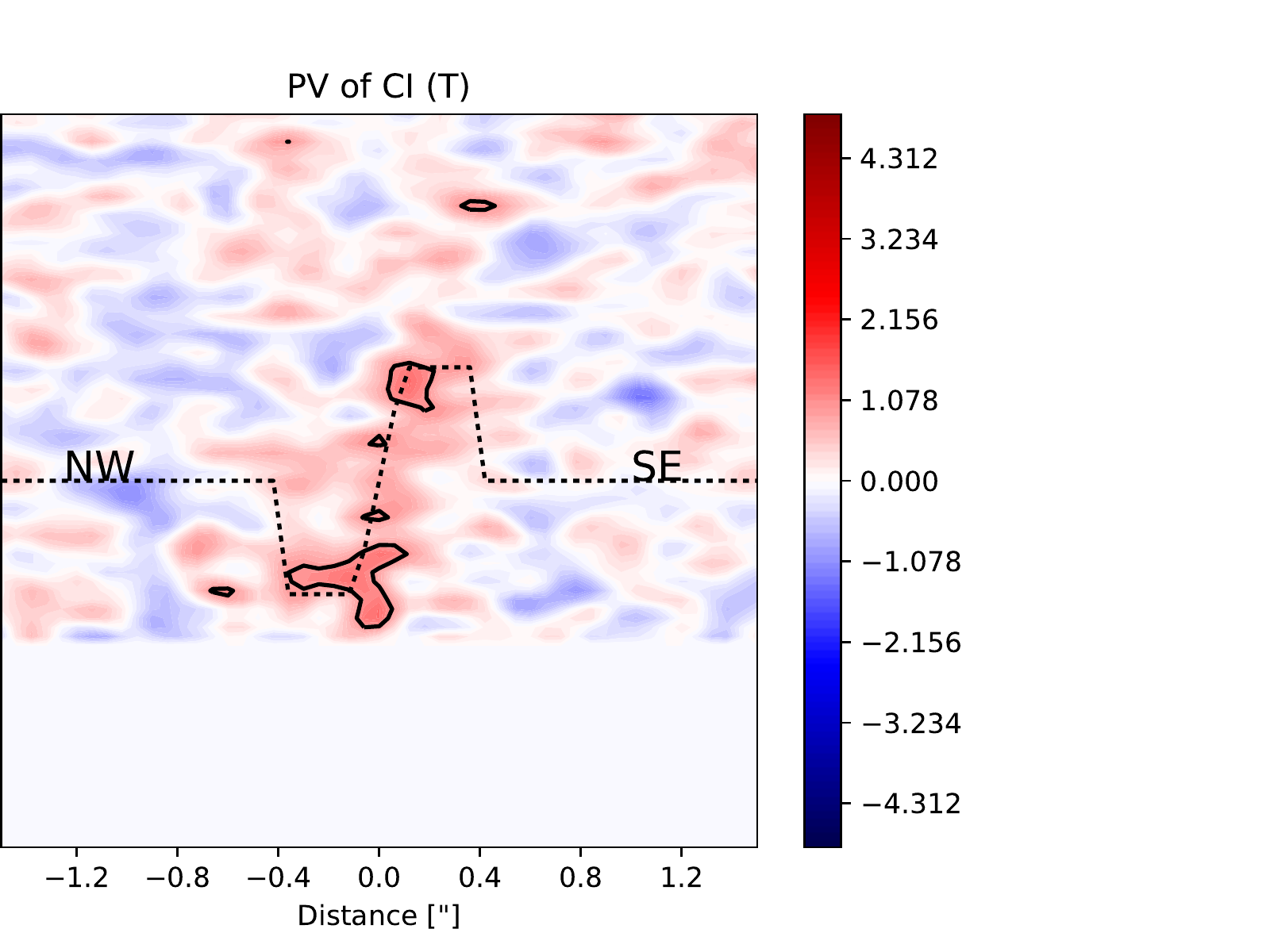}\\   \vspace{-5.158cm}
   \hspace*{12.9cm}   \includegraphics[scale=0.42]{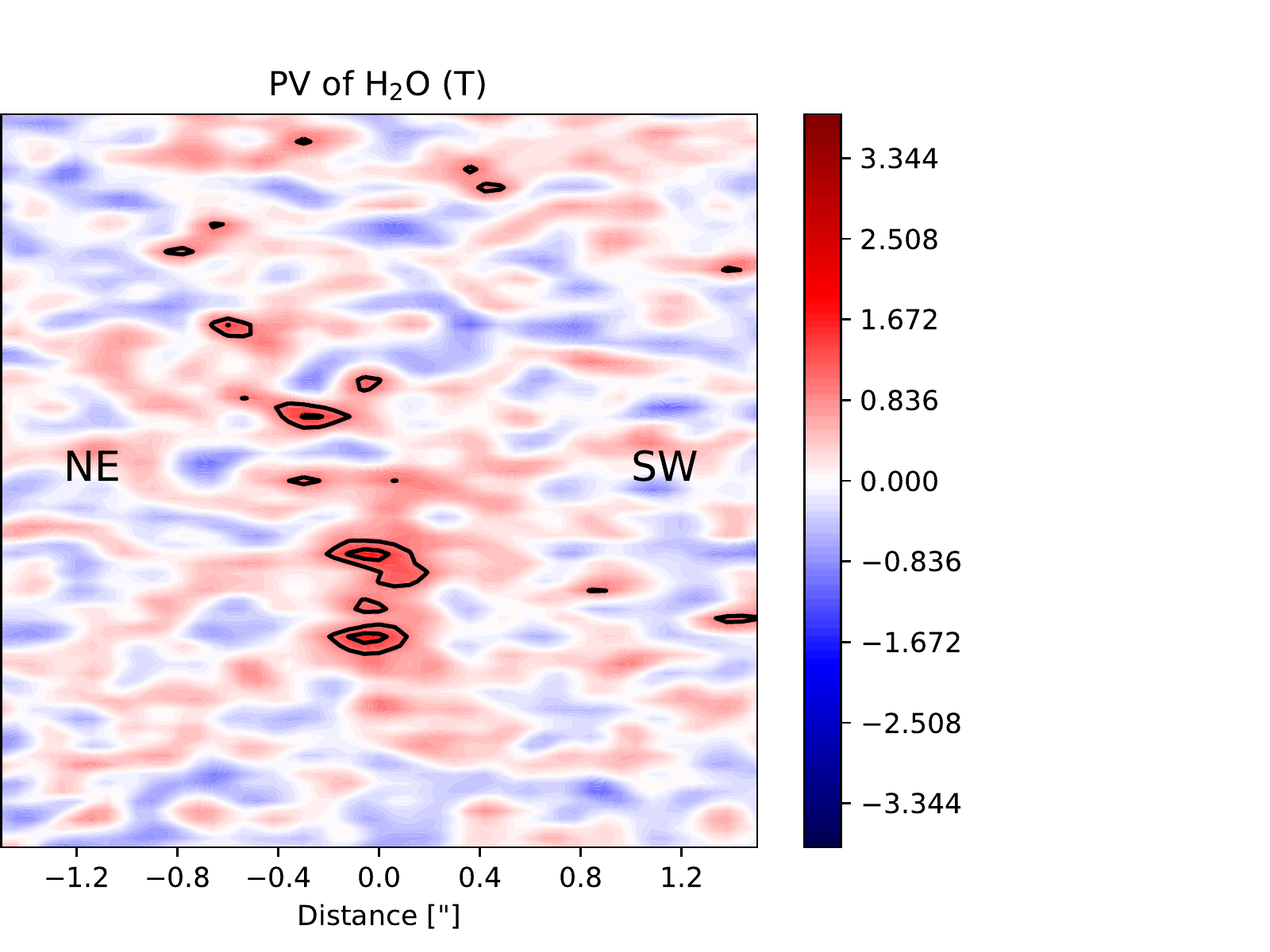}\\
   
   \hspace*{-15.7cm} \includegraphics[scale=0.42]{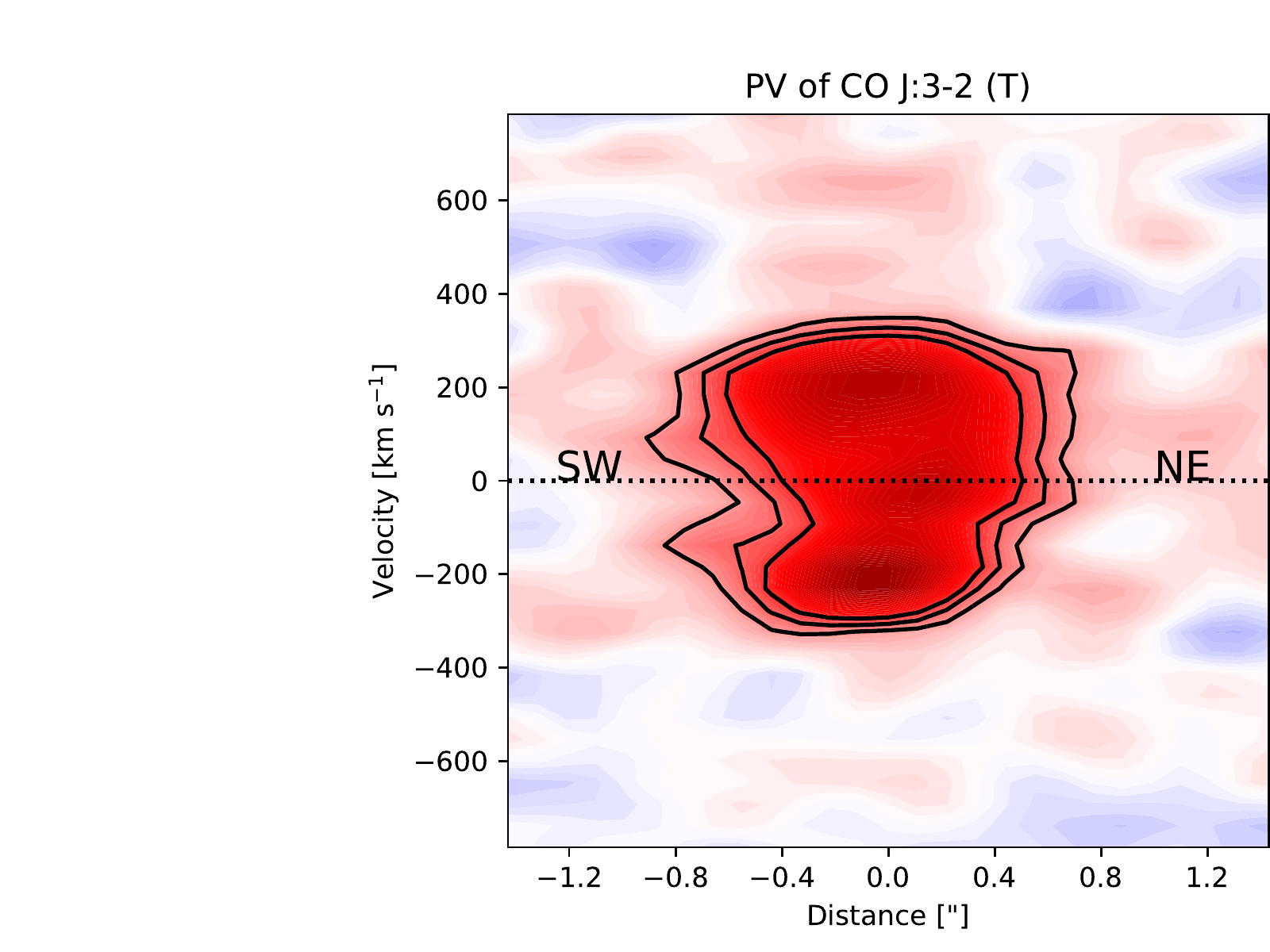}\\ \vspace{-5.158cm}
   \hspace*{-2.08cm} \includegraphics[scale=0.42]{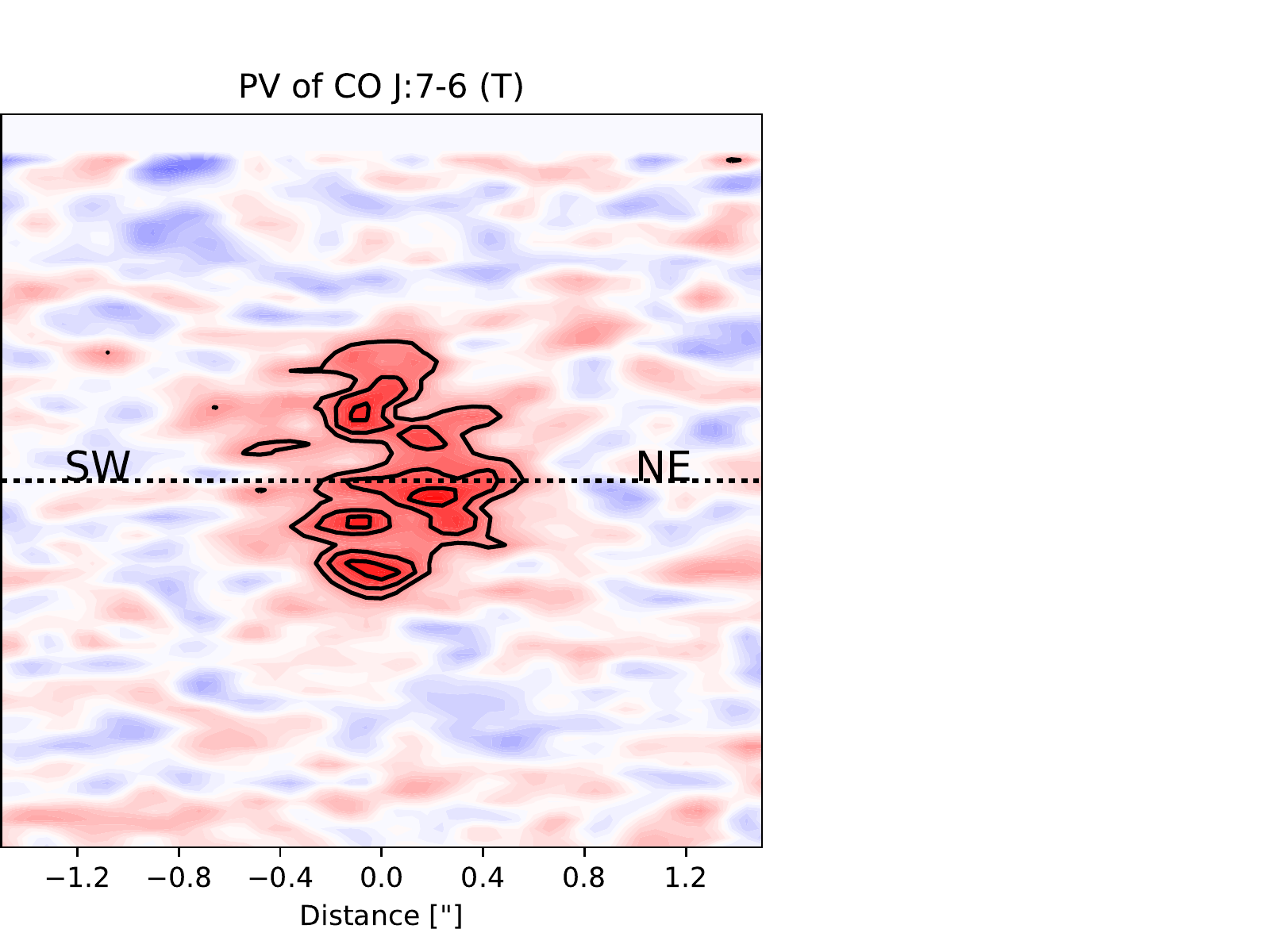}\\ \vspace{-5.158cm} 
   \hspace*{6.1cm}   \includegraphics[scale=0.42]{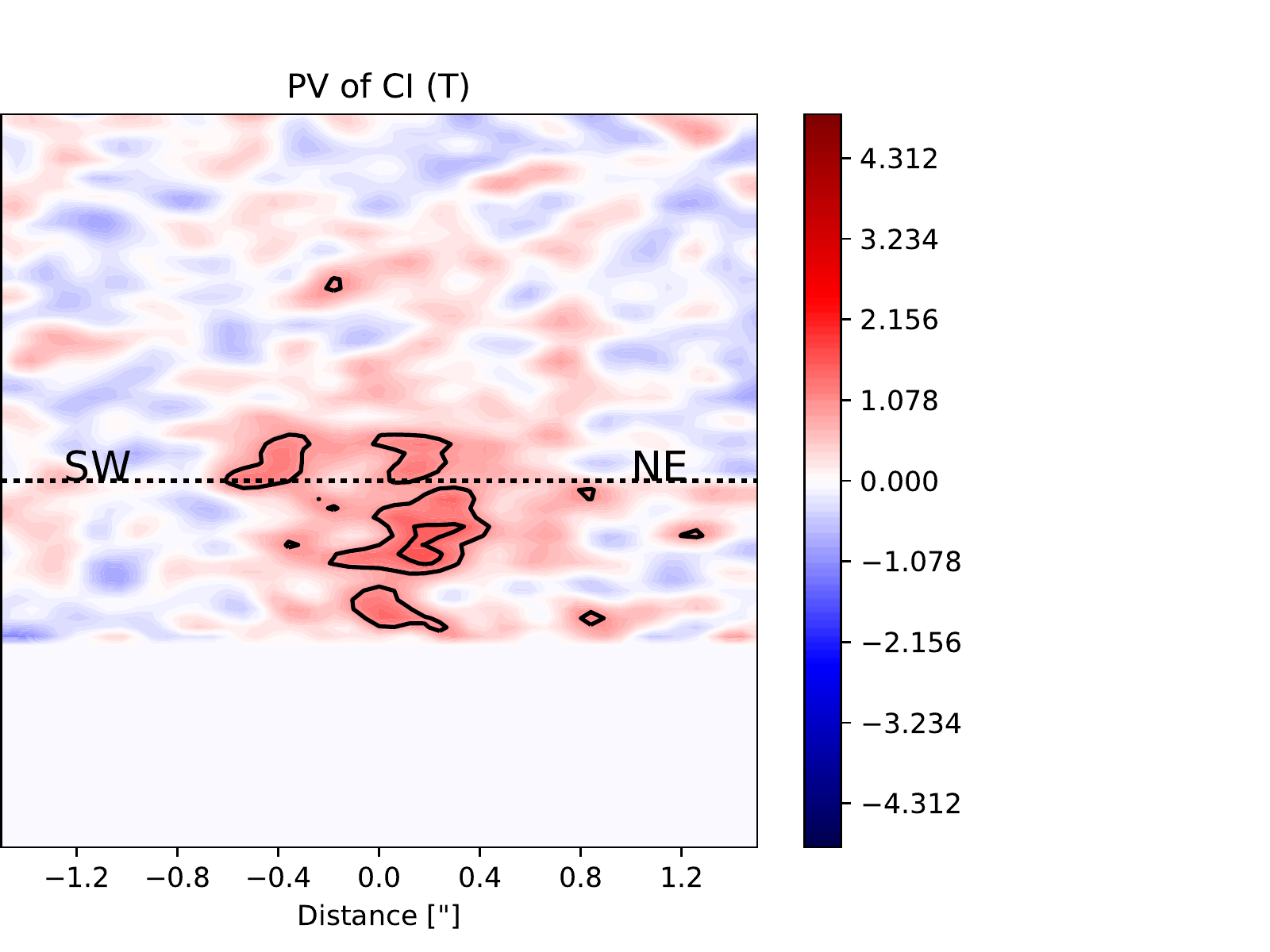}\\   \vspace{-5.158cm}
   \hspace*{12.9cm}   \includegraphics[scale=0.42]{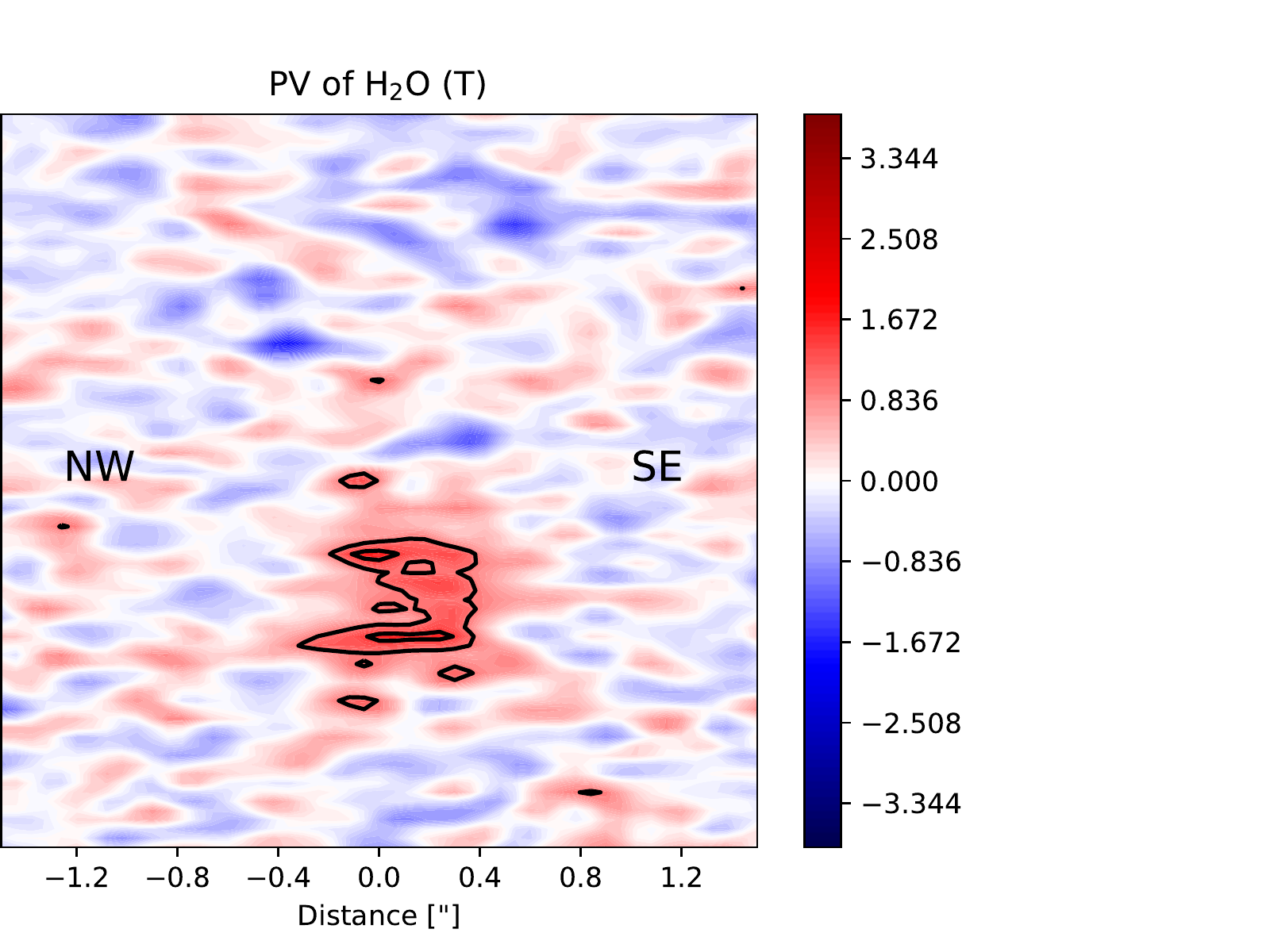}\\
   
   \caption{As in Fig.~\ref{figpvw}, but for component T.}
   \label{figpvt}
   \end{figure*}
   
Position-velocity (PV) diagrams of all four detected lines are shown in Figs~\ref{figpvw} and \ref{figpvt})
for components W and T, respectively. The PV diagrams were extracted along the kinematic major 
and minor axes discussed in the previous section: kinematic major axes PAs of 55\arcdeg\ and 135\arcdeg\ for W and T, respectively; see Figs~\ref{figvelmodelw} and \ref{figvelmodelt}),
using a pseudo-slit of width 0.06\arcsec\  for CO J:7--6, and [C\,{\sc i}] 2--1, and 0.11\arcsec\ for CO J:3--2; they are thus 
limited in spatial resolution by the intrinsic spatial resolution of the images.
On each PV diagram we have overlaid the predictions 
of our toy rotation model (black dashed line)  described in Section~\ref{sectrot}.
Along the major axis, in both W and T, the toy rotation model follows well the velocity structure
of the CO and \cishort\ gas: in fact, especially in CO J:3--2, one can discern the point at which the presumed rotation
changes from solid body to flat. The gas velocities do show interesting wiggles away from the prediction
of the rotation model -- see e.g.\ the radii close to the nucleus and to the SW in the CO J:3--2 major axis
PV diagram -- but a higher spatial resolution and signal-to-noise is required to model these deviations.
The kinematic signatures of W and T in their PV diagrams along the posited kinematic major axis, specifically
the velocity gradient seen close to the nucleus, is a strong argument 
that the kinematics is rotation-dominated. Outflows within the disk would produce a more abrupt change in
velocity from one side of the disk to the other.
Along the kinematic minor axes of W and T, the data are more difficult to interpret: we expect to see zero velocities
at all offsets but smearing from the relatively large synthesised beam means that multiple velocity components
are seen in all spatial offsets along the minor axis. This is unfortunate since any outflow within the plane of
the disk would be most apparent along the minor axis. In W, the minor axis PV diagram of the CO J:3--2 line 
shows a $\sim$100 \kms\ feature to the SE. In fact this distortion in the zero-velocity line is also clearly
seen in the equivalent velocity field (e.g.\ top row of Fig.~\ref{figvelmodelw}). The equivalent, but fainter,
blueshifted component is seen to the NW. This could trace a molecular outflow in the disk if the SE is the far
side of the disk. However, many alternative explanations are possible, including bar driven distortions in the gas kinematics and thus we cannot at this point present evidence for outflows or inflows.

The PV diagrams presented here are also useful to highlight the different kinematics seen in the water
vapour line. In W, the water vapour line appears to follow the expectations of the rotation model, but in the
case of T the water line does not appear to follow rotation along the major axis and its velocities along
the minor axis are marginally larger than that seen in the other lines. A potential explanation
is that this line comes from outflowing gas: given the PV diagrams presented here, and the velocity and dispersion maps presented in the
previous sections,  the outflow velocities are likely to be predominantly  spherical in component T.
   
   \begin{figure*}
   \includegraphics[scale=0.557]{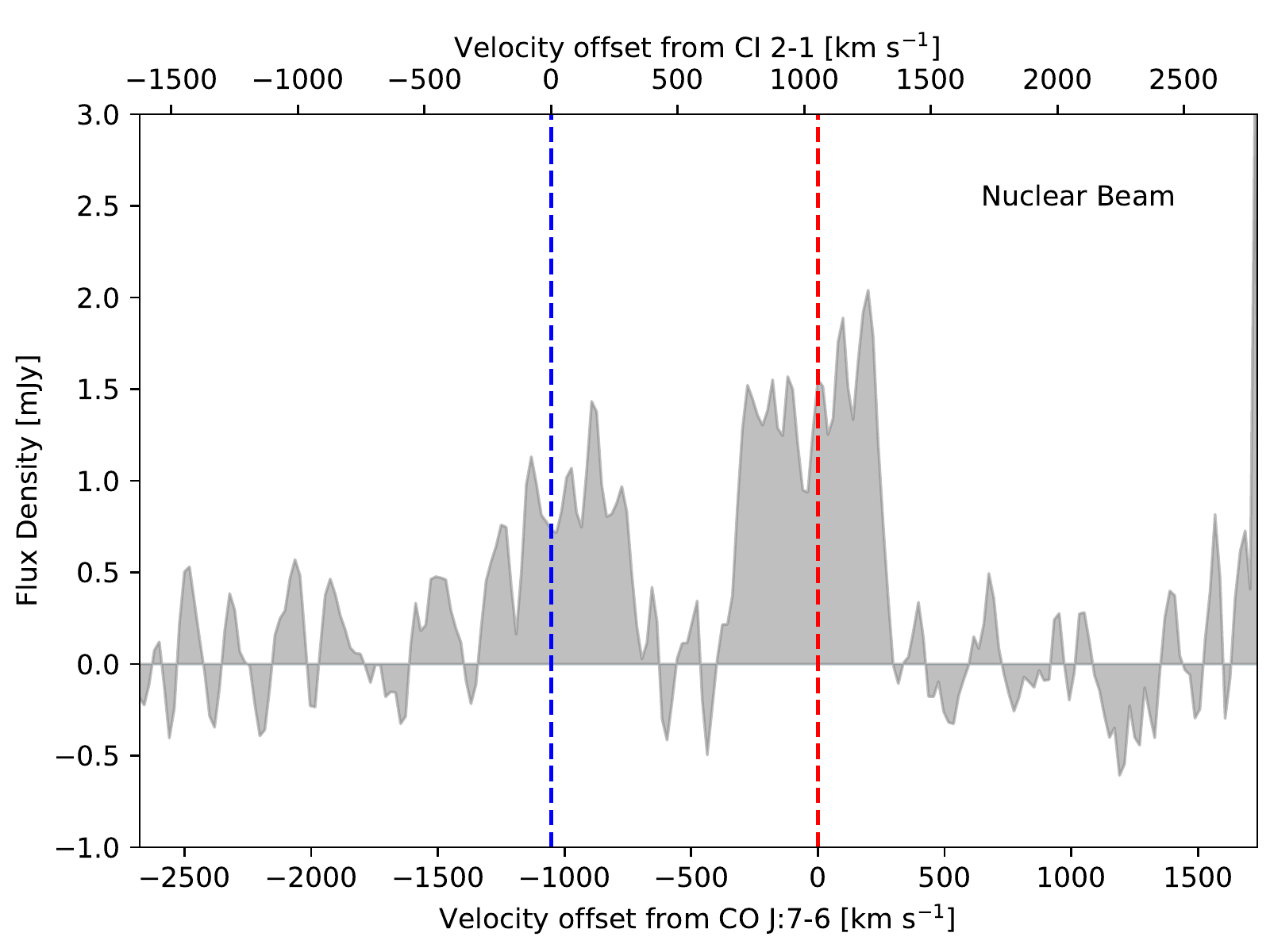}
   \includegraphics[scale=0.557]{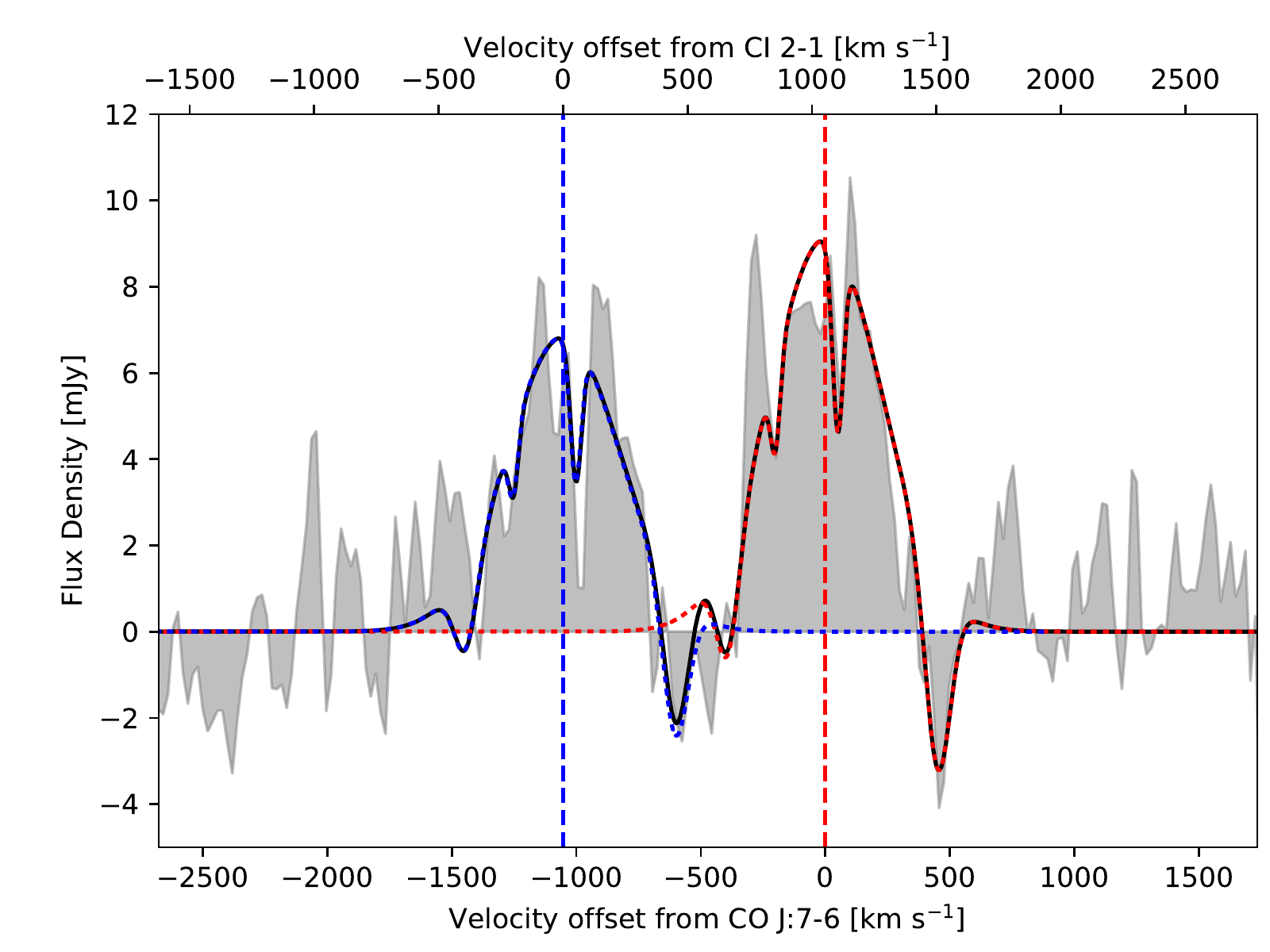}
   \caption{Potential absorption components in CO and [C\,{\sc i}]: each panel shows
   a spectrum of component T covering the  CO J:7--6 (left) and [C\,{\sc i}] 2--1 (right) 
   emission lines at a resolution of 19.7 \kms\ per channel (grey filled region). 
   The dashed blue line marks the 0-velocity of the [C\,{\sc i}] 2–-1 frequency at $z = 2.41$ (scale on the top
   $y$-axis),
   and the dashed red line marks the zero-velocity for the CO J:7–-6 line at the same redshift (scale on 
   the bottom $y$-axis). The left panel shows a nuclear spectrum of T while the right panel shows a galaxy-integrated
   spectrum of T.
   In the right panel, the solid red line shows the best fit spectrum to the CO J:7--6 line profile: the starting model was the sum of a 
   Gaussian emission component at the rest frequency of the line (0 \kms) plus four Gaussian absorption 
   components at velocities fixed (by eye) to $-$385, $-$225, +80 and +455 \kms. The amplitudes and widths
   of each of the five components were then varied to achieve the best fit using r.m.s.\ minimisation.
   The equivalent solid blue line for [C\,{\sc i}] 2--1  is 
   obtained from the best fit CO J:7--6 spectrum but using a  displacement of $-1051$ \kms\ 
   (based only on the frequency difference between the lines) and an amplitude scaling of 
   0.7 for each emission and absorption component. 
   The black solid line represents the sum of the best fit to the \co7-6\ and \cishort\ profiles (sum of 
   the red and blue spectra; this summed spectra is often not visible as it often lies under the blue and red solid 
   lines). The only relatively consistent and  significant absorption component is that at +455 \kms.
   }
   \label{figrp-cygni}
   \end{figure*}

P-Cygni (or simple absorption-line) profiles are potentially detected at similar velocity offsets (from
systemic) in  all of \co7-6, \cishort\ and \water. To quantify if any of these can be fit with a single velocity
absorption line component, we modelled the combined galaxy-wide \co7-6\ and \cishort\ spectrum of T with
a sum of emission line and absorption line components.
The best-fit model is created as follows: 
we use the sum of five Gaussian components to fit the CO J:7--6 line profile:   
one emission line centred at systemic velocity and 4 absorption lines centred at the most promising (by eye)
potential absorption lines. The free parameters are thus the amplitude and width ($\sigma$) of each
of the five Gaussians.  Since the \co7-6\ and \cishort\ lines overlap we are forced to fit them together.
We thus add to the model five more Gaussians to recreate the \cishort\ line, whose parameters are completely tied 
to the first five Gaussians : the offset between each pair of emission and absorption 
Gaussians is fixed to 1051 km/s, which corresponds to the velocity offset between the 
CO J:7--6 and [C\,{\sc i}] 2--1 lines, the amplitude ratio between each Gaussian pair is a single new free
parameter, and the widths of the new five Gaussians are fixed to 
the widths of the original five Gaussians. Thus the summed model of the \cishort\ profile is merely a
shifted (at a fixed velocity) and scaled (a single amplitude scaling for all five new Gaussians) 
copy of the summed model used for the \co7-6\ line,
and the total number of free parameters is 11 (the widths and amplitudes of the first five Gaussians, and
the global amplitude scaling factor between the five Gaussians used to recreate \co7-6\ and the five Gaussians
used to recreate \cishort)).
We then use a minimised  Chi-square ($\chi^2$) method to determine the model which best fits the observed line
profile. 

The best fit model profile obtained is compared to the observed spectrum of T in
the right panel of Fig.~\ref{figrp-cygni}.
The best-fit parameters for the CO line are as follows: (a) emission line with a FWHM of 534 km/s  (fitted)
and amplitude 9.11 mJy (fitted); (b) absorption lines centred at velocities of 
$-$390, $-$200, +50 and +450 km/s (fixed) with amplitudes of 2.59, 2.01, 4.27, and 4.45 mJy (fitted)
and FWHMs of 106, 47, 52 and 118 km/s  (fitted), respectively.
Finally a value of 0.75 (fitted) is obtained as the scaling factor in converting the above (CO) Gaussians to
the Gaussians used to form the \cishort\ line profile; this scaling also matches that obtained in component W.

The only consistent absorption line is that at 450 km/s, whose  FWHM is
118 \kms. This is present in both emission line profiles, even when these are smoothed to spectral
resolutions of $\approx$100 km/s (see figure \ref{figmomt}). The spectrum of the nuclear aperture
(left panel of Fig.~\ref{figrp-cygni}) also shows these absorption components but they
are not as strong as in the global spectrum. If true, this would imply
a P-Cygni-like spectrum but tracing absorption in an inflow, which is highly unlikely. In any
case, given the many other potential absorption 
features seen in the profiles we cannot make any definitive statement on the reality of this absorption line.

\subsection{Resolved Line Ratios: gas excitation and \cishort\ as a tracer of molecular gas}

   \begin{figure*}
   \centering
   \hspace*{-7cm}\includegraphics[scale=0.45]{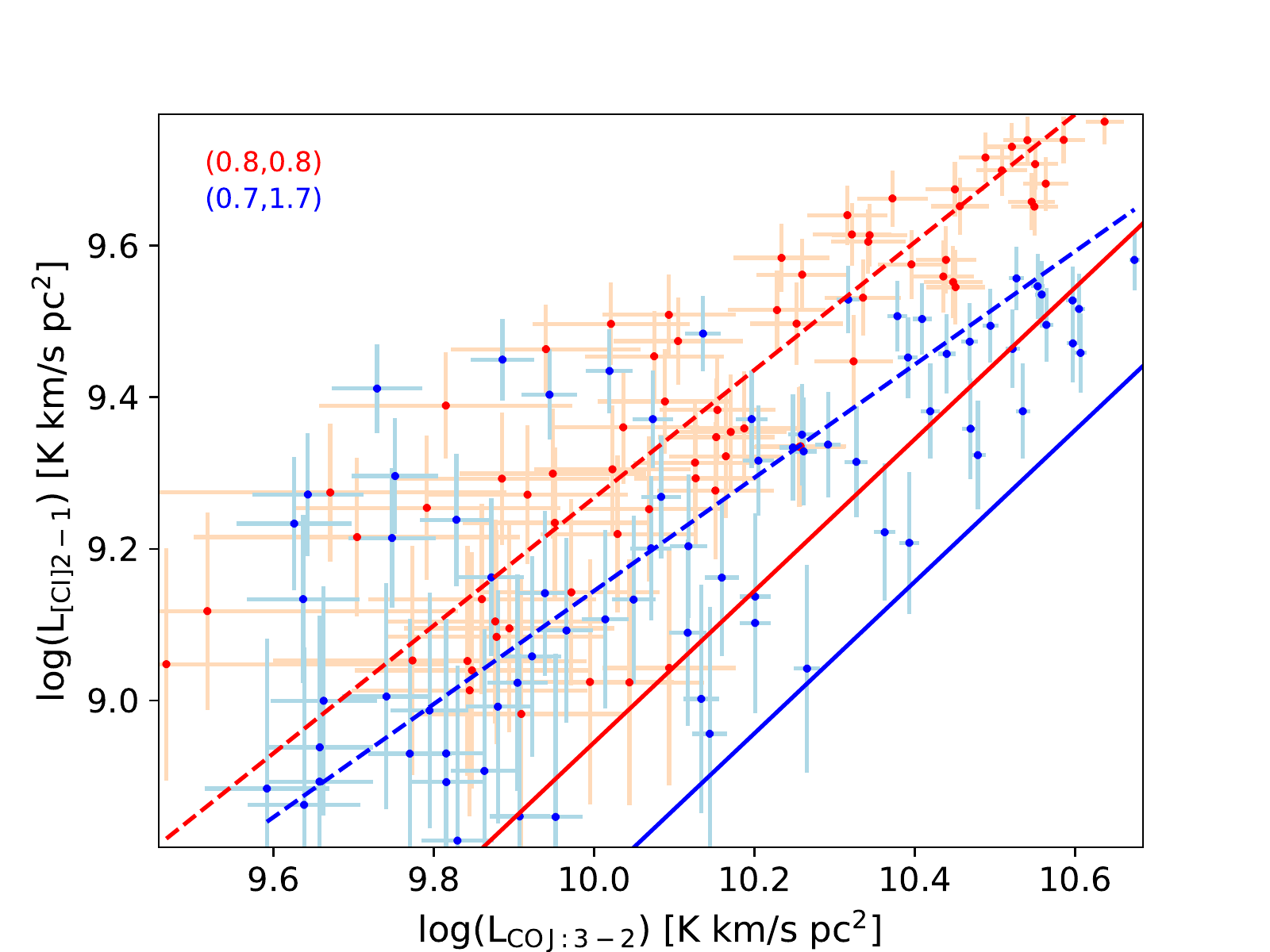}  \\ \vspace{-5.55cm}
   \hspace*{ 7cm}\includegraphics[scale=0.45]{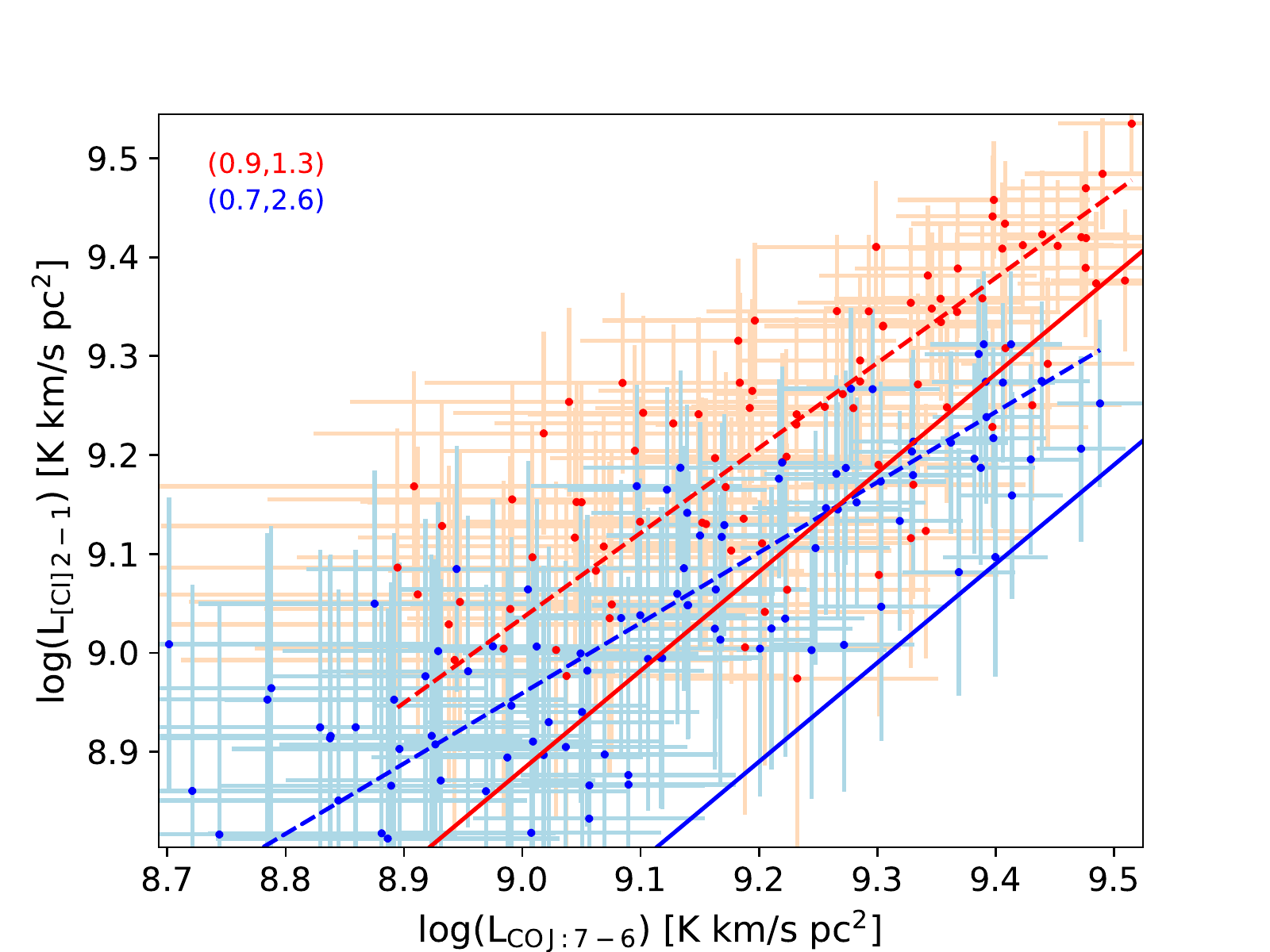}  \\    
   \hspace*{-7cm}\includegraphics[scale=0.45]{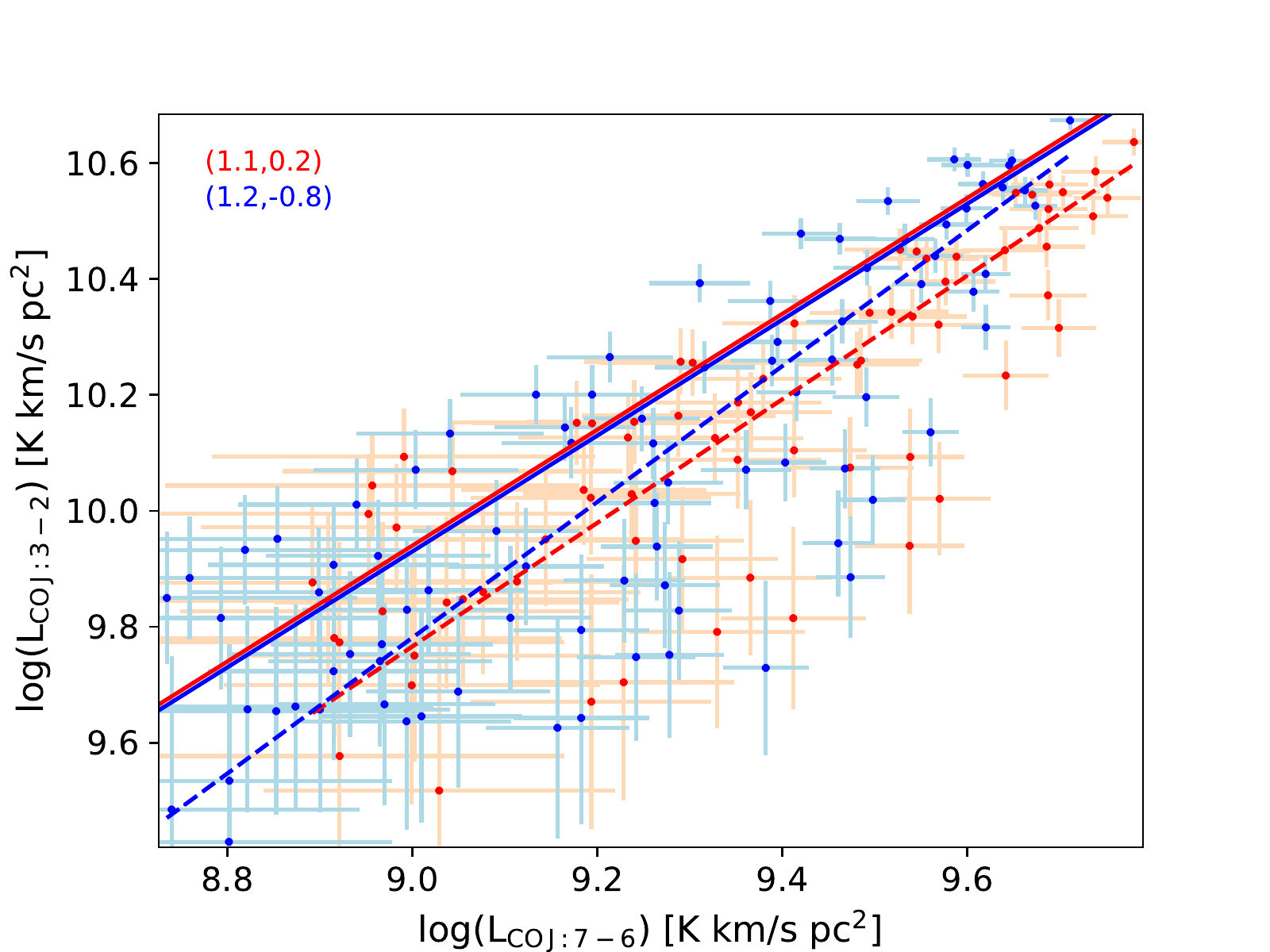}\\ \vspace{-5.55cm}
   \hspace*{ 7cm}\includegraphics[scale=0.45]{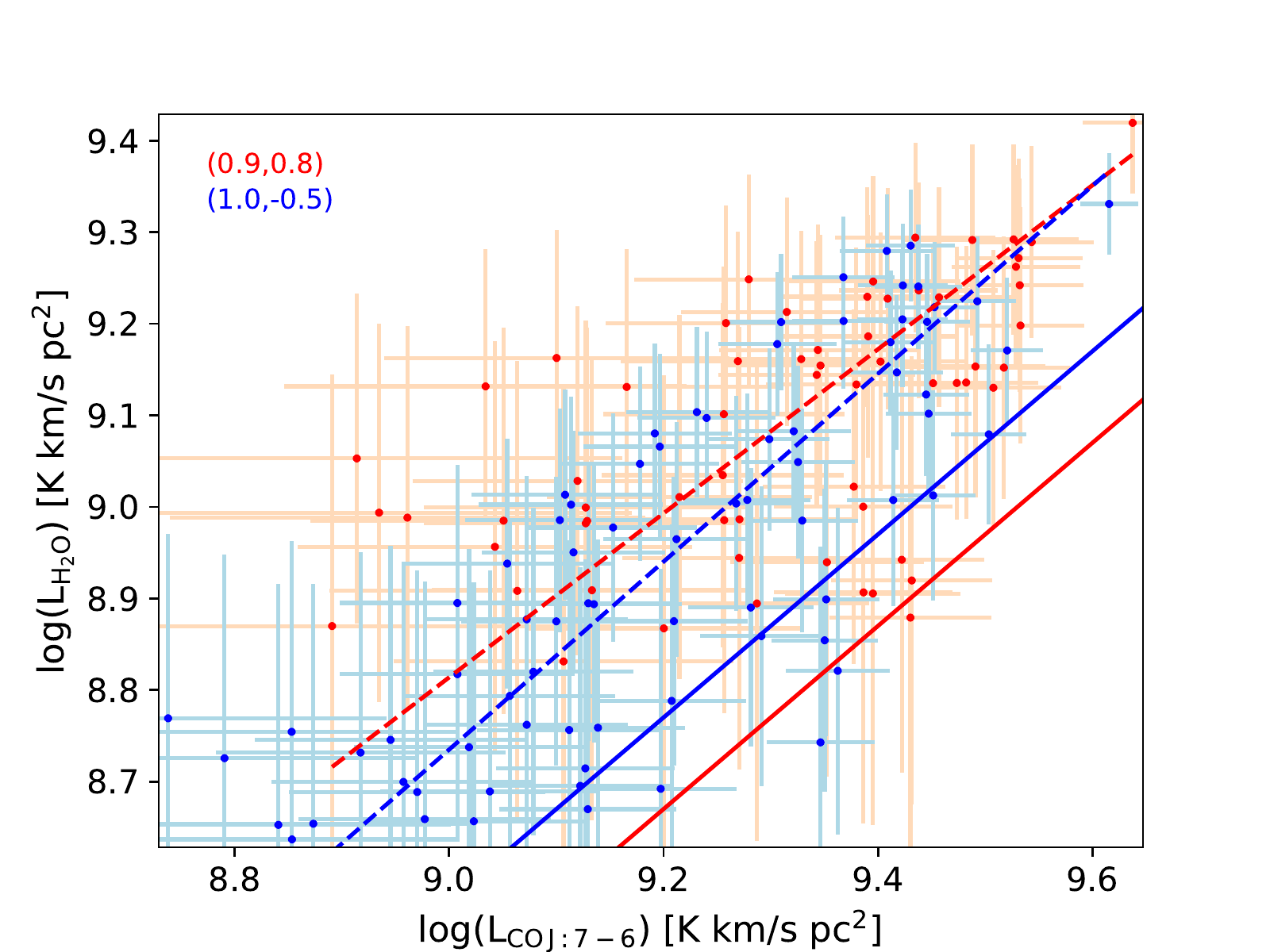} \\
   \caption{As in Fig.~\ref{figratecont}, but
   here we compare the emission line luminosities ($L'_{\mathrm{line}}$ from eqn.~3) of 
   CO J:3--2, CO J:7--6, [C\,{\sc i}] 2-1, and H$_2$O. In each panel the solid lines in the corresponding colour show
   the prediction if the galaxy integrated line flux ratio was valid over all individual apertures.}
   \label{figratelines}
   \end{figure*}

Resolved line luminosities for CO J:3--2, CO J:7--6, [C\,{\sc i}] 2--1, and H$_2$O
were calculated as in Section~\ref{secglobal} but instead of galaxy- and
velocity-integrated line fluxes, we use velocity-integrated fluxes extracted in apertures equivalent to the synthesised beam size (FWHM), 
and spaced by half a synthesised beam.  As before, for the two line emission two maps involved, the higher resolution map was
convolved with a Gaussian in order to degrade the resolution to that of the lower resolution map. Note that latter three lines have
very similar resolutions; however, when comparing CO J:3--2 and CO J:7--6, the synthesised beam
sizes are highly different.
Thus, extra caution is required when comparing CO J:3--2 and CO J:7--6, and thus interpreting the CO
ladder excitation.
Figure~\ref{figratelines} shows the relationship between the kpc-scale resolved \co7-6\ 
luminosity and the luminosities of the CO J:3--2, \cishort, and H$_2$O emission lines. The best
fits to the data from individual apertures (dashed lines) and the prediction of the galaxy-integrated ratios (solid lines)
are shown for easy comparison. The galaxy-integrated \cishort\ to CO ratios are driven by
a few dominant apertures: the resolved ratios of \cishort\ to CO J:3--2 show a dependence which is
significantly sub-linear.
The resolved ratios of \cishort\ to CO J:7--6 are also sub-linear, though not as much as in the case of using CO J:3--2: 
the resolved \cishort\ 
luminosity is typically $\sim$0.1 dex (W) or 0.2 dex (T) lower than the value which would be estimated
by the galaxy-integrated line ratio. This is loosely consistent with our finding that the \cishort\ line emission
is slightly  more extended (at relatively low fluxes) than CO J:7--6. Further, recall that the 
[C\,{\sc i}] 1--0 line is typically used to trace molecular (or CO) gas. Using the \cishort\ line requires assuming an
excitation temperature (where a higher temperature implies more \cishort\ emission per unit C\,{\sc i} mass). Thus converting
a CO J:7--6 to \cishort\ ratio to a CO to C\,{\sc i} gas mass ratio requires information on temperature and density, and 
the sub-linear relationship we see may thus be mainly an effect of physical conditions in the gas.

The resolved CO J:7--6 to CO J:3--2 ladder also shows deviations from the galaxy-integrated value, but
the relatively large error bars, and the highly mismatched synthesised beams of the two maps makes interpretation 
of this figure  difficult. The relationship between CO J:7--6 and water vapour 
shows a lot of scatter as expected, though surprisingly the best fit to the scattered points gives an
almost  linear relationship between the two luminosities.

Thus overall, \cishort\ is a relatively good, almost linear, tracer of  dense/warm (as traced by CO J:7--6) molecular gas in both W and T,
though as discussed here and in previous sections, there are mismatches in the morphology of the C\,{\sc i} and CO J:7--6 in both W and T.
Due to assumptions of the CO ladder and excitation temperature of C\,{\sc i}, we are unable to evaluate the relationship between C\,{\sc i} and CO mass.

\subsection{A resolved (2\,kpc-scale) warm/dense gas Schmidt-Kennicutt Relationship}

   \begin{figure*}
   \centering
   \hspace*{-12cm}\includegraphics[scale=0.4]{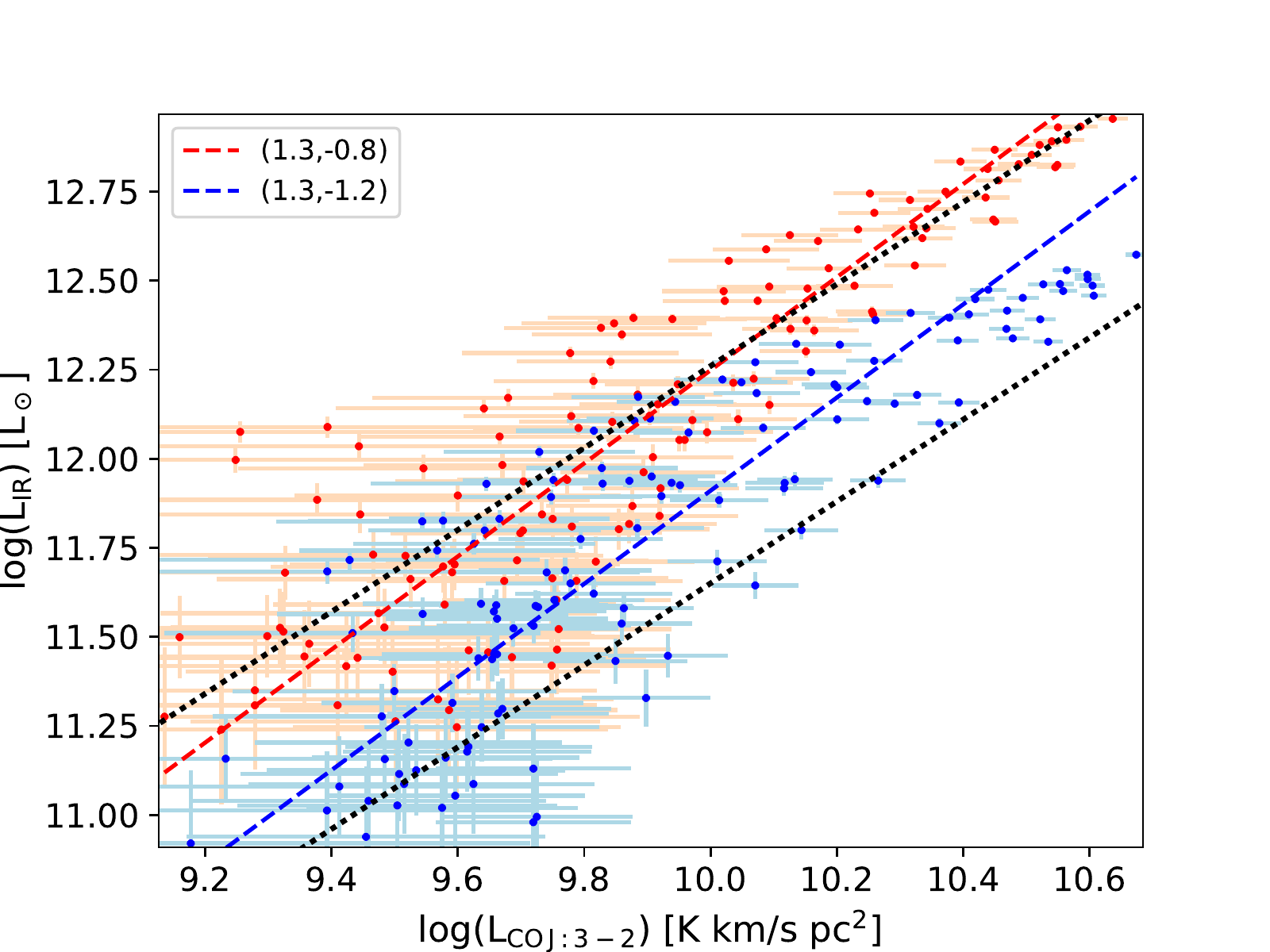}\\ \vspace{-4.92cm}    
   \hspace*{0.1cm}\includegraphics[scale=0.4]{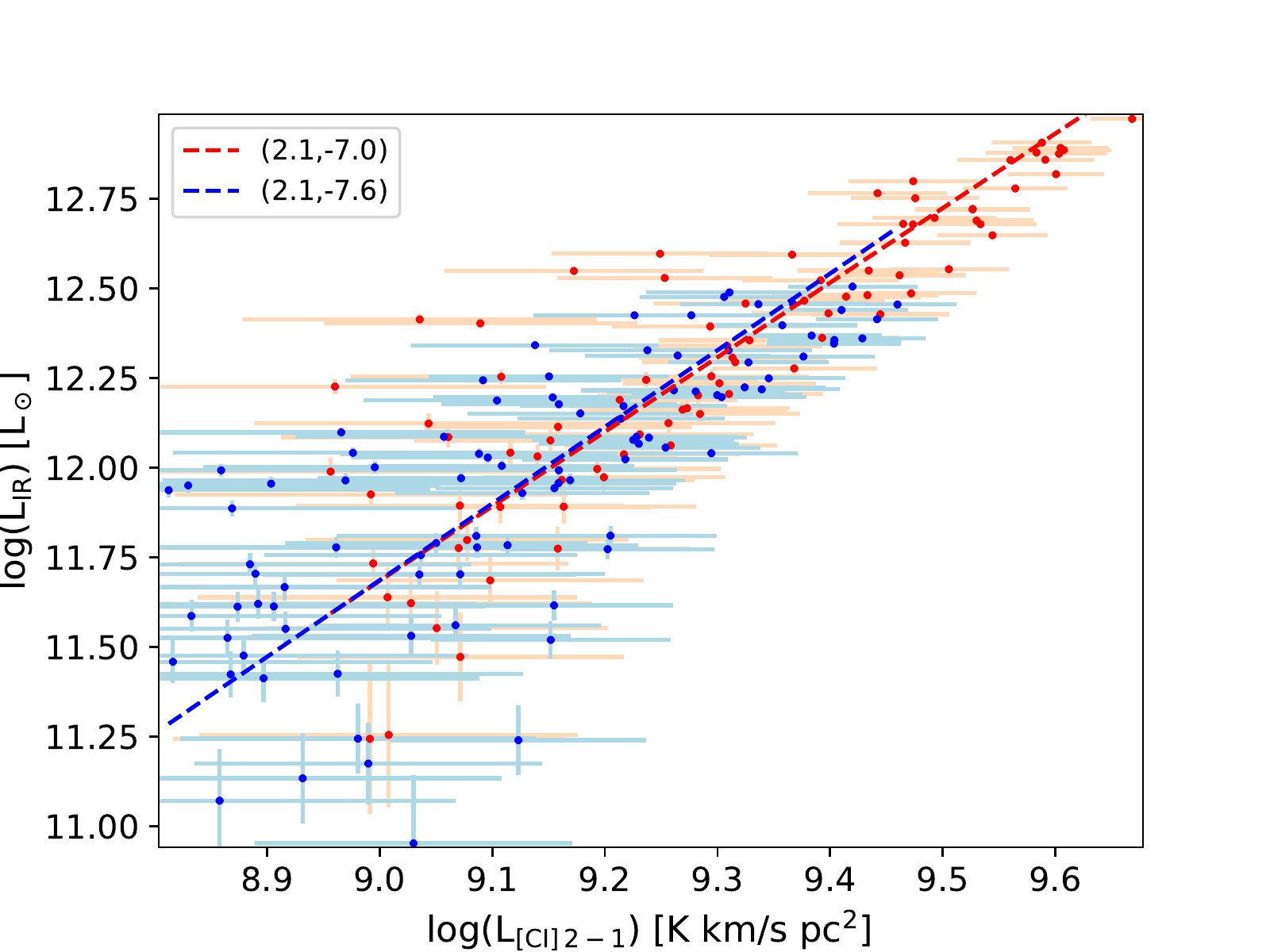}\\ \vspace{-4.92cm} \hspace*{12cm}
   \includegraphics[scale=0.4]{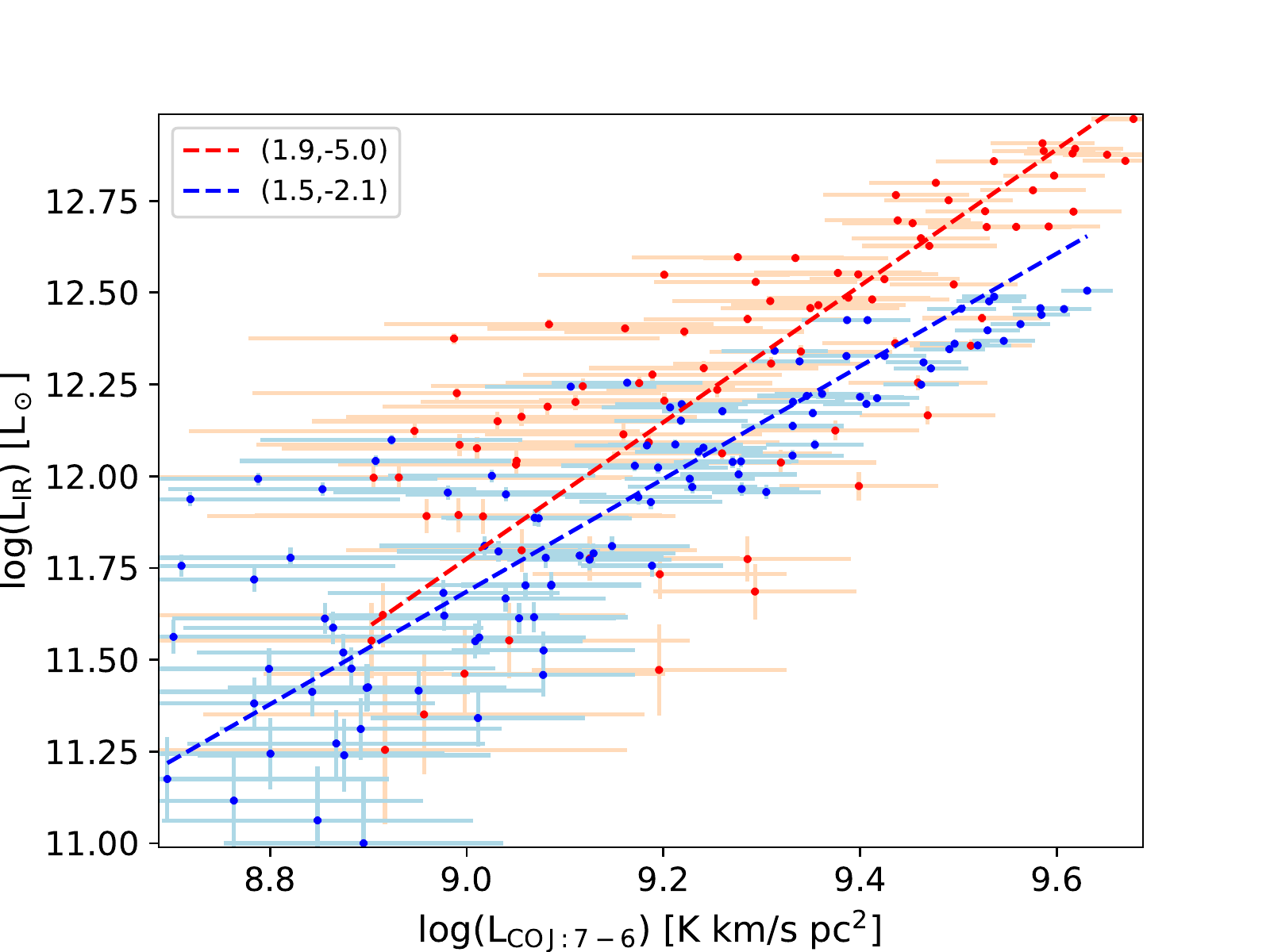}\\
   \caption{As in Figs~\ref{figratecont} and \ref{figratelines}, 
   but here we show the correlation between the  
   resolved IR luminosity (estimated from the resolved rest-frame 1160 GHz continuum map; see text) and the   
   luminosities ($L'_{\mathrm{line}}$) of the CO J:7--6, [C\,{\sc i}], and CO J:3--2 lines. In the left 
   panel the black dashed lines show the galaxy-integrated relationships derived by \citet{dadet10} and \citet{genet10}
   for normal star-forming (lower line) and `starbursting' (upper line) galaxies; here we assumed 
   thermally excited gas, i.e. $L'_{\mathrm{line}_{\rm CO J:3--2}}$/$L'_{\mathrm{line}_{\rm CO J:1-0}}$ = 1.}
   \label{figratelir}
   \end{figure*}
 

   \begin{figure}
   \centering
   \hspace*{-0.2cm} \includegraphics[scale=0.6]{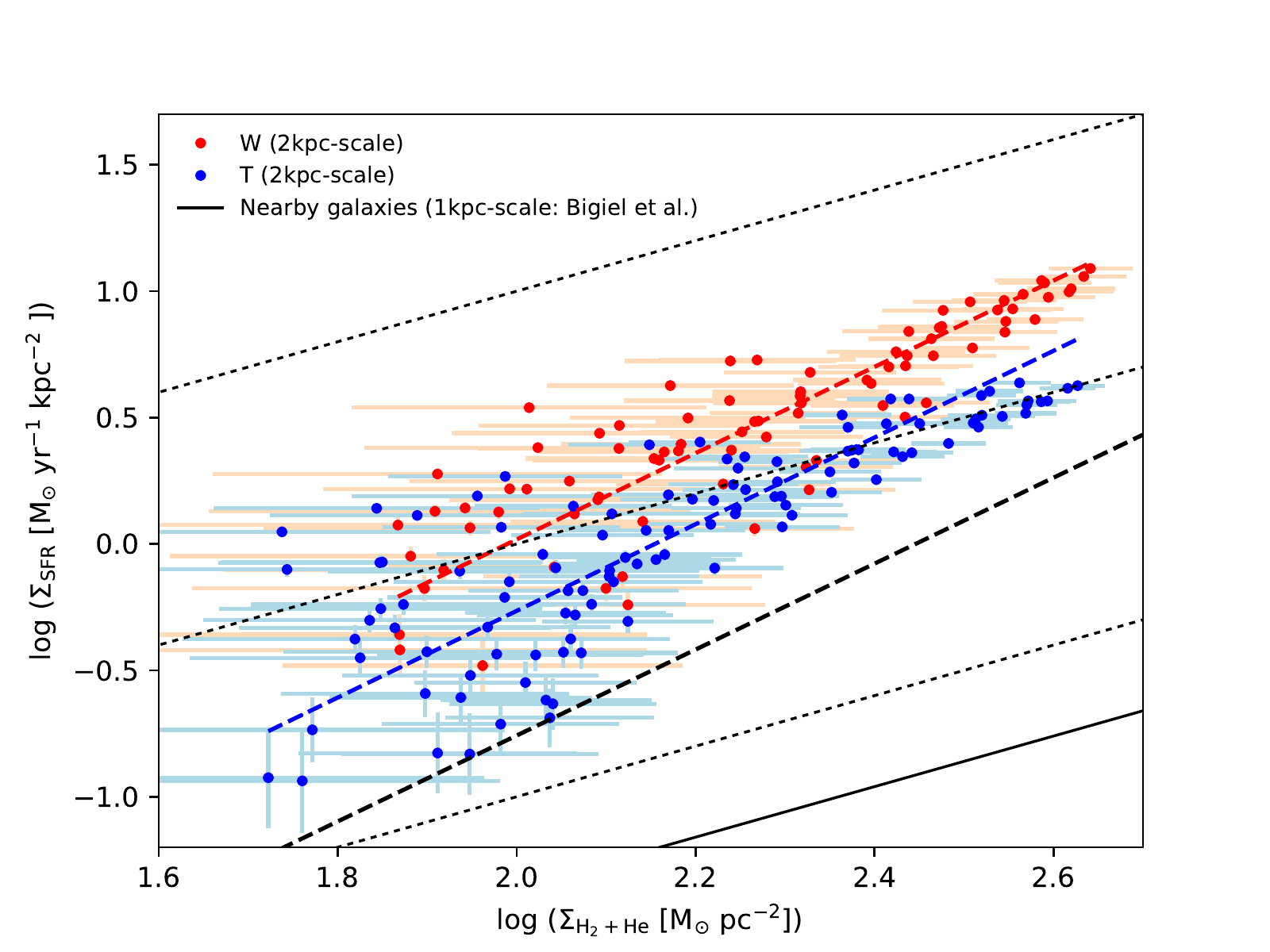}\\
   \caption{The resolved Schmidt-Kennicutt relationship equivalent for warm/dense molecular gas, i.e.\ the relationship between the  
   surface densities of star formation and warm/dense molecular gas (as traced by CO J:7--6), for components W (red points) and T (blue points). 
   Each data point was calculated over an aperture equivalent to the (FWHM) synthesised beam (roughly
   2~kpc),  with a spacing of half a synthesised
   beam between points, i.e.\ roughly a quarter of the points are independent measurements.
   The star-formation rate is estimated from the resolved rest-frame 
   1160~GHz continuum map and the molecular gas 
   mass from the resolved CO J:7--6 emission line map and standard conversions of this to CO J:1--0 and thus gas mass (see text).
   The solid black line at the bottom-right is the kpc-scale resolved (cold gas, i.e.\ as derived from CO J:1--0) 
   SK relationship in nearby 'normal'
   galaxies \citet{biget11}. The dotted lines delineate gas exhaustion times of 
   (top to bottom) 0.1, 1, and 10 Gyr 
   and the blue and red dashed lines show illustrative SK relationships with slopes $1.72 \pm 0.24$ and $1.71 \pm 0.26$,
   respectively (see text).
   For component T we used the observed (image plane) linear sizes corresponding to the synthesised beam;
   if the image plane is spatially stretched
   due to lensing then the individual points will move towards the top right, parallel to
   the dashed lines (constant gas exhaustion times) in the figure, further differentiating them from 
   the data points for component W.}
   \label{figfs-k}
   \end{figure}
   
Given that we have both resolved estimators of molecular gas mass (CO J:3--2, CO J:7--6 and \cishort) and
resolved estimators of the IR luminosity, thus SFR (the rest-frame 1160~GHz continuum
luminosity), it is highly relevant to test the relationship between the two both as global quantities
and as surface densities.

We first outline the approximations and estimations required to be made. First, we use the values derived
by I13 for the galaxy-integrated molecular
gas mass (derived from CO J:1--0 imaging) and total (rest-frame 8--1000 \micron m) IR luminosity (from detailed SED fitting) for both W and T.
Our resolved CO J:7--6 (or \cishort) and rest-frame 1160~GHz continuum images 
are used to distribute these
global values over individual resolved apertures, i.e.\ a relative, and not absolute, distribution. 
In other words, we assume that the rest-frame  1160 GHz  (260\micron; close to the IR peak; see I13) luminosity 
closely traces -- in a relative sense -- the distribution of the IR luminosity, and that the  CO J:7--6 map closely traces 
-- in a relative sense -- the distribution of molecular gas.
Once more,  as for the continuum maps in Fig.~\ref{figratecont}, we use apertures with size equal to the ALMA 
synthesised beam in the CO J:7--6 and \cishort\ maps, spaced by half a beam width.
The resulting conversion factors between total continuum luminosity at 1160GHz (in units of $\mathrm{L_{\sun}\, Hz^{-1}}$)
and total $\mathrm{L_{IR}}$ (in units of $\mathrm{L_{\sun}}$) for the galaxies W and T are 
\begin{equation}
\mathrm{L_{IR} = \frac{2.414\times 10^{-14}}{[Hz^{-1}]} L_{1160GHz}}
\end{equation}
and
\begin{equation}
\mathrm{L_{IR} = \frac{2.256\times 10^{-14}}{[Hz^{-1}]} L_{1160GHz}},
\end{equation}
respectively. The conversions are different at the 10\% level as each galaxy has a different global IR (from I13) to L$_{1160\rm GHz}$ 
(this work) ratio, reflecting the different shapes of their SEDs.

We then use this value of L$_\mathrm{IR}$ to estimate the resolved SFR using the 
relationship of \citet{ken98}, which assumes a Salpeter IMF:
\begin{equation}
\mathrm{SFR \left[M_\sun yr^{-1} \right] = \dfrac{L_{IR}}{5.8 \times 10^9 [L_\sun]}}
\end{equation}
The SFR surface density, when required, is calculated using the linear size of the synthesised beam.

For molecular gas masses, we have three alternatives to derive the gas mass in each aperture, all
of which use the galaxy-integrated CO mass derived by I13 together with an estimated resolved map of 
the CO J:1--0 emission:
\begin{enumerate}
    \item using the resolved CO J:7--6 flux map and the galaxy-integrated CO J:7--6 to CO J:1--0 ratio 
    to derive a CO J:1--0 luminosity map 
    (i.e.\ assuming that the galaxy-integrated CO ladder is valid for each resolved aperture of a given galaxy);
    \item using the resolved CO J:7--6  luminosity map, together with the linear fit 
    to the relationship  between the resolved (see Fig.~\ref{figratelines}) CO J:7--6  and CO J:3--2
    line luminosities, to derive an estimated resolved CO J:3--2 luminosity map. 
     This is then converted to
    a CO J:1--0 luminosity map assuming thermally excited gas, as evidenced in Fig.~\ref{figcoladderT}.
    This method should be more reliable than (a) above since we know that the CO J:7--6 to CO J:3--2
    ratio varies over the galaxy (Fig.~\ref{figratelines}) while the  CO J:3--2 to CO J:1--0
    ratio is more likely to be constant in SMGs;
    \item using the resolved \cishort\ luminosity map, together with the galaxy-integrated \cishort\ to CO J:1--0 
    luminosity ratio to create an estimated map of the CO J:1--0 luminosity.
\end{enumerate}

All three alternatives give consistent results, so here we show the results of using method (a) above.
We use the galaxy-integrated CO 'ladder' of the individual components (see Fig.~\ref{figcoladderT}) 
to obtain the following for component W: 
$S_{\mathrm{CO\, J:7-6}}/S_{\mathrm{CO\, J:1-0}} \sim 5.52 \pm 1.64$, 
thus $L'_{\mathrm{CO}} = (8.87  \pm 2.57) \times L'_{\mathrm{CO\, J:7-6}}$ 
and $M_{\mathrm{mol}} = (7.10 \pm 2.06)  \times L'_{\mathrm{CO\, J:7-6}}$. 
For the other components, the equivalent conversions are as follows.
Component T: $M_{\mathrm{mol}} =( 8.06  \pm 1.77) \times L'_{\mathrm{CO\, J:7-6}}$, 
Component M : $M_{\mathrm{mol}} = 4.85  \pm 0.87 \times L'_{\mathrm{CO\, J:7-6}}$, 
and Component C: $M_{\mathrm{mol}} = 6.06 \pm 0.13  \times L'_{\mathrm{CO\, J:7-6}}$. 

Fig.~\ref{figratelir} shows the dependence of the resolved IR luminosities on the 
CO and \cishort\ emission line luminosities. When using CO J:3--2 on the $x$-axis (as a proxy for
gas mass), both W and T follow
relatively well the (galaxy-integrated) relationships (slope $\sim1.2$) found by \cite{dadet10} and \citet{genet10} \citep[cf.][]{ivison11}
for nearby and high-redshift star-forming galaxies: in this scenario, most apertures in W are
consistent with being on the 'luminous starburst' sequence while most apertures in T lie between the supposed sequences
for luminous starbursts and 'normal star-forming' galaxies. 
The dependence of the IR luminosity on warm and dense gas mass (using CO J:7--6 or \cishort\ as a proxy) is 
significantly steeper than the case of CO J:3--2: the slope of the relationship is now between $\sim1.5$ and $\sim2.5$, and W and 
T are relatively indistinguishable in these plots. This is unusually steep given that previous works have found
that the dependence of IR luminosity on dense gas mass (via HCN J:1--0) is linear when galaxy-integrated
quantities are taken into account \citep{gaosol04,shiet17,oteet17},
though this relationship is not necessarily universal: \citet{liuet16} find that the relationship is
sub-linear in resolved clumps in the Galaxy. 
Since we are tracing both warm and dense gas with CO J:7--6, rather than only dense gas as in the case of
HCN J:1--0, the steep dependence is potentially an effect of varying temperature and density across the apertures, i.e.\ in higher temperature dense regions
the dependence of CO J:7--6 luminosity on SFR is steeper (the sub-mm emission would not be expected to change by a large
factor, given the relatively uniform sub-mm spectral slopes seen over the galaxy). The slopes seen here are not the result of systematic
differences in the SNR and beam shape in the individual maps. Constructing the same figures but using regions between the galaxies (i.e.\ `empty
sky' in the ALMA maps) shows that the data cluster around [0,0] without any noticeable systematics.

Converting the relationships between IR luminosity and gas (CO or \cishort) luminosity into
a resolved Schmidt-Kennicutt relationship is trivial in our case since all data have the
same aperture size: the conversion to gas mass (instead of luminosity) and surface density
(division by linear aperture size) only involves a change in the axis units. Nevertheless,
for easier understanding, we show, in a new figure (Fig.~\ref{figfs-k}) the 
resolved Schmidt-Kennicutt relationship ($\mathrm{\Sigma_{SFR} \propto \Sigma_{H_2}^N}$)
for components W and T in the case of using CO J:7--6 to predict the gas surface density, i.e.\
using a warm/dense gas tracer, instead of a cold molecular gas tracer like CO J:1--0. 
Not surprisingly the resolved warm/dense gas
SK relationship in W and T are $\sim1$ dex higher than the cold SK relationship seen in nearby galaxies
(solid back line in the figure). Our data cover similar parameter space as those of the few
previous determinations of the resolved SK relationship in high-redshift galaxies \citep{thomet15,freet13}.
Component W shows slightly higher star-formation efficiencies than T,
with most apertures within 0.5 dex of a gas exhaustion time of $\sim 1$~Gyr. Not surprisingly,
the highest efficiencies are in the nuclei.

SK-type power-law fits to the data for W and T in Fig.~\ref{figfs-k} result in significantly different slopes depending on the
specific fitting routine used. This is due both to the large spread of data points in the bottom left quadrant of
the figure, and the fork shape seen in the red points at higher gas surface densities. 
Fits to the W (red) data points typically give slopes of $1.71 \pm 0.26$, while fits to the T (blue) data points give slopes
of $1.72 \pm 0.24$. In all cases the slopes are greater than 1.5. The intercepts (log A in the SK equation) are relatively invariant with values of $\sim -3.4$ and $\sim -3.7$ for W and T, respectively.
Thus, in both W and T, the resolved warm/dense gas SK relationship follows a power law with slope $\sim1.7$ (red and blue dashed line), significantly
steeper than seen previously in cold gas SK relationships in low- or high-redshift galaxies. 
 
How reliable is this finding of a steep slope in the warm/dense gas SK relationship? The variations in the slopes  discussed above result 
from the measurement and linear fitting errors only, without taking systematic uncertainties into account.  It is thus relevant to examine the 
systematic uncertainties
in deriving the two quantities used in the figure: SFR and gas mass. The total IR luminosity is the most reliable estimator
of SFR \citep[e.g.,][]{ken_ev12} -- more reliable than, e.g., an SFR estimation from the H$\alpha$ emission line 
luminosity as used by \citet{freet13}, and we emphasise that  
the galaxy-integrated IR luminosities of W and T were derived from the comprehensive SED analysis of I13. 
Ideally, the SFR is derived from a combination of IR and UV luminosities (2.2 L$_{\rm UV}$
+ L$_{\rm IR}$), where the UV luminosity can be obtained by, e.g., 1.5 $\nu$ L$_\nu$ at 2800\AA\ \citep{belet05}. 
Using the galaxy-integrated rest-frame-UV (2600\AA) fluxes listed in I13 for W and T (see their Table 1 and Fig.~2),
L$_{\rm UV}$  comes out to $\sim1$ to 2\% of L$_{\rm IR}$ for both W and T. The UV luminosity is thus unlikely
to cause significant differences in any aperture of W and T and is thus safe to ignore. 
An additional systematic in the derivation of our resolved SFRs is the use of our rest-frame 250\micron\ 
(which is longward of the IR peak) ALMA map to distribute the total SFR into individual apertures. 
For dust at a known single temperature, 
which appears to be the case in component W (Fig.~\ref{figratecont} and related discussion),
the 250\micron\ flux is a good tracer of the IR luminosity in an absolute and relative sense \citep[e.g.,][]{oreet17}.
For multiple dust temperatures, as is likely the case in component T (Fig.~\ref{figratecont} and related discussion), 
systematics would be expected. 
If the hotter dust component(s) is due to star formation, then we are underestimating the SFR in the nuclear regions
(so that the true SK relationship in T would have an ever steeper slope); if the hotter dust component(s) is due to an AGN (for which deep {\it XMM-Newton} imaging provides no evidence -- Ivison et al., in preparation)
then we are over-estimating the SFR in the nuclear regions, so that the intrinsic SK relationship in T is shallower than $\sim 2$.
Resolved maps of the IR continuum shortward of the IR peak would be required to resolve this.
Thus, in comparison to equivalent
relevant studies at high redshift, our resolved SFR estimations are among the most reliable.

For resolved gas masses, the systematics are likely larger, though here too we have the advantage of having multiple
tracers of molecular gas (CO and \cishort). The galaxy-integrated gas mass has been derived in I13 from CO J:1--0 imaging
and an \alphaco\ of 0.8. This total mass has been distributed into resolved apertures using our 
resolved CO J:7--6 or \cishort\ maps. A varying \alphaco\ within an individual galaxy, or a highly varying CO ladder over the
individual apertures, would be required to
change the slope of the resolved warm/dense gas SK relationship that we find.

\section{Summary}   

We present new resolved imaging of the continuum and emission lines in the four known components of the
binary HyLIRG, HATLAS\,J084933.4+021443 at $z=2.41$. The new imaging allows us to further extend the
comprehensive characterisation of this system presented in I13.

Our main results are:
\begin{enumerate}
\item All four component galaxies of \src\ (W, T, M and C) are spatially ($\sim$0\farcs3 or 2.5~pc) resolved in 
CO J:7--6, \cishort, and the sub-mm continuum.  Components W and T are also resolved in the \water line.
\item The internal kinematics of CO and \cishort\ are clearly dominated by rotation. While the kinematics of the \water\ line in W could be
consistent with rotation, the \water\ line in T shows different, disturbed, kinematics.
\item Component T is significantly more extended, in gas and continuum, along its kinematic minor axis, 
likely the result of spatial magnification due to lensing.
\item Spatially resolved sub-mm SEDs  show that component W is well fit with greybody emission from dust at a single temperature over the  full extent of the galaxy, but component T requires both an additional component of significantly hotter nuclear dust and additional sources of emission in the mm.
\item We confirm that, in a rough sense, the \cishort\ line can be used as a warm/dense molecular gas tracer in such extreme systems.
However, there are several caveats: an excitation temperature assumption is required for \cishort,
the resolved dependence of \cishort\ on CO J:7--6 is  slightly flatter than linear, and the morphology of  \cishort\ and CO J:7--6 are 
different in W and T, with the \cishort\ slightly more extended and the CO J:7--6 emission. 
\item We obtain an exquisite and unprecedented 2.5~kpc-scale resolved warm/dense gas Schmidt-Kennicutt relationship for
components W and T, i.e.\ an SK relationship in which CO J:7--6 -- a tracer of warm/dense gas -- is used to trace molecular gas mass spatially.  
Typical gas exhaustion times for all apertures in W are within 0.5~dex of 1~Gyr; 
in T the gas exhaustion timescales are about 0.4~dex slower than those in W. 
Both W and T follow a  resolved warm/dense gas SK relationship with power law $n\sim 1.7$, significantly steeper than the $n\sim 1$ found 
previously, using `cold' (i.e.\ as traced by, e.g., CO J:1--0) molecular gas, in nearby normal star-forming galaxies.

\end{enumerate}


NN acknowledges funding support from CONICYT CATA/BASAL PFB-06/Etapa II, Fondecyt 1171506 and PIA ACT172033. 
GO acknowledges the support provided by CONICYT(Chile) through FONDECYT postdoctoral research grant no 3170942. RJI acknowledges support from ERC in the form of Advanced Grant, COSMICISM, 321302.

%

\end{document}